\documentclass[12pt, a4paper]{article}
\usepackage{amsmath,amssymb,bm,epsfig,afterpage}
\usepackage{cite}
\usepackage{bbm}
\usepackage{here}
\usepackage{color}
\usepackage{colortbl}
\usepackage{xcolor}
\usepackage{tabularx}
\usepackage{subcaption}
\usepackage{float}
\usepackage{cancel}
\usepackage{graphicx}
\usepackage{listings}
\usepackage{longtable}
\usepackage[utf8]{inputenc}

\makeatletter
    
    \@addtoreset{equation}{section}
  \makeatother

\definecolor{Black}{cmyk}{0.99,0.99,0.99,0.99}
\definecolor{Orange}{cmyk}{0,0.61,0.87,0}
\definecolor{JungleGreen}{cmyk}{0.99,0,0.52,0}
\definecolor{OliveGreen}{cmyk}{0.64,0,0.95,0.40}
\definecolor{Brown}{cmyk}{0,0.81,1,0.60}
\definecolor{RoyalBlue}{cmyk}{0.71,0.53,0,0.12}
\definecolor{Gray}{cmyk}{0,0,0,0.40}
\definecolor{LightPink}{cmyk}{0.0,0.25,0,0}
\definecolor{LLightPink}{cmyk}{0.0,0.10,0,0}
\definecolor{LightBlue}{cmyk}{0.25,0,0,0}
\definecolor{LightGray}{cmyk}{0,0,0,0.2}

\renewcommand{\thefootnote}{\fnsymbol{footnote}}
\newcommand{\vev}[1]{{\langle{#1}\rangle}}

\newcommand{\val}[2]{{#1}\times 10^{#2}}
\newcommand{\br}[2]{\text{BR}({#1}\to{#2})}
\newcommand{\mhc}[1]{ {\left({#1}\right)^\dagger} }   
\newcommand{\ol}[1]{\overline{#1}}
\newcommand{\ti}[1]{\tilde{#1}}
\newcommand{{\lag}}{\mathcal{L}}
\newcommand{\mcal}{\mathcal{M}}
\newcommand{\Mcal}{\mathcal{M}}

\newcommand{\ket}[1]{|{#1}\rangle}
\newcommand{\bra}[1]{\langle{#1}|}

\newcommand{\re}[1]{\text{Re} \left( {#1} \right)}
\newcommand{\abs}[1]{\left|{#1}\right|}

\newcommand{\la}{\lambda}

\newcommand{\eps}{\epsilon}
\newcommand{\ups}{\upsilon}
\newcommand{\gam}{\gamma}
\newcommand{\sig}{\sigma}
\newcommand{\zv}{\boldsymbol{z}}
\newcommand{\nv}{\boldsymbol{n}}

\newcommand{\gp}{{g^\prime}}
\newcommand{\Zp}{{Z^\prime}}

\newcommand{\tM}{\tilde{M}}
\newcommand{\ty}{\tilde{y}}
\newcommand{\tla}{\tilde{\la}}

\newcommand{\define}{:=}
\newcommand{\Hcal}{\mathcal{H}}
\newcommand{\Lcal}{\mathcal{L}}
\newcommand{\Ncal}{\mathcal{N}}

\newcommand{\Ocal}{\mathcal{O}}
\newcommand{\vz}{\boldsymbol{z}}
\newcommand{\vn}{\boldsymbol{n}}

\newcommand{\hg}{\hat{g}}
\newcommand{\hY}{\hat{Y}}

\newcommand{\Pfb}{P_{\ol{5}}}
\newcommand{\Uop}{\mathrm{\U1'}}
\newcommand{\id}{\mathbbm{1}}

\newcommand{\order}[1]{\mathcal{O}\left( {#1}\right)}
\newcommand{\bsll}{b\rightarrow s\ell^+\ell^-}
\newcommand{\MMbar}[2][]{ {#2}_{#1} \text{-} \ol{#2}_{#1}  }
\newcommand{\Heff}{\mathcal{H}_\text{eff}}

\newcommand{\bzero}{{\boldsymbol{0}}}
\newcommand{\U}[1]{\ensuremath{\mathrm{U}(#1)}}

\newcommand{\SU}[1]{\ensuremath{\mathrm{SU}(#1)}}

\allowdisplaybreaks[3]
\newcolumntype{Y}{&gt;{\centering\arraybackslash}X}

\usepackage[colorlinks=true, linkcolor=OliveGreen, 
citecolor=RoyalBlue,
urlcolor=RoyalBlue]{hyperref}
\definecolor{darkgreen}{HTML}{109930}

\newcommand{\AT}[1]{\textcolor{darkgreen}{#1}}

\begin{document}

\begin{titlepage}

\begin{flushright}
{\tt
}
\end{flushright}

\vskip 1.35cm
\begin{center}

{\Large
{\bf
Complete Vector-like Fourth Family with 
\texorpdfstring{$\boldsymbol{\mathrm{U(1)^\prime}}$}{$\mathrm{U(1)}^\prime$}: \\
A Global Analysis
}
}

\vskip 1.5cm

Junichiro~Kawamura$^{a,b,}$\footnote{%
\href{mailto:kawamura.14@osu.edu}{\tt kawamura.14@osu.edu}},
Stuart~Raby$^{a,}$\footnote{%
\href{mailto:raby.1@osu.edu}{\tt raby.1@osu.edu}},
and
Andreas Trautner$^{c,}$\footnote{%
\href{mailto: trautner@mpi-hd.mpg.de}{\tt trautner@mpi-hd.mpg.de}}
\vskip 0.8cm

{\it $^a$Department of Physics, Ohio State University, Columbus, Ohio
 43210, USA}
\\[3pt]
{\it $^b$Department of Physics, Keio University, Yokohama 223-8522, Japan}
\\[3pt]
{\it $^c$Max-Planck-Institut f$\ddot{\text{u}}$r Kernphysik, Saupfercheckweg 1, 69117 Heidelberg, Germany}
\\[3pt]

\date{\today}

\vskip 1.5cm

\begin{abstract}
In this paper we present an in-depth analysis of a recently proposed Standard Model extension
with a complete fourth generation of quarks and leptons, which are vector-like with respect to the Standard Model gauge group
and charged under a new spontaneously broken vector-like $\mathrm{U(1)^\prime}$ gauge symmetry. 
The model is designed to explain the known muon anomalies, i.e.\ the observed deviations from Standard Model
predictions in the anomalous magnetic moment of the muon, $\Delta a_\mu$, and in $b \rightarrow s \ell^+ \ell^-$ processes. 
We perform a global $\chi^2$ analysis of the data with $65$ model parameters 
and including $98$ observables.
We find many points with $\chi^2$ per degree of freedom $\leq 1$. 
The vector-like leptons and the new heavy $Z^\prime$ are typically much lighter than a TeV and would, thus, be eminently visible at the HL-LHC.
\end{abstract}

\end{center}

 \clearpage
 \tableofcontents
 \thispagestyle{empty}
 \addtocontents{toc}{\protect\thispagestyle{empty}}
\end{titlepage}

\renewcommand{\thefootnote}{\arabic{footnote}}
\setcounter{footnote}{0}

\section{Introduction}
In the search for physics beyond the Standard Model (SM) one certainly looks for new particles produced at the LHC.
Absent any new discoveries, one then considers experimental discrepancies with SM predictions. There will always be $2 \sigma$ or $3 \sigma$ discrepancies, 
some of which are real and some of which are just statistical fluctuations. In an attempt to distinguish these two possibilities,
one might focus on a cluster of discrepancies which can all be resolved with the same new physics. This is what we have done in a previous
letter \cite{Kawamura:2019rth}.  We have shown that the muon anomalies associated with the anomalous magnetic moment of the muon, $\Delta a_\mu$,
and the angular and lepton non-universality anomalies in $b \rightarrow s \ell^+ \ell^-$ decays can simultaneously be resolved by the addition to the SM
of a complete vector-like (VL) family of quarks and leptons together with a new $\U1^\prime$ gauge symmetry carried only by the new VL states. 
VL leptons are introduced for $\Delta a_\mu$ 
in Refs.~\cite{Czarnecki:2001pv,Kannike:2011ng,Dermisek:2013gta,Lindner:2016bgg}. 
The $\bsll$ anomaly is addressed by introducing a $\Zp$ boson~\cite{Altmannshofer:2014cfa,Crivellin:2015mga, King:2017anf,Falkowski:2018dsl}  
and new particles which induce box diagram contributions~\cite{Gripaios:2015gra,Arnan:2016cpy,Grinstein:2018fgb,Arnan:2019uhr, Chiang:2017zkh,Cline:2017qqu,Kawamura:2017ecz,Barman:2018jhz,Cerdeno:2019vpd}.   
Both anomalies can be explained simultaneously in models with VL fermions and a $\Zp$ boson~\cite{Allanach:2015gkd,Altmannshofer:2016oaq,Megias:2017dzd,Raby:2017igl,Darme:2018hqg}.

In the present paper, we perform a global $\chi^2$ analysis of these phenomena with the addition of all SM processes that might feel the effects of mixing
between the VL family and the SM families. We find many solutions with $\chi^2/N_{\mathrm{d.o.f.}} \leq 1$. 
Moreover, the model is highly testable
via both direct observation of new physics at the LHC or via the improved analysis of SM processes. 
For example, 
precision measurements of $K$-$\bar K$ and $B_d$-$\bar B_d$ mixing may be sensitive to the new physics.  
Finally, the tight observational constraints on the $\mu \rightarrow e \gamma$ branching ratio hinders a 
simultaneous fit to $\Delta a_\mu$ and $\Delta a_e$. 
We can only fit one but not both, and we choose to fit the former.

The paper is organized as follows. In Section \ref{model} we briefly review the details of the model and describe the mass mixing between the SM and VL states.
In Section \ref{observables} we provide the theoretical formulae for calculating the 98 observables in our analysis. In Section \ref{results}
we present the results of the $\chi^2$ analysis with many plots illustrating the range of VL masses and the quality of the fits.  
In addition we choose four ``best fit points'' to illustrate some of the new physics processes which are, 
in principle, observable at the LHC. Finally, in Section \ref{summary} we summarize our results.
More details of the fits are presented in the Appendices.

\section{Model \label{model}}
\subsection{Matter Content and Fermion Mass Matrices}
\begin{table}
\centering
\caption{\label{tab-contentsSM}
Quantum numbers of SM particles. Here, $i=1,2,3$ runs over the three SM families.
The electromagnetic charge of a fermion $f$ is $Q_f =T^3_f + Y_f/2$.
}
 \begin{tabular}{c|cccccc|c}\hline
                  & ${q_L}_i$&${\ol{u}_R}_i$&${\ol{d}_R}_i$&${l_L}_i$&${\ol{e}_R}_i$
                  &${\ol{\nu}_R}_i$&$H$ \\ \hline\hline
 $\SU3_{\mathrm{C}}$&$\bf{3}$&$\ol{\bf{3}}$& $\ol{\bf{3}}$&$\bf{1}$&$\bf{1}$& $\bf{1}$&$\bf{1}$\\
 $\SU2_{\mathrm{L}}$&$\bf{2}$&$\bf{1}$& $\bf{1}$& $\bf{2}$& $\bf{1}$& $\bf{1}$&$\bf{2}$ \\
 $\U1_{\mathrm{Y}}$ &$1/3$&$\text{-}4/3$&$2/3$&$\text{-}1$&$2$&$0$&$\text{-}1$ \\ \hline
 $\U1'$     & 0&0&0&0&0&0&0 \\ \hline
 \end{tabular} \\
\vspace{0.5cm}
\caption{\label{tab-contentsEXO}
Quantum numbers of new fermion and scalar fields.
}
 \begin{tabular}{c|cccccc|cccccc|cc}\hline
         & $Q_L$&$\ol{U}_R$&$\ol{D}_R$&$L_L$&$\ol{E}_R$&$\ol{N}_R$&
         $\ol{Q}_R$&$U_L$&$D_L$&$\ol{L}_R$&$E_L$&$N_L$&
         $\phi$ &$\Phi$\\ \hline\hline
 $\SU3_{\mathrm{C}}$&$\bf{3}$&$\ol{\bf{3}}$& $\ol{\bf{3}}$&$\bf{1}$&$\bf{1}$& $\bf{1}$&
                      $\ol{\bf{3}}$&$\bf{3}$& $\bf{3}$&$\bf{1}$&$\bf{1}$& $\bf{1}$&$\bf{1}$&$\bf{1}$\\
 $\SU2_{\mathrm{L}}$&$\bf{2}$&$\bf{1}$& $\bf{1}$& $\bf{2}$& $\bf{1}$& $\bf{1}$&
                      $\bf{2}$&$\bf{1}$& $\bf{1}$& $\bf{2}$& $\bf{1}$& $\bf{1}$&$\bf{1}$&$\bf{1}$   \\
 $\U1_{\mathrm{Y}}$ &$1/3$&$\text{-}4/3$&$2/3$&$\text{-}1$&$2$&$0$&
                    $\text{-}1/3$&$4/3$&$\text{-}2/3$&$1$&$\text{-}2$&$0$&
                    $0$&$0$ \\ \hline
 $\U1'$     & $\text{-}1$ & $1$ & $1$ & $\text{-}1$ & $1$ & $1$ & $1$ & $\text{-}1$ & $\text{-}1$ & $1$ & $\text{-}1$ & $\text{-}1$ & $0$ & $\text{-}1$ \\ \hline
 \end{tabular}
\end{table}
We study a model with a complete VL fourth family and $\Uop$ gauge symmetry.
The quantum number of all particles are listed in Tables~\ref{tab-contentsSM} and~\ref{tab-contentsEXO}.
The $\mathrm{SU(2)_L}$ doublets in the SM are defined as
${q_L}_i = ({u_L}_i,\ {d_L}_i)$, $ {l_L}_i=({\nu_L}_i,\ {e_L}_i),\ H=(H_0,\ H_-)$.
Here, $i=1,2,3$ runs over the three SM families.
The exotic doublets are $Q_L = (U'_L,\ D'_L)$, $L_L = (N'_L,\ E'_L)$,
$\ol{Q}_R = (-\ol{D}'_R,\ \ol{U}'_R)$, $\ol{L}_R = (-\ol{E}'_R,\ \ol{N}'_R)$.
Only the VL fermions and $\Uop$ breaking scalar $\Phi$ are charged
under the $\Uop$ gauge symmetry.
This model is trivially anomaly-free since the $\Uop$ charges are vector-like.
A singlet real scalar $\phi$ is introduced to model mass terms for the VL fermions.

In the gauge basis, the Yukawa couplings are given by
\begin{align}
\Lcal_\mathrm{Yukawa} =&\ \Lcal_\mathrm{SM} +\Lcal_\mathrm{H} +\Lcal_\phi +\Lcal_\Phi+\mathrm{h.c.}\;,
\end{align}
with 
\begin{align}
\label{eq-yukSM}
\Lcal_\mathrm{SM} :=&\
          {\ol{u}_R}_i y^u_{ij} {q_L}_{j}\ti{H}  +{\ol{d}_R}_i y^d_{ij} {q_L}_j H +{\ol{e}_R}_i y^e_{ij} {l_L}_j H
                    + {\ol{\nu}_{R}}_i  y^n_{ij} {l_L}_j \ti{H}\;, \\
\Lcal_\mathrm{H}    :=&\ \la_{u} \ol{U}_R Q_L \ti{H} + \la_{d} \ol{D}_R Q_L H+ \la_{e} \ol{E}_R L_L H
                          + \la_{n}  \ol{N}_R L_L \ti{H}  \notag \\
               &\ + \la'_{u} \ol{Q}_R H U_L - \la'_{d} \ol{Q}_R \ti{H} D_L - \la'_{e} \ol{L}_R \ti{H} E_L
                          + \la'_{n}  \ol{L}_R H N_L\;, \\
\Lcal_\phi  :=&\ \phi \left( \la^Q_{V} \ol{Q}_R Q_L-\la^U_{V} \ol{U}_R U_L-\la^D_{V} \ol{D}_R D_L \right. \notag\\
               &\ + \left.\la^L_{V} \ol{L}_R L_L-\la^E_{V}\ol{E}_R E_L-\la^N_{V}\ol{N}_R N_L \right)\;, \\
\Lcal_\Phi :=&\ \Phi \left( \la^Q_{i}\ol{Q}_R {q_L}_i +\la^L_{i}\ol{L}_R {l_L}_i \right) \notag \\
               &\ - \Phi^* \left( \la^U_{i} {\ol{u}_R}_i U_L  + \la^D_{i}{\ol{d}_R}_i D_L
                                        + \la^E_{i} {\ol{e}_R}_i E_L + \la^N_{i} {\ol{\nu}_R}_i N_L
                                            \right) \;.
\label{eq-yukPhi}
\end{align}
Here, $\ti{H} \define i\sigma_2 H^* = (H_-^*,\ -H_0^{*})$ and $i,j=1,2,3$ 
label the SM generations.

The neutral scalar fields acquire Vacuum Expectation Values (VEVs) given by
$v_H:=\vev{H_0}$, $v_\Phi:=\vev{\Phi}$, $v_\phi:=\vev{\phi}$.
The $5\times5$ Dirac mass matrices are given by\footnote{ 
Of course, for VL fermions there is always the possibility of rigid, Lagrangian-level mass terms.
However, for all our purposes the effect of those terms is completely equivalent to the masses induced by the VEV of $\phi$
and so we do not include such terms here.}
\begin{align}
\label{eq:Me}
\ol{e}_R \mcal^e e_L
 \define &\
\begin{pmatrix}
{\ol{e}_R}_i & \ol{E}_R&  \ol{E}^\prime_R
\end{pmatrix}
\begin{pmatrix}
y^e_{ij} v_H         &   0_i & \la^E_{i} v_\Phi  \\
 0_j                         & \la_e v_H &  \la_V^E v_\phi \\
 \la^L_j v_\Phi     & \la_V^L v_\phi & \la^\prime_e v_H
 \end{pmatrix}
\begin{pmatrix}
{e_L}_j  \\ E^\prime_L \\  E_L
\end{pmatrix}, \\
\label{eq:mDneutrino}
 \ol{n}_R \mcal^n n_L
 \define &\
\begin{pmatrix}
{\ol{\nu}_R}_i & \ol{N}_R&  \ol{N}^\prime_R
\end{pmatrix}
\begin{pmatrix}
y^n_{ij} v_H         &   0_i & \la^N_{i} v_\Phi  \\
 0_j                         & \la_n v_H &  \la_V^N v_\phi \\
 \la^L_j v_\Phi     & \la_V^L v_\phi & \la^\prime_n v_H
 \end{pmatrix}
\begin{pmatrix}
{\nu_L}_j  \\ N^\prime_L \\  N_L
\end{pmatrix},  \\
\ol{u}_R \mcal^u u_L
 \define &\
\begin{pmatrix}
{\ol{u}_R}_i & \ol{U}_R&  \ol{U}^\prime_R
\end{pmatrix}
\begin{pmatrix}
y^u_{ij} v_H         &   0_i & \la^U_{i} v_\Phi  \\
 0_j                         & \la_u v_H  &  \la_V^U v_\phi\\
 \la^Q_j v_\Phi     &  \la_V^Q v_\phi &  \la^\prime_u v_H
 \end{pmatrix}
\begin{pmatrix}
{u_L}_j  \\ U^\prime_L \\  U_L
\end{pmatrix}, \\ 
\label{eq:Md}
\ol{d}_R \mcal^d d_L
 \define &\
\begin{pmatrix}
{\ol{d}_R}_i & \ol{D}_R&  \ol{D}^\prime_R
\end{pmatrix}
\begin{pmatrix}
y^d_{ij} v_H         &   0_i & \la^D_{i} v_\Phi  \\
 0_j                         & \la_d v_H  &  \la_V^D v_\phi \\
 \la^Q_j v_\Phi     &   \la_V^Q v_\phi & \la^\prime_d v_H
 \end{pmatrix}
\begin{pmatrix}
{d_L}_j  \\ D^\prime_L \\  D_L
\end{pmatrix}.  
\end{align}
For electrically charged fermions, $f=u,d,e$, the mass basis is defined as
\begin{align}
 \hat{f}_{L} \define  \mhc{U^f_{L}}  f_{L},
\quad
\hat{f}_{R} \define  \mhc{U^f_{R}}  f_{R},
\end{align}
where unitary matrices diagonalize the Dirac matrices as
 \begin{align}
  \mhc{U^e_R} \mcal^e {U^e_L} =& \;
\mathrm{diag}(m_{e}, m_{\mu}, m_{\tau}, m_{E_1}, m_{E_2}), \\
  \mhc{U^u_R} \mcal^u {U^u_L} =& \;
\mathrm{diag}(m_{u}, m_{c}, m_{t}, m_{U_1}, m_{U_2}), \\
  \mhc{U^d_R} \mcal^d {U^d_L} =& \;
\mathrm{diag}(m_{d}, m_{s}, m_{b}, m_{D_1}, m_{D_2}).
 \end{align}
Here, $m_{E_a}, m_{U_a}, m_{D_a}$ $(a=1,2)$
are masses for the new charged leptons, up and down quarks, respectively.
These are predominantly the VL fermions of the gauge basis.

We consider the type-I see-saw mechanism to explain the tiny neutrino masses.
Since the $\Uop$ gauge symmetry prohibits Majorana masses for the VL family,
only three families of right-handed neutrinos have Majorana masses,
\begin{align}
\lag_\mathrm{Maj} = - \frac{1}{2}  {\ol{\nu}_R}_{i}  M_\mathrm{Maj}^{ij} {\nu^c_R}_j.
\end{align}
The Majorana masses are assumed to be $M^{ij}_\mathrm{Maj} \sim 10^{14}$ GeV.
The mass matrix for the neutrinos is then a $10\times 10$ matrix,
\begin{align}
\label{eq-Mn10}
\ol{\Ncal}_R \mcal^{\Ncal} \Ncal_L \define
\begin{pmatrix}
\ol{n}_R  &  \ol{n}^c_L
\end{pmatrix}
\begin{pmatrix}
 \mcal^R &  \mcal^n \\
\left({\mcal^n}\right)^{\mathrm{T}} & 0_{5\times 5}
\end{pmatrix}
\begin{pmatrix}
n^c_R  \\ n_L
\end{pmatrix},
\end{align}
where the Dirac mass matrix $\Mcal^n$ is defined in Eq.~\eqref{eq:mDneutrino}.
The $5\times 5$ Majorana mass matrix is given by
\begin{align}
 \mcal^R =
\begin{pmatrix}
\label{eq:Mneutino10}
 M^{ij}_\mathrm{Maj} & 0_i &0_i \\
 0_j & 0 & 0 \\
 0_j & 0 & 0
\end{pmatrix}.
\end{align}
After diagonalizing this matrix,
there are three left-handed Majorana neutrinos $\hat{\nu}_{L_i}$ with mass
of $\order{v_H^2/M_\mathrm{Maj}}$,  and
three right-handed Majorana neutrinos $\hat{\nu}_{R_i}$ with mass of $\order{M_\mathrm{Maj}}$
as in the ordinary type-I see-saw mechanism.
In addition to these,
there are two Dirac neutrinos $N_1, N_2$ with mass of $\order{v_{\phi}}$
which are predominantly the VL neutrinos of the gauge basis.
Mixing between the left- and right-handed neutrinos is suppressed by the huge Majorana mass terms.
An approximate mass basis is then defined as\footnote{ 
In principle, the rotation that diagonalizes the neutrino mass matrix mixes left- and charge conjugated right-handed states.
However, the effects of this mixing are suppressed by $\mathcal{O}\left(v_H / M_\mathrm{Maj}\right)$ and so we will neglect
them here and in the following, see Appendix \ref{sec-anal} for details.}
\begin{align}
 \hat{n}_L \define \left(U_L^n\right)^\dag n_L, \quad
 \hat{n}_R \define \left(U_R^n\right)^\dag n_R,
\end{align}
with unitary matrices $U_L^n$ and $U_R^n$, given in Appendix~\ref{sec-anal} where we discuss
the diagonalization of the neutrino mass matrix more explicitly.

There are three electrically neutral scalar fields in this model. We expand them around their VEVs as
\begin{align}
 H_0 =&\ v_H+\frac{1}{\sqrt{2}}\left(h + i a_H \right), \quad 
 \Phi =  v_\Phi +\frac{1}{\sqrt{2}} \left( \chi + i a_\chi\right), 
\quad\text{and}\quad \phi =  v_\phi +\sigma.
\end{align}
Here, $h$, $\chi$, and $\sigma$ are physical real scalar fields while the pseudo-scalar components $a_H$ and $a_\chi$ are eaten by the gauge bosons.
We introduce \textit{effective} quartic couplings of the scalars $\chi$ and $\sigma$,
which parametrize the scalar masses as
\begin{align}
m_\chi^2 = \la_\chi v_\Phi^2,\quad m_\sigma^2 = \la_\sigma v_\phi^2.
\end{align}
In this paper, we do not specify a scalar potential in this model
and treat their VEVs and the effective quartic couplings as input parameters.
For an effective quartic coupling $\la_\chi$ in a perturbative regime
and a sizable gauge coupling $g'$, the masses of 
$\chi$ and the $Z'$ gauge boson should roughly be of the same order.
This is important, since the Yukawa couplings of $\chi$ together with $v_\Phi$ set the scale of 
mass mixing between VL particles and the SM families.
The fact that this mixing is necessary for an explanation of the muon anomalies means that also 
$\chi$ will give sizable contributions in our fits.
In contrast, $v_\phi$ could be very large, and therefore $m_\sigma$ very heavy,
as long as the $\phi$-Yukawa couplings are small enough to prevent a complete
decoupling of the VL fermions.
The scalar $\sigma$, thus, can be very heavy and its contributions irrelevant for all discussed observables.
In fact, this limit resembles the case of rigid tree-level VL masses.
Consistent with that, contributions from $\sigma$ are always negligible at the best fit points we discuss below.

\subsection{Gauge and Yukawa Couplings}
In the gauge basis, there are no interactions between the $Z'$ boson and the SM fermions.
In order to explain the muon anomalies, the SM families are required to mix with the VL families.
Consequently, also the SM gauge couplings of quarks and leptons will receive corrections 
from these mixing effects. 
Of course, these couplings must remain consistent with all SM observables and we shall verify this.
For this discussion we combine left- and right-handed fermions to Dirac fermions as 
\begin{equation}
 f \define \left(f_L, f_R\right), \qquad\text{where}\qquad f=u,d,e,n\;.
\end{equation}

\subsubsection{Neutral Gauge Couplings}
The fermion couplings to the $\Zp$ boson are given by
\begin{align}
\lag_{Z^\prime} =&\ g^\prime Z^\prime_\mu  \sum_{f=u,d,e,n}  \ol{f} \gamma^\mu
                         \left(  {Q}^\prime_{f_L} P_L  + Q^\prime_{f_R}  P_R  \right) f  \\
 =&\  Z^\prime_\mu  \sum_{f=u,d,e,n}  \ol{\hat{f}}  \gamma^\mu
                         \left(  \hat{g}^{Z^\prime}_{f_L} P_L  + \hat{g}^{Z^\prime}_{f_R}  P_R  \right) \hat{f},
\end{align}
where the charge matrices in the gauge basis are
\begin{align}
{Q}^\prime_{f_L}= {Q}^\prime_{f_R}= \mathrm{diag}\left(0,0,0,-1,-1\right).
\end{align}
The coupling matrices in the mass basis are
\begin{align}
\hat{g}^{Z^\prime}_{f_L} = g^\prime \left(U^f_L \right)^\dag Q^\prime_{f_L} U^f_L
\qquad \text{and} \qquad
\hat{g}^{Z^\prime}_{f_R} = g^\prime \left(U^f_R \right)^\dag Q^\prime_{f_R} U^f_R.
\end{align}
The $Z$ boson couplings are given by
\begin{align}
\lag_{Z} =&\ \frac{g}{c_W} Z_\mu  \sum_{f=u,d,e,n}  \ol{f} \gamma^\mu
                         \left[  \left(T^f_3 \Pfb - Q_f s_W^2  \id_5  \right) P_L
                                +\left(T^f_3 P_5-Q_f s_W^2  \id_5\right) P_R  \right] f \\
=&\   Z_\mu  \sum_{f=u,d,e,n}  \ol{\hat{f}} \gamma^\mu  \left(
                                               \hg^{Z}_{f_L} P_L + \hg^Z_{f_R} P_R
                                             \right)  \hat{f},
 \end{align}
where $Q_f$ and $T^f_3$ are the electromagnetic charge
and third component of the weak isospin for a fermion $f$, respectively, 
while $s_W (c_W)$ denotes the (co)sine of the weak mixing angle.
The flavor space projectors are defined as
 \begin{align}
  P_5 \define \mathrm{diag}\left(0,0,0,0,1 \right) \qquad \text{and} \qquad 
  \Pfb \define \id_5- P_5 = \mathrm{diag}\left(1,1,1,1,0 \right).
 \end{align}
The coupling matrices in the mass basis are given by
 \begin{align}
\label{eq-Zcoup}
\hg^Z_{f_L} = \frac{g}{c_W} \left[T^f_3 \left(U^f_{L}\right)^\dag\Pfb U^f_{L} - Q_f s_W^2 \id_5 \right],
\quad
\hg^Z_{f_R} = \frac{g}{c_W} \left[T^f_3 \left(U^f_{R}\right)^\dag P_5 U^f_{R}-Q_fs_W^2 \id_5 \right].
 \end{align}
In general, this model has tree-level flavor changing neutral vector currents
mediated by the $Z$ boson. However, for the SM generations
these are automatically suppressed by $\mathcal{O}(m_{f_\mathrm{SM}}^2/M_{F_\mathrm{VL}}^2)$ 
coefficients which we show analytically in Appendix~\ref{sec-anal}. 
Here, $m_{f_\mathrm{SM}}$ and $M_{F_\mathrm{VL}}$ 
denote the mass of the SM fermion $f_\mathrm{SM}$, 
as well as the mass of the VL fermion $F_\mathrm{VL}$.

\subsubsection{CKM and PMNS Matrices}
The fermion couplings to the $W$ boson are given by
\begin{align}
\label{eq:Wcouplings}
\lag_{W} =&\ \frac{g}{\sqrt{2}} W^+_\mu
               \left[
              \ol{u} \gamma^\mu \left( \Pfb P_L + P_5 P_R  \right) d  +
              \ol{n} \gamma^\mu \left( \Pfb P_L + P_5 P_R \right)  e
                 \right] + \mathrm{h.c.}  \\
 =&\ W^+_\mu \left[
              \ol{\hat{u}} \gamma^\mu \left( \hg^W_{q_L} P_L + \hg^W_{q_R} P_R  \right) \hat{d}
            +\ol{\hat{n}} \gamma^\mu \left( \hg^W_{\ell_L} P_L + \hg^W_{\ell_R} P_R  \right) \hat{e}
       \right]+\mathrm{h.c.}
\end{align}
In the mass basis, the coupling matrices are
\begin{align}
 \hg^W_{q_L} =&\ \frac{g}{\sqrt{2}} \left(U_L^u\right)^\dag \Pfb {U_L^d} ,&\quad
 \hg^W_{q_R} =&\ \frac{g}{\sqrt{2}} \left(U^u_R\right)^\dag P_5 U^d_R, \\
 \hg^W_{\ell_L} =&\ \frac{g}{\sqrt{2}} \left(U_L^n\right)^\dag \Pfb {U_L^e} ,&\quad
 \hg^W_{\ell_R} =&\ \frac{g}{\sqrt{2}} \left(U^n_R\right)^\dag P_5 U_R^e.
\end{align}
The extended $5\times 5$ CKM matrix $\hat{V}_\mathrm{CKM}$ is defined as
 \begin{align}
  \hat{V}_\mathrm{CKM} \define  \left(U_L^u\right)^\dag \Pfb {U_L^d}\ .
\end{align}
Since one of the left-handed quarks is an $\mathrm{SU(2)_L}$ singlet
this extended CKM matrix has only rank $4$ and is, therefore, clearly non-unitary. 
Correspondingly, there exist right-handed charged current interactions 
which are completely absent in the SM.
Also the $3\times 3$ sub-matrix, which corresponds to the SM CKM matrix, is generally non-unitary 
due to mixing with the VL quarks.
Again these effects are suppressed 
by $\mathcal{O}(m_{f_\mathrm{SM}}^2/M_{F_\mathrm{VL}}^2)$ coefficients.  
In addition, the right-handed current interactions to the SM quarks 
are also suppressed and at most $\order{10^{-4}}$ as shown in Eq.~\eqref{eq-WR}, 
see Appendix~\ref{sec-anal}. These are negligible against experimental sensitivities.

The mixing between the SM and VL neutrinos are suppressed by the huge Majorana mass term. 
In Appendix~\ref{sec-anal}, 
we found that non-unitarity of the $3\times 3$ PMNS matrix 
is induced by the Yukawa coupling $\la_n$ 
and tiny mixing angles between the SM and VL charged leptons, that is $U_{e_L}^2$ in Eq.~\eqref{eq-U2fL}. 
In other words, there would be non-zero mixing between the SM neutrinos and the Dirac neutrinos 
if $\la_n \sim 1$.  This is an interesting possibility, but is beyond the scope of this paper. 
We consider a parameter space where $\la_n \ll 1$ and the PMNS matrix is approximately unitary. 
The details of neutrino mass diagonalization as well as the definition of the PMNS matrix 
are shown in Appendix~\ref{sec-anal}.

\subsubsection{Yukawa Couplings}
The Yukawa interactions with the real scalars are given by
\begin{align}
 -\Lcal_\mathrm{Yukawa} =&\ \frac{1}{\sqrt{2}}\sum_{f=u,d,e,n} \ol{f}_R \left[
h Y_f^h + \chi  Y_f^\chi  + \sigma  Y_f^\sigma
\right] f_L   + \mathrm{h.c.} \\
= &\ \frac{1}{\sqrt{2}}  \sum_{S=h,\chi, \sigma} \sum_{f=u,d,e,n}
   S  \ol{\hat{f}}_R   \hY_f^S  \hat{f}_L   + \mathrm{h.c.}
\end{align}
Here the Yukawa matrices for the fermions in the gauge basis are given by
\begin{align}
Y^h_f \define \frac{d\,\mathcal{M}^f}{d v_H},\quad Y^\chi_f \define \frac{d\,\mathcal{M}^f}{d v_\Phi},\quad Y^\sigma_f \define\sqrt{2}\ \frac{d\,\mathcal{M}^f}{d v_\phi}\;,
\end{align}
with the mass matrices of Eqs.\ \eqref{eq:Me}-\eqref{eq:Md}.
In the mass basis, these are
\begin{align}
 \hat{Y}_f^S = \left(U_R^f\right)^\dagger Y_f^S U_L^f,\qquad\text{for}\qquad S=h, \chi, \sigma.
\end{align} 
As in the $Z$ and $W$ boson couplings, 
the flavor violating couplings to the Higgs boson is automatically suppressed 
by $\order{m_{f_\mathrm{SM}}^2/M_{F_\mathrm{VL}}^2}$,
see Appendix~\ref{sec-anal}.

\subsubsection{Landau Pole Constraints on the \texorpdfstring{$\boldsymbol{\U1'}$}{U(1)'} Gauge Coupling} 
\label{sec-RGE}
As already discussed in more detail in \cite{Kawamura:2019rth}, the $\U1^\prime$ gauge coupling $g'$ diverges at a Landau pole at the scale
\begin{align}
 \Lambda_{g'} \simeq \mu_{Z'} \exp\left( \frac{24\,\pi^2}{65 \, {g'(\mu_{Z'})}^2} \right)\;.
\end{align}
Here $\mu_{Z'}\sim 1\,\mathrm{TeV}$ is the scale where we define the couplings of our model.
In order for the model to be consistent up to a typical GUT scale of $\Lambda_{g'} \sim 10^{16}\,\mathrm{GeV}$,
we require $g'(1\ \mathrm{TeV}) < 0.35$ in our analysis.

\section{Observables}
\label{observables}
In our model, $\Delta a_\mu$ is explained by chirally enhanced 1-loop corrections involving the $\Zp$ boson and VL leptons.
At the same time, tree-level $\Zp$ exchange induces new contributions to $\bsll$.
An explanation for $\Delta a_\mu$ requires sizable $\Zp$ couplings to muons, in agreement
with those necessary to explain deviations in $b\to s\mu\mu$.
The required mixing of the SM and VL fermions may, thus, induce new physics effects 
in various observables in both, lepton and quark sectors.
We have already shown that this model can explain the muon anomalies
without spoiling other observables in Ref.~\cite{Kawamura:2019rth}.
The purpose of the present paper is to completely map out the parameter space
where muon anomalies are explained consistently with all other observables,
and at the same time discuss the consequences for future experiments.
In this section, we introduce the 98 observables included in our $\chi^2$ analysis.
In our analysis, $1\sigma$ uncertainties for observables which only have upper bounds are set  
so that $1.64\sigma$ ($1.96\sigma$) deviation gives a value of 90\% (95\%) C.L. limit.

\subsection{Charged Leptons}
Here we study the charged lepton masses, lepton decays, as well as lepton anomalous magnetic moments.
Central values and uncertainties of the observables are listed in Table~\ref{tab-obsL}.

We fit the charged lepton masses to the values calculated from Yukawa couplings
at $m_Z=91.2$ GeV~\cite{Antusch:2013jca}.
Charged lepton masses are experimentally known with accuracy much better 
than what makes sense to provide in our analysis.
We, thus, assume $0.01\%$ relative uncertainties for the lepton masses.
\begin{table}[t] 
 \centering
\caption{\label{tab-obsL}
Central values and uncertainties of charged lepton observables.
Uncertainties stated in percent are understood as uncertainties relative to their central values.
}
\begin{tabular}{c|c|c|c} \hline
Name                        & Center      & Uncertainty    & Remark   \\ \hline\hline
$m_e(m_Z)$     [MeV] & 0.486576 & 0.01 $\%$    & Ref.~\cite{Antusch:2013jca}  \\
$m_\mu(m_Z)$ [MeV] & 102.719 &  0.01 $\%$    & Ref.~\cite{Antusch:2013jca}  \\
$m_\tau(m_Z)$ [GeV] & 1.74618   & 0.01 $\%$   & Ref.~\cite{Antusch:2013jca} \\ \hline\hline
$\br{\mu}{e\nu\ol{\nu}}$     & 1.0000 & 0.01 $\%$   &  SM  \\ 
$\br{\mu}{e\gamma}$    & 0.    & 2.6$\times 10^{-13}$ & Ref.~\cite{Tanabashi:2018oca} \\
$\br{\mu^-}{e^-e^+e^-}$  & 0. & 6.1$\times 10^{-13}$ & Ref.~\cite{Tanabashi:2018oca} \\  \hline
$\br{\tau}{e\nu\ol{\nu}}$      & 0.179 & 0.01 $\%$   & SM \\ 
$\br{\tau}{\mu\nu\ol{\nu}}$ & 0.174 &  0.01 $\%$    & SM \\ 
$\br{\tau}{e\gamma}$        & 0. & 2.0 $\times 10^{-8}$ & Ref.~\cite{Tanabashi:2018oca} \\
$\br{\tau}{\mu\gamma}$    & 0. & 2.7 $\times 10^{-8}$ & Ref.~\cite{Tanabashi:2018oca} \\  \hline
$\br{\tau^-}{e^-e^+e^-}$  & 0. & 1.6$\times 10^{-8}$ & Ref.~\cite{Tanabashi:2018oca} \\
$\br{\tau^-}{e^-\mu^+\mu^-}$  & 0. & 1.6$\times 10^{-8}$ & Ref.~\cite{Tanabashi:2018oca} \\
$\br{\tau^-}{\mu^-e^+\mu^-}$  & 0. & 1.0$\times 10^{-8}$ & Ref.~\cite{Tanabashi:2018oca} \\
$\br{\tau^-}{\mu^-e^+e^-}$  & 0. & 1.1$\times 10^{-8}$ & Ref.~\cite{Tanabashi:2018oca} \\
$\br{\tau^-}{e^-\mu^+e^-}$  & 0. & 1.0$\times 10^{-8}$ & Ref.~\cite{Tanabashi:2018oca} \\
$\br{\tau^-}{\mu^-\mu^+\mu^-}$  & 0. & 1.3$\times 10^{-8}$ & Ref.~\cite{Tanabashi:2018oca} \\ \hline\hline
$\Delta a_e$ & -8.7 $\times 10^{-13}$& 3.6 $\times 10^{-13} $ & Ref.~\cite{Davoudiasl:2018fbb}  \\
$\Delta a_\mu$ & 2.68$\times 10^{-9}$& 0.76 $\times 10^{-9}$ & Ref.~\cite{Tanabashi:2018oca} \\
 \hline
\end{tabular}
\end{table}

\subsubsection{Anomalous Magnetic Moment}
\label{sec-AMM}
There are discrepancies between experiments and SM predictions for both, the electron and muon magnetic moment.
The current size of the discrepancies are~\cite{Tanabashi:2018oca,Davoudiasl:2018fbb}
\begin{align}
 \Delta a_e \define&\ a_e^\mathrm{exp}-a_e^\mathrm{SM}
                     =(-87\pm 36)\times 10^{-14},  \\
 \Delta a_\mu \define&\ a_\mu^\mathrm{exp}-a_\mu^\mathrm{SM}
                     = (2.68\pm 0.76)\times 10^{-9}.
\end{align}
Simultaneous explanations for these two anomalies are studied
in Refs.~\cite{Giudice:2012ms,Davoudiasl:2018fbb,Crivellin:2018qmi,Liu:2018xkx,
Dutta:2018fge,Parker:2018vye,Han:2018znu,Endo:2019bcj,
Bauer:2019gfk,Badziak:2019gaf,CarcamoHernandez:2019ydc}.

The 1-loop beyond the SM corrections involving $\Zp$, $Z$ and $W$ bosons to $\Delta a_\mu$
are given by~\cite{Jegerlehner:2009ry,Dermisek:2013gta}
\begin{align}\label{eq:daMuZp}
 \delta_{\Zp} a_\mu =& -\frac{m_\mu}{8\pi^2m_\Zp^2} \sum_{B=1}^5\left[
\left(\abs{\left[\hg^{\Zp}_{e_L}\right]_{2B} }^2
  +\abs{\left[\hg^{\Zp}_{e_R}\right]_{2B}  }^2 \right) m_\mu F_{Z}(x^{\Zp}_{e_B})  \right. \\  \notag
&\hspace{5.0cm} \left. + \mathrm{Re}\left(
  \left[\hg^{Z'}_{e_L}\right]_{2B} \left[{\hg^{Z'}_{e_R}}\right]^*_{2B} \right)
       m_{e_B} G_{Z}(x^\Zp_{e_B})
\right],  \\
 \delta_Z a_\mu =& -\frac{m_\mu}{8\pi^2m_Z^2} \sum_{b=4}^5 \left[
\left(\abs{\left[\hg^{Z}_{e_L}\right]_{2b} }^2+\abs{\left[\hg^{Z}_{e_R}\right]_{2b}  }^2 \right)
 m_\mu F_Z(x^{Z}_{e_b}) \right. \\  \notag
&\hspace{5.0cm} \left.
+ \mathrm{Re}\left( \left[\hg^{Z}_{e_L}\right]_{2b} \left[{\hg^{Z}_{e_R}}\right]^*_{2b}  \right) m_{e_b}
G_Z(x^Z_{e_b}) \right],  \\
 \delta_W a_\mu =& -\frac{m_\mu}{16\pi^2 m_W^2} \sum_{b=4}^{5} \left[ \left(
\abs{\left[\hg_{\ell_L}^{W}\right]_{b2}}^2 + \abs{\left[\hg_{\ell_R}^{W}\right]_{b2} }^2  \right)
m_\mu F_W(x^W_{n_b}) \right. \\  \notag
&\hspace{5.0cm} \left.
+\text{Re}\left(\left[\hg_{\ell_L}^{W}\right]_{b2}
\left[\hg_{\ell_R}^{W}\right]_{b2}^*  \right) m_{N_b} G_W(x^W_{n_b})
\right],
\end{align}
where $x^{V}_{e_B} \define m^2_{e_B}/m_{V}^2$ $(V=Z, \Zp)$ and $x^{W}_{n_b} \define m^2_{n_b}/m_W^2$.
Here, $m_{e(n)_B}$ is the mass of the $B$-th generation charged(neutral) lepton.
The flavor index $B=1,\dots,5$ runs over all five families, while $b=4,5$ runs only over the VL family.
The loop functions are defined as
\begin{align}
  F_Z(x) = &\ \frac{5x^4-14x^3+39x^2-38 x-18x^2 \ln{(x)}+8 }{12(1-x)^4},  \\
\label{eq-GZ}
 G_Z(x) = &\ \frac{x^3+3x-6x \ln{(x)}-4}{2(1-x)^3},  \\
 F_W(x) =&\ -\frac{4x^4-49x^3+78x^2-43x+18x^3\ln{(x)}+10}{6(1-x)^4},  \\
 G_W(x) =&\ -\frac{x^3-12x^2+15x +6x^2 \ln{(x)}-4}{(1-x)^3}.
\end{align}

The scalar 1-loop beyond the SM contributions to $\Delta a_\mu$ are given by~\cite{Dermisek:2013gta,Jegerlehner:2009ry}
\begin{align}
 \delta_h a_\mu =& -
 \frac{m_\mu}{32\pi^2 m_h^2} \sum_{b=4,5} \left[
\left(\abs{\left[{\hY^h_e}\right]_{2b}}^2+\abs{\left[\hY^h_e\right]_{b2}}^2 \right)
m_\mu F_S(y^h_{e_b})    \right. \\ \notag
&\hspace{5.0cm} \left.
+ \re{\left[\hY^h_e\right]_{2b} \left[\hY^h_e\right]_{b2}} m_{e_b} G_S(y^h_{e_b})
\right],  \\ \label{eq:daMuS}
 \delta_S a_\mu =& -
 \frac{m_\mu}{32\pi^2 m_S^2} \sum_{B=1}^5 \left[
\left(\abs{  \left[{\hY^S_e}\right]_{2B}}^2+\abs{\left[\hY^S_e\right]_{B2}}^2 \right)
m_\mu F_S(y^S_{e_B})    \right. \\ \notag
&\hspace{5.0cm} \left.
+ \re{\left[\hY^S_e\right]_{2B} \left[\hY^S_e\right]_{B2}} m_{e_B} G_S(y^S_{e_B})
\right],
\end{align}
where $S=\chi, \sigma$.
Here, $y^h_{e_b} \define m_{e_b}^2/m_h^2 $ and $y^S_{e_B} \define m_{e_B}^2/m_S^2$.
The loop functions are defined as
\begin{align}
 F_S(y) =&\ -\frac{y^3-6y^2+3y+6y \ln{(y)} + 2}{6(1-y)^4}, \\
 G_S(y) =&\ \frac{y^2-4y+2\ln{(y)}+3}{(1-y)^3}.
\label{eq-GS}
\end{align}
Analogous formulae for $\Delta a_e$ are obtained by formally replacing $2\to 1$ and $\mu \to e$.

\begin{figure}[t]
\begin{minipage}[c]{0.5\hsize}
\centering
\includegraphics[height=65mm]{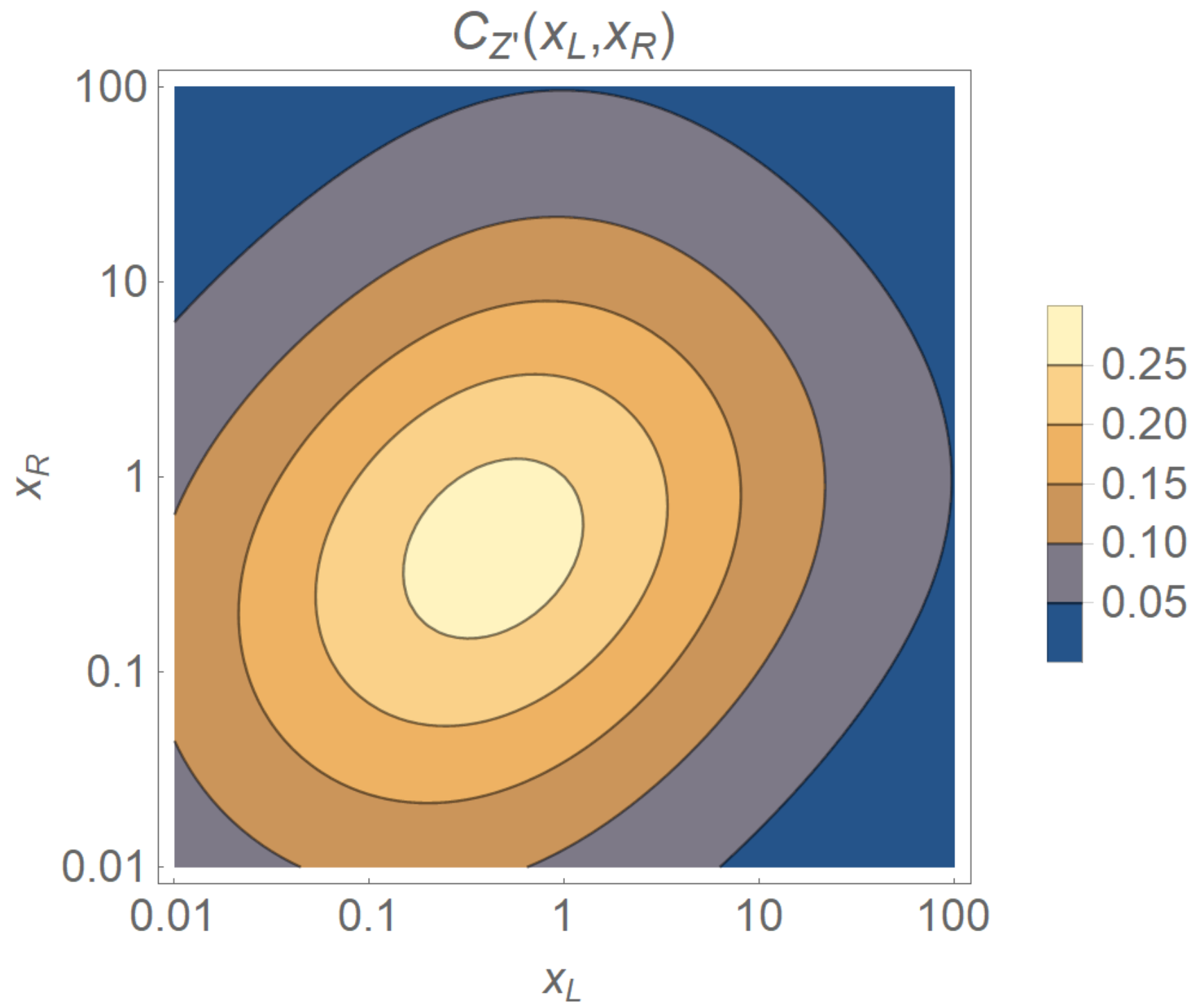}
\end{minipage}
\begin{minipage}[c]{0.5\hsize}
\centering
\includegraphics[height=65mm]{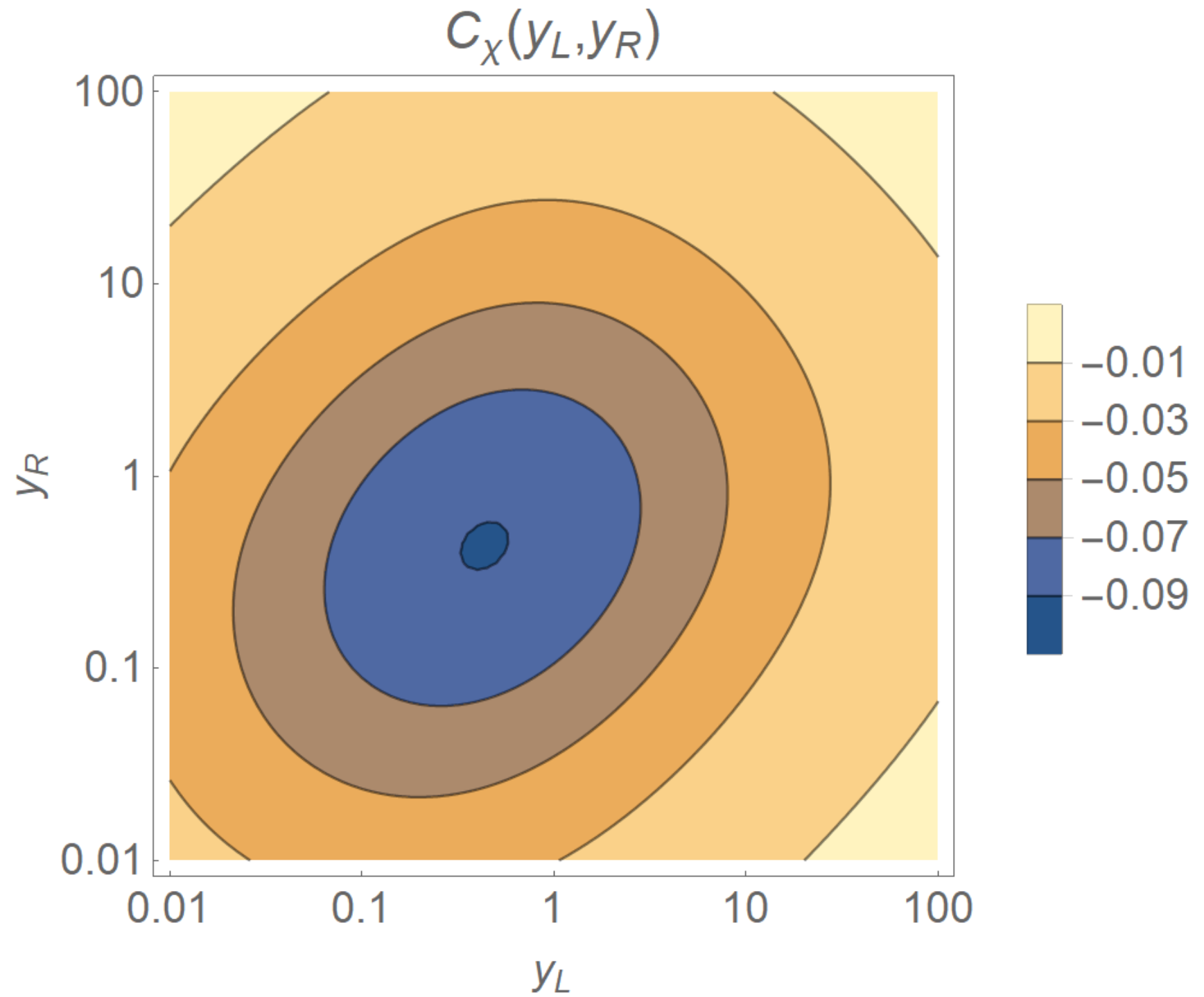}
\end{minipage}
 \caption{
Contour plot of the functions $C_{\Zp}(x_L,x_R)$ and $C_{\chi}(y_L,y_R)$ defined in Eq.\ \eqref{eq-CLR}.
}
\label{fig-CLR}
\end{figure}

We now discuss the leading contributions to $\Delta a_\mu$ analytically. 
This will be important to understand upper bounds on $v_\Phi$ and the masses of VL leptons stated below. 
The new physics contribution to $\Delta a_\mu$ is dominated
by chirally enhanced effects proportional to $\la^\prime_e v_H$, namely the Higgs-VEV induced mixing between 
left- and right-handed VL leptons. 
Contributions involving the SM bosons are very suppressed.
The leading contribution can be estimated as
\begin{align}
\label{eq:aMu_anal}
 \Delta a_\mu \sim&\
- \frac{m_\mu \la^\prime_e v_H}{64\pi^2 v_\Phi^2} s_{2 \mu_L} s_{2 \mu_R} C_{LR}
\sim 2.9\times10^{-9} \times
                     \left(\frac{ s_{2\mu_L} s_{2\mu_R}  C_{LR}}{0.1} \right)
                     \left(\frac{\la^\prime_e}{-1.0} \right) \left(\frac{1\ \text{TeV}}{v_{\Phi}} \right)^2,
\end{align}
where $s_{2\mu_{L(R)}}$ are mixing angles between the muon and VL leptons and $C_{LR}$ is a function
of the mass ratios that we will define shortly. 
The mixing angles are approximately given by (see Appendix~\ref{sec-anal} for details)
\begin{align}
\label{eq:aMu_anal2}
 s_{2\mu_L} = 2 s_{\mu_L} c_{\mu_L} \sim
                  2 \frac{\la^L_V v_\phi}{M_{E_L}}  \frac{\la^L_2 v_\Phi}{M_{E_L}},
\quad
s_{2\mu_R} = 2 s_{\mu_R} c_{\mu_R} \sim
                  2 \frac{\la^E_V v_\phi}{M_{E_R}} \frac{\la^E_2 v_\Phi}{M_{E_R}  }.
\end{align}
Here, $M_{E_{L}}$ and $M_{E_R}$ the masses of the doublet- and singlet-like VL leptons 
which can  be approximated as $M_{E_L}^2 \sim \left(\la^L_Vv_\phi\right)^2+\left(\la^L_2 v_\Phi\right)^2$ and 
$M_{E_R}^2 \sim \left(\la^E_Vv_\phi\right)^2+\left(\la^E_2 v_\Phi\right)^2$.
The mixing angles are maximized at $\la_V^L v_\phi = \la^L_2 v_\Phi$, and $\la_V^E v_\phi = \la^E_2 v_\Phi$.

The dimensionless function $C_{LR}$ is defined in \eqref{eq:CLRApp} in Appendix~\ref{sec-anal}. 
It can be approximated as 
\begin{align}
\label{eq-CLR}
C_{LR} := &\ C_{\Zp}(x_L, x_R) + C_{\chi}(y_L, y_R) \\ \notag
             \approx &\  \sqrt{x_L x_R} \frac{G_Z(x_L)-G_Z(x_R) }{x_L-x_R}
                     + \frac{1}{2} \sqrt{y_L y_R}
                      \frac{y_L G_{S}(y_L)-y_R G_{S}(y_R)  }{y_L-y_R},
\end{align}
with $x_{L,R} \define M_{E_{L,R}}^2/m_{\Zp}^2$ and $y_{L,R} \define M_{E_{L,R}}^2/m_{\chi}^2$.
Figure~\ref{fig-CLR} shows contour plots of the functions $C_{\Zp}$ and $C_\chi$.
$C_\Zp$ has a maximum value $\sim 0.272$ at $x_L = x_R \sim 0.433$.
$C_\Zp({C_\chi})$ is always positive (negative) and $\abs{C_\Zp} > \abs{C_\chi}$
at most parts of the parameter space.
Altogether, an upper bound on $\Delta a_\mu$ is given by
\begin{align}
\label{eq-g2upper}
 \Delta a_\mu \lesssim \frac{m_\mu v_H}{64\pi^2 v_\Phi^2} C_{\Zp} \define \Delta a_\mu^\mathrm{max}.
\end{align}
The equality is saturated when $\la^\prime_e \sim -1.0$, $s_{2\mu_L} \sim s_{2\mu_R} \sim 1$ and $C_{\chi} \sim 0$.
Inserting the maximal value of $C_\Zp$ one finds
\begin{align}
\label{eq:amu_upperbound}
 \Delta a_\mu \lesssim 2.74 \times 10^{-9} \times \left(\frac{1.7\ \mathrm{TeV}}{v_\Phi} \right)^2.
\end{align}
Thus $v_\Phi \lesssim 1.7$ TeV is required to explain $\Delta a_\mu$.
Moreover, 
\begin{equation}
M_{E_L}\sim M_{E_R}\sim \sqrt{0.433} \cdot m_{\Zp} \lesssim 550\,\mathrm{GeV}\;, 
\end{equation}
is required to maximize $C_\Zp$.
\begin{figure}[t]
 \centering
\includegraphics[height=100mm]{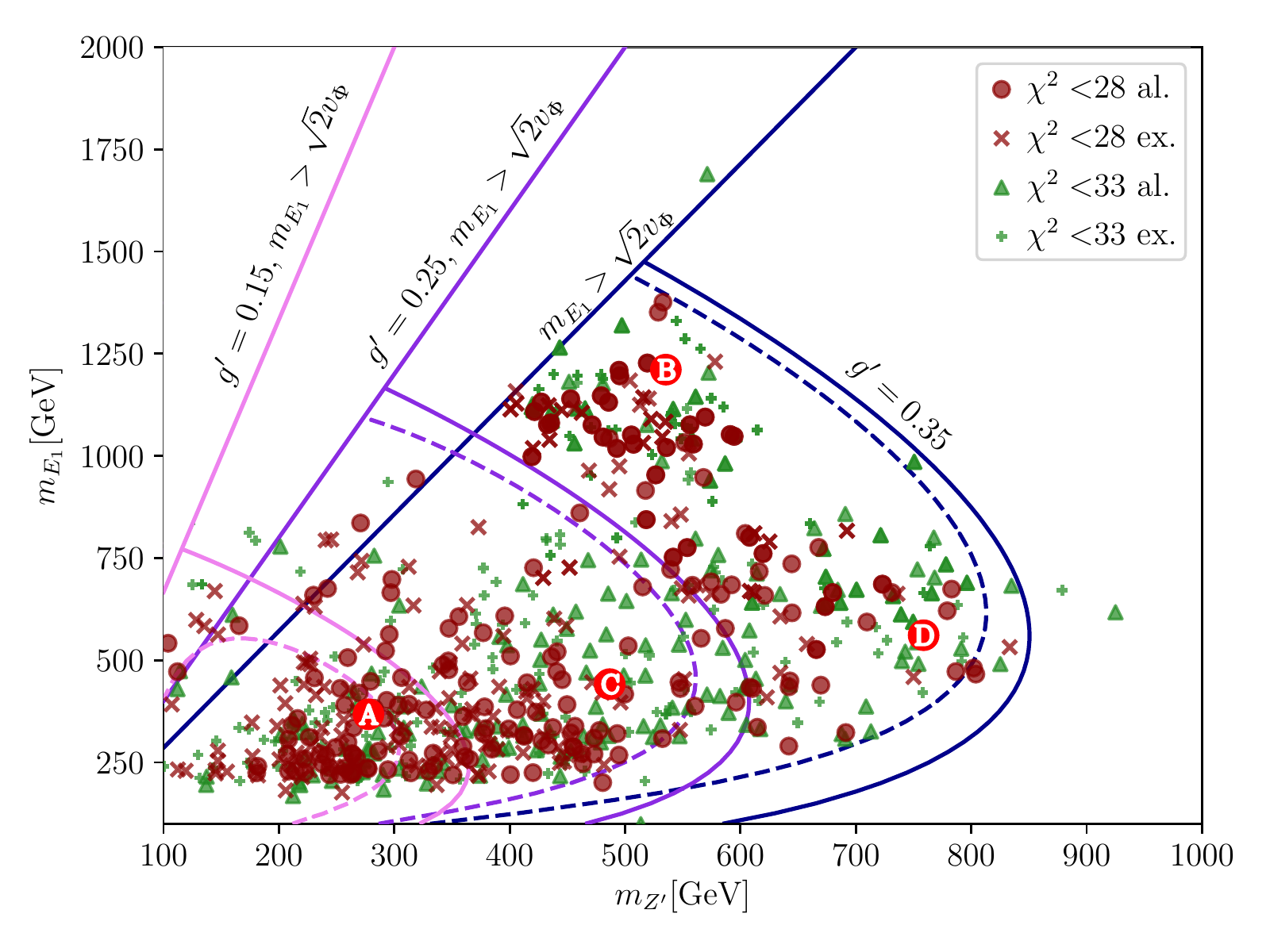}
 \caption{\label{fig-E1bound}
Contours in the ($m_\Zp$, $m_{E}$) plane which realize the 
maximal possible value of $\Delta a_\mu = 2.68\times10^{-9}$ for several values of $\gp$.
The solid (dashed) lines are obtained without (with) the scalar contribution $C_\chi$,
where for the purpose of the plot $m_\chi = 100\,\mathrm{GeV}$.
$M_E > \sqrt{2} v_\Phi$ on the left of the straight lines.
The dots and crosses represent best fit points (which in particular fit $\Delta a_\mu$ in the $1\sigma$ allowed region).
Details will be explained in detail in Section~\ref{results}. Red (green) dots have $\chi^2 < 28~(33)$, while points which are excluded by other data are 
shown as crosses or pluses.}
\end{figure}
We are interested in upper bounds on the lightest VL charged lepton.
For a fixed lightest VL charged lepton mass,
the function $C_\Zp$ is maximized if the heavier state has the same mass,
i.e. $M_{E_L}=M_{E_R} \define M_E$.
Then $x_L = x_R \define x$, and
\begin{align}
\label{eq-CZpsym}
C_{\Zp}(x,x) = x \frac{d G_Z(x)}{d x}
                                           = \frac{3 x \left[x^2+4x-5-2(2x+1)\log{x} \right]}{2 (1-x)^4}.
\end{align}
Figure~\ref{fig-E1bound} shows contours of $\Delta a_\mu^\mathrm{max} =2.68\times10^{-9}$
in the ($m_\Zp$, $M_{E}$) plane where $C_\Zp$ is replaced by Eq.~\eqref{eq-CZpsym}
and the gauge coupling constant $\gp$ is fixed.  Different colors correspond to different values of $\gp$.
$\Delta a_\mu = 2.68\times10^{-9}$ can be realized only inside the contours for a given $\gp$.
We further restrict the VL masses by $M_E < \sqrt{2} v_\Phi$,
because the condition $s_{2 \mu_L} = s_{2 \mu_R} = 1$ requires
$M_{E_L}=M_{E_R}=\sqrt{2} \la^{E}_2 v_\Phi \le \sqrt{2}v_\Phi$.
The last inequality comes from our requirement for perturbativity, $\la^E_2 \le 1$.
This condition is depicted by the straight lines in Fig.~\ref{fig-E1bound}.
Altogether, the upper bound on the VL lepton mass is about 1.4 TeV
where $m_\Zp \sim 500$ GeV and $\gp = 0.35$.
Note that $m_{E_1} \sim 1.4$ TeV is realized only if all of the conditions are satisfied:
(1) $s_{2\mu_L} \sim s_{2\mu_R} \sim 1$, (2) $\la^\prime_e \sim -1.0$, (3) $C_{\chi} \sim 0$
(4) $m_\Zp \sim 500$ GeV, (5) $M_{E_L} \sim M_{E_R}$ and (6) $\gp \sim 0.35$.
Consequently, the upper bound is hardly ever saturated.
The dashed lines in Fig.~\ref{fig-E1bound} show
the same contour but the destructive $\chi$ contribution with $m_\chi=100$ GeV
is added to $C_\Zp$.
$m_\chi = 100$ GeV is chosen to minimize the $C_\chi$ contribution.
Including the scalar contribution,
the upper bound on $m_E$ is tightened to 1.3 TeV.
Clearly the actual upper bound will be lower if some of the conditions (1)-(6) are not satisfied.
The points shown in Fig.~\ref{fig-E1bound} are results of our fit. 
As anticipated, good fits are only obtained for points within the contours. 
The details of our analysis will be shown in the next section.

On a different note, a lighter scalar $\chi$ allows one to explain $\Delta a_\mu$ with smaller $\Zp$ contribution (see Figure~\ref{fig-CLR}),
especially when the VL leptons are heavy.
$y_{L,R} \ll 1$ or $y_{L,R} \gg 1$ is favored
to suppress the destructive contributions from $C_{\chi}$.
In the phenomenologically viable parameter space,
$M_{E_{L,R}} \gtrsim 250$ GeV and a lower bound on $y_{L,R}$ is
\begin{align}
y_{L,R} \gtrsim 0.0625 \times \left(\frac{1\ \mathrm{TeV} }{m_\chi} \right)^2. 
\end{align}
For $y_{L,R} \sim 0.0625$ with $m_\chi = $ 1 TeV,  we have  $C_\chi \sim -0.06$ .
On the other hand, $y_{L,R} \sim 100$, corresponding to $C_\chi \sim 0.004$, is possible
if $m_\chi \sim 100$ GeV and $M_{E_{L,R}} \sim 1$ TeV.

Another important consequence of $\Delta a_\mu$
is that the Higgs Yukawa coupling $\la^\prime_e$ should be $\sim -1$.
Of course, chiral enhancement proportional to the VL lepton mass is absent
if there is no mixing between the left-handed and right-handed VL leptons.
In other words, $\Delta a_\mu$ is enhanced by the left-right mixing induced by the 
Higgs VEV $v_H$ and not directly by the VL lepton masses.
Hence, $\Delta a_\mu$ is proportional to $\la^\prime_e$.
For this reason, the $\Uop$ charge assignment in our model must not be universal 
for $(L_L, \ol{E}_R)$, but must be flipped as in Table~\ref{tab-contentsEXO}.
Importantly, such a charge assignment is incompatible with $SO(10)$ unification.
However, it is still consistent with unification in the Pati-Salam gauge group, 
$\mathrm{SU(4)}\times\mathrm{SU(2)_L}\times\mathrm{SU(2)_R}$.

\subsubsection{\boldmath \texorpdfstring{$\ell_i \to \ell_j \nu \ol{\nu}$}{l->l' nu nu}\unboldmath}
\begin{table}[t]
 \centering
\caption{\label{tab-consL}
Values of constants for charged lepton decays~\cite{Tanabashi:2018oca}.} 
\begin{tabular}[t]{c|c|c|c}\hline\hline
$m_e$     [MeV] & 0.5109989  & $\Gamma_e$       [GeV] &  0 \\
$m_\mu$ [MeV] & 105.65837  & $\Gamma_\mu$  [GeV] & 2.99598 $\times 10^{-19}$ \\
$m_\tau$ [GeV] & 1.77686      & $\Gamma_\tau$   [GeV] & 2.26735 $\times 10^{-12}$ \\ \hline
$g$                     & 0.65290      & $m_W$                [GeV]   & 80.379  \\
 \hline\hline
\end{tabular}
\end{table}
The dominant decay modes of the charged leptons are
three-body decays via a $W$ boson.
The branching fraction for a lepton $\ell_i$ to decay into a lighter lepton $\ell_j$ is given by
\begin{align}
\label{eq:BRlnunu}
 \br{\ell_i}{\ell_j \nu \ol{\nu}}
= \sum_{k,l=1}^3 \frac{
               \abs{\left[ \hg^W_{\ell_L} \right]_{ki}}^2 \cdot
               \abs{ \left[ \hg^W_{\ell_L} \right]_{lj}}^2 }{1536\pi^3 \Gamma_{\ell_i}}
              \frac{m_{\ell_i}^5}{m_W^4}
       F_\ell\left(\frac{m_{\ell_j}^2}{m_{\ell_i}^2}\right),
\end{align}
where $m_{\ell_i}$ and $\Gamma_{\ell_i}$ are the mass and decay width of the lepton $\ell_i$, respectively.
The function $F_\ell$ is given by
\begin{align}
 F_\ell(y) = 1-8y+8y^3-y^4-12y^2 \log{y}.
\end{align}
Experimental values of the lepton masses and decay widths are listed in Table~\ref{tab-consL}.
For the muon decay rate, the tree-level branching fractions are multiplied by a QED correction factor $\eta_\text{QED} = 0.995802$~\cite{Tanabashi:2018oca}.
This factor is less important for tau decay. 
Just as for the charged lepton masses, the charged lepton decay rates are measured 
more precisely than the accuracy of our numerical analysis.
We assume $0.01\%$ relative uncertainties for these observables, remarking that we 
could always fit them by increasing the numerical accuracy of our analysis. 
Branching fractions could deviate from their SM predictions if the mixing between the 
SM families and VL family affects the $W$ couplings.
The values obtained in our model are compared with the tree-level SM values,
that are given by replacing $\left[\hg^W_{\ell_L}\right]_{ij} \to g \left[V_\text{PMNS}\right]_{ij}/\sqrt{2}$ in Eq.~\eqref{eq:BRlnunu}.

\subsubsection{\boldmath\texorpdfstring{$\ell_i \to \ell_j \gamma$}{}\unboldmath}
\label{sec-LFV-lgl}
Lepton Flavor Violating (LFV) processes are severely constrained by experiments.
We follow \cite{Lavoura:2003xp} to calculate one-loop corrections including general gauge and Yukawa interactions.
The Lagrangian for general gauge and Yukawa interactions is given by
\begin{align}
 \lag_{\ell_i\to\ell_j\gamma}
           =\sum_{k=i,j} \sum_{F}  \ol{F} \left[\sum_{V}
                 V_\mu \gamma^\mu \left(L^{VF}_k P_L + R^{VF}_k P_R \right)
                   -   \sum_{S}  S  \left( L^{SF}_k P_L + R^{SF}_k P_R\right)
 \right]  \ell_k   +\mathrm{h.c.},
\end{align}
where $\ell_k$ are external charged leptons, $F$ internal fermions, $V$ vector bosons and $S$ scalars.
The gauge couplings in our model are identified as
\begin{align}
  L_k^{\Zp\hat{e}_A} =&\  \left[\hat{g}^\Zp_{e_L} \right]_{Ak}, \quad
  L_k^{Z\hat{e}_A} =  \left[\hat{g}^Z_{e_L} \right]_{Ak}, \quad
  L_k^{W \hat{n}_B} =  \left[\hat{g}^W_{\ell_L} \right]_{kB}, \\
  R_k^{\Zp\hat{e}_A} =&\       \left[\hat{g}^\Zp_{e_R}\right]_{Ak},\quad
  R_k^{Z\hat{e}_A} =       \left[\hat{g}^Z_{e_R}\right]_{Ak},\quad
  R_k^{W N_b} =      \left[\hat{g}^W_{\ell_R}\right]_{kb},
\end{align}
where $A,B = 1,\dots,5$, while $b=4,5$  runs only over the VL neutrinos.\footnote{%
The sum for the W-boson coupling $R_k^{WN_b}$ only runs over the Dirac neutrinos because the light 
neutrinos are left-handed.}
The Yukawa couplings are given by
\begin{align}
 L_k^{S\hat{e}_A} =&\ \frac{1}{\sqrt{2}} \left[\hat{Y}^S_e \right]_{Ak}, \quad
 R_k^{S\hat{e}_B} =   \frac{1}{\sqrt{2}} \left[\hat{Y}^S_e\right]^*_{kB},\quad
  S= h, \chi, \sigma.
\end{align}
The branching fraction is then given by
\begin{align}
\br{\ell_i}{\ell_j \gamma} =
\frac{e^2}{16\pi \Gamma_{\ell_i}}
\left(m_{\ell_i}-\frac{m_{\ell_j}^2}{m_{\ell_i}}\right)^3
 \left( \abs{\sigma_L}^2 + \abs{\sigma_R}^2 \right),
\end{align}
where 
\begin{align}
\label{eq-sigmaL}
 \sigma_L=&\  \sum_{F} \Big[
    \rho^{WF} \ol{y}_1+ \la^{WF} \ol{y}_2+ \ups^{WF} \ol{y}_3+\zeta^{WF} \ol{y}_4
         \notag  \\
            +&\ \sum_{V=Z, \Zp} \left(
                \rho^{VF}  y_1+\la^{VF} y_2+\ups^{VF}y_3+\zeta^{VF} y_4
                  \right)  \notag  \\
        &\  +\sum_{S=h, \chi, \sigma}
                \left( \rho^{SF}  k_1+\la^{BF} k_2+\ups^{BF}k_3 \right)
          \Big].
\end{align}
Here, $y_i$, $\ol{y}_i$, and $k_j$ (with $i=1,\dots,4$ and $j=1,2,3$) 
are loop functions defined in Ref.~\cite{Lavoura:2003xp} while combinations of couplings are defined as
 \begin{align}
  \la^{BF} =&\ \left(L_j^{BF}\right)^* L^{BF}_i,\quad \rho^{BF}=\left(R_j^{BF}\right) ^* R^{BF}_i, \\
  \zeta^{BF} =&\ \left(L_j^{BF}\right)^* R^{BF}_i,\quad \ups^{BF}=\left(R_j^{BF}\right) ^* L^{BF}_i,
 \end{align}
with $B=S,V,W$. $\sigma_R$ is obtained from $\sigma_L$ by formally replacing $\rho^{BF} \leftrightarrow \la^{BF}$
and $\zeta^{BF} \leftrightarrow \ups^{BF}$.

Just as for $\Delta a_\mu$, the dominant contribution to $\ell_i \to \ell_j \gamma$ is again a chirally enhanced effect. 
To a good approximation, $\sigma_L$ is given by
\begin{align}
\sigma_L \simeq &\   \sum_{b=4,5} \left(\frac{m_{e_b}}{m_{Z'}^2}
               \left[\hg^{Z'}_{e_R} \right]_{jb}\left[\hg^{Z'}_{e_L} \right]_{b i} G_Z(x_{e_b}^{\Zp})
                  +   \frac{m_{e_b}}{4m_{\chi}^2} \left[\hY^\chi_e \right]_{jb} \left[\hY^\chi_e \right]_{b i}
                      G_S(y^\chi_{e_b})
                     \right),
\end{align}
where the loop functions $G_{Z,S}$ are the same as in Eqs.~\eqref{eq-GZ} and~\eqref{eq-GS}.
Using the results from Appendix~\ref{sec-anal}, analytic expressions for the branching 
fractions of $\mu\to e\gamma$ and $\tau \to \mu \gamma$ are given by
\begin{align}
\label{eq-muegam}
 \br{\mu}{e\gamma} \sim&\  \frac{\alpha_e m_\mu^3}{1024\pi^4\Gamma_\mu }
                      \left( \frac{ \la^\prime_e v_H}{2v_\Phi^2} c_{\mu_L} c_{\mu_R}  C_{LR} \right)^2
                       \left( \eps_{e_R}^2 s_{\mu_L}^2 + \eps_{e_L}^2 s_{\mu_R}^2 \right) \notag \\
                                \sim&\  2.3\times 10^{-14} \times
                                     \left(  \frac{ c_{\mu_L} c_{\mu_R}C_{LR}} {0.1} \right)^2
                                      \left( \frac{\la^\prime_{e}}{1.0} \right)^2
                                      \left( \frac{1.0\ \mathrm{TeV}}{v_\Phi} \right)^4
                                      \left( \frac{ \eps_{e_R}^2 s_{\mu_L}^2 + \eps_{e_L}^2 s_{\mu_R}^2 }{10^{-12}} \right), \\
 \br{\tau}{\mu\gamma} \sim&\
                      \frac{\alpha_e m_\tau^3}{1024\pi^4\Gamma_\tau}
                      \left( \frac{ \la^\prime_e v_H}{2v_\Phi^2} c_{\mu_L} c_{\mu_R}  C_{LR} \right)^2
                       \left(\eps_{\tau_L}^2 s_{\mu_R}^2 + \eps_{\tau_R}^2 s_{\mu_L}^2  \right) \notag \\
                                \sim&\  1.5 \times 10^{-9} \times
                                     \left(  \frac{ c_{\mu_L} c_{\mu_R}C_{LR}} {0.1} \right)^2
                                      \left( \frac{\la^\prime_{e}}{1.0} \right)^2
                                      \left( \frac{1.0\ \mathrm{TeV}}{v_\Phi} \right)^4
                                 \left( \frac{ \eps_{\tau_R}^2 s_{\mu_L}^2 + \eps_{\tau_L}^2 s_{\mu_R}^2 }{10^{-4}} \right),
\label{eq-tamgam}
\end{align}
with $C_{LR}$ given in Eq.~\eqref{eq-CLR}.
Here, $\eps_{e_{L,R}}$ ($\eps_{\tau_{L,R}}$) are the mixing angles
between the left- and right-handed electrons (tauons) and the VL leptons, respectively.
$\eps_{e_{L,R}} \lesssim 10^{-6}$ and $\eps_{\tau_{L,R}} \lesssim 10^{-2}$ are required to suppress 
$\mu\to e\gamma$ and $\tau \to \mu\gamma$, respectively. 
Once both of these processes are suppressed, $\tau \to e \gamma$ is automatically suppressed as well.

\subsubsection{\boldmath\texorpdfstring{$\ell_i \to \ell_j \ell_k \ell_l$}{}\unboldmath}
\label{sec-LFV-l3l}
The neutral bosons also mediate LFV three-body decays,
such as $\mu^- \to e^-e^+e^-$, $\tau^-\to \mu^- \mu^+ \mu^-$ and so on.
Effective interactions relevant for a decay $\ell_i^- \to \ell^-_j \ell^+_k \ell^-_j$ are
\begin{align}
 \lag_{\ell_i^- \to \ell^-_j \ell^+_k \ell^-_j}
=&\  B^{ijkj}_{LL} \left(\ol{\ell}_i P_L \ell_j \right)\left( \ol{\ell}_k P_L \ell_j \right)
      +B^{ijkj}_{LR} \left(\ol{\ell}_i P_L \ell_j \right)\left( \ol{\ell}_k P_R \ell_j \right)
    \notag \\
    &+ C^{ijkj}_{LL} \left(\ol{\ell}_i \gamma^\mu P_L \ell_j \right)
                          \left(\ol{\ell}_k \gamma_\mu P_L \ell_j  \right)
      + C^{ijkj}_{LR} \left(\ol{\ell}_i \gamma^\mu P_L \ell_j  \right)
                           \left(\ol{\ell}_k \gamma_\mu P_R \ell_j  \right)
   \notag \\
     & + (L \leftrightarrow R) + \mathrm{h.c.}\; .
\end{align}
The branching fraction is given by~\cite{Kuno:1999jp,Okada:1999zk} 
\begin{align}
 \br{\ell^-_i}{\ell^-_j \ell^+_k \ell^-_j} =&\ 
       \frac{m_{\ell_i}^5}{1536\pi^3 \Gamma_{\ell_i}} \\ \notag  
             &\times  \left[ 2\left(\abs{C^{ijkj}_{LL}}^2+\frac{\abs{B^{ijkj}_{LL}}^2}{16} \right)
                     +\abs{C^{ijkj}_{LR}-\frac{1}{2}B^{ijkj}_{RL}}^2
                      + (L\leftrightarrow R) \right],
\end{align}
where masses of daughter leptons 
are neglected.
Interactions relevant for $\ell_i^- \to \ell_j^- \ell_j^+ \ell_k^- $, $k\ne j$ (see Table \ref{tab-obsL}) are given by
\begin{align}
 \lag_{\ell_i^- \to \ell_j^- \ell_j^+ \ell_k^-}  =&\
 B_{LL}^{ikjj} \left(\ol{\ell}_i P_L \ell_k \right) \left(\ol{\ell}_j P_L \ell_j\right)
+ B_{LL}^{ijjk} \left(\ol{\ell}_i P_L \ell_j \right) \left(\ol{\ell}_j P_L \ell_k\right)
 \notag \\
&\quad
    +B_{LR}^{ikjj}\left(\ol{\ell}_i P_L \ell_k \right) \left(\ol{\ell}_j P_R \ell_j\right)
     +  B_{LR}^{ijjk} \left(\ol{\ell}_i P_L \ell_j \right) \left(\ol{\ell}_j P_R \ell_k\right)
\notag \\
&\quad\quad
  + C_{LL}^{ikjj} \left(\ol{\ell}_i\gamma^\mu P_L\ell_k \right)
                         \left(\ol{\ell}_j\gamma_\mu P_L \ell_j \right)
   +  C_{LL}^{ijjk} \left(\ol{\ell}_i\gamma^\mu P_L \ell_j\right)
                         \left(\ol{\ell}_j\gamma_\mu P_L \ell_k \right)
\notag \\
&\quad\quad\quad  +
      C_{LR}^{ikjj} \left(\ol{\ell}_i \gamma^\mu P_L \ell_k\right)
                          \left(\ol{\ell}_j \gamma_\mu P_R \ell_j \right)
    +  C_{LR}^{ijjk} \left(\ol{\ell}_i \gamma^\mu P_L\ell_j \right)
                         \left(\ol{\ell}_j\gamma_\mu P_R \ell_k \right)
\notag \\
&\quad\quad\quad\quad +(L \leftrightarrow R) + \mathrm{h.c.}\; .
\end{align}
The branching fraction is given by~\cite{Brignole:2004ah}  
\begin{align}
\br{\ell_i^-}{\ell_j^- \ell_j^+ \ell_k^-}
 =&\ \frac{m_{\ell_i}^5}{1536\pi^3 \Gamma_{\ell_i}}
 \left[
\abs{C_{LL}^{ikjj}+C_{LL}^{ijjk}}^2
        + \abs{C_{LR}^{ijjk}-\frac{1}{2}B_{RL}^{ikjj} }^2
        + \abs{C_{LR}^{ikjj}-\frac{1}{2}B_{RL}^{ijjk} }^2  \right.
 \notag \\
    &\left.\quad +\frac{1}{4}\left(
    \abs{B_{LL}^{ikjj}}^2+\abs{B_{LL}^{ijjk}}^2+
                  \text{Re}\left(B_{LL}^{ijjk}{B_{LL}^{ikjj}}^*\right)
  \right)
\right]  + (L\leftrightarrow R).
\end{align}
In this model, the Wilson coefficients are given by
\begin{align}
&\ B^{ijkl}_{LL}= - \sum_{S=h,\chi,\sigma}
 \frac{1}{2m_{S}^2}\left[\hat{Y}_e^{S}\right]_{ij}\left[\hat{Y}_e^{S} \right]_{kl},
\quad
   B^{ijkl}_{LR}=- \sum_{S=h,\chi,\sigma} \frac{1}{2m_{S}^2}
       \left[\hat{Y}_e^{S}\right]_{ij}\left[\hat{Y}_e^{S} \right]^*_{lk},
\\ \notag
&\ B^{ijkl}_{RL}=- \sum_{S=h,\chi,\sigma}\frac{1}{2m_{S}^2}
      \left[\hat{Y}_e^{S} \right]^*_{ji} \left[\hat{Y}_e^{S} \right]_{kl},
\quad
 B^{ijkl}_{RR}=- \sum_{S=h,\chi,\sigma} \frac{1}{2m_{S}^2}
      \left[\hat{Y}_e^{S} \right]^*_{ji} \left[\hat{Y}_e^{S}\right]^*_{lk},
\\ \notag
&\ C^{ijkl}_{XY}= \sum_{V=Z, Z'} \frac{1}{m_{V}^2}
  \left[\hat{g}_{e_X}^V \right]_{ij}\left[\hat{g}_{e_Y}^{V} \right]_{kl},
\end{align}
where $X,Y = L, R$.

These LFV three body decays are dominated by $\Zp$ boson exchange.
Using the result in Appendix~\ref{sec-anal}, 
the $\Zp$ contributions to $\mu^-\to e^-e^+e^-$ and $\tau^- \to \mu^-\mu^+\mu^-$ are estimated as
\begin{align}
 \br{\mu^-}{e^-e^+e^-} \sim&\   \frac{m_\mu^5}{1536 \pi^3\Gamma_\mu}
                                   \frac{s_\mu^2 \eps_e^6 }{2 v_\Phi^4}  
                                   =
                                       2.3 \times 10^{-40} \times
                                      \left( \frac{ s_\mu}{1/\sqrt{2} } \right)^2
                                      \left( \frac{  \eps_e}{ 10^{-6} } \right)^6
                                      \left( \frac{1.0\ \mathrm{TeV}}{v_\Phi} \right)^4,    \\
 \br{\tau^-}{\mu^-\mu^+\mu^-} \sim&\  \frac{m_\tau^5}{1536 \pi^3\Gamma_\tau}
                                              \frac{s_\mu^6 \eps_{\tau}^2 }{2 v_\Phi^4}
                                   =  1.0 \times 10^{-9} \times
                                      \left( \frac{ s_\mu}{1/\sqrt{2} } \right)^6
                                      \left( \frac{  \eps_\tau}{ 10^{-2} } \right)^2
                                      \left( \frac{1.0\ \mathrm{TeV}}{v_\Phi} \right)^4,
\end{align}
where $s_\mu$, $\eps_e$, and $\eps_\tau$ are
the maximum values of $s_{\mu_{L,R}}$, $\eps_{e_{L,R}}$ and $\eps_{\tau_{L,R}}$, respectively.
$\br{\mu}{3e}$ is strongly suppressed by $\eps_e^6$
and will be much smaller than $\br{\mu}{e\gamma}$.
On the other hand,  $\br{\tau}{3\mu}$ scales as $\eps_\tau^2$,
and therefore in the same way as $\br{\tau}{\mu\gamma}$.
$\br{\tau}{3\mu} \gtrsim \br{\tau}{\mu\gamma}$ is expected
because the former is proportional to an absolute sum of different chirality structures,
while the latter is dominated by the left-right mixing effect.
All other $\tau$ decays are suppressed by additional factors of $\eps_e$.

\subsection{EW Bosons}
\label{sec-EWBoson}
The fermion couplings to the SM bosons, namely Higgs, $W$ and $Z$ bosons,
are also modified by the mixing to VL fermions.
This might affect their decays.
For instance, LFV Higgs boson decays are predicted in models with VL leptons
studied in Refs.~\cite{Dermisek:2013gta,Altmannshofer:2016oaq,Poh:2017tfo}.
All observables here are calculated at tree-level, except for $h\to \gamma\gamma$.
All formulae that we use to compute two-body decays are summarized in Appendix~\ref{sec-twobody}.
Table~\ref{tab-EWconst} summarizes experimentally determined values of constants used in the EW boson observables.
Experimental central values and uncertainties of the relevant observables are summarized in Table~\ref{tab-obsB}.

\begin{table}[t]
 \centering
\caption{\label{tab-EWconst}
Values of constants for EW boson sector.
Masses and widths are from Ref.~\cite{Tanabashi:2018oca}.  
}
\begin{tabular}[t]{cc|cc} \hline\hline
$m_W$ [GeV]& 80.379   $\pm 0.012$  
& $\Gamma_W$ [GeV] & 2.085 $\pm 0.042$  \\
$m_Z$ [GeV]&  91.1876$\pm 0.0021$ 
& $\Gamma_Z$ [GeV]&2.4952$\pm 0.0023$    \\
$m_h$ [GeV]& 125.18$\pm 0.16$ 
& $\Gamma_h$ [MeV]& 4.07$\pm 0.16$    \\ \hline
$g(m_Z)$~\cite{Antusch:2013jca} & 0.65184$\pm 0.00018$ & 
$\alpha_s(m_Z)$ & 0.1181$\pm 0.0011 $ \\
$s_W^2$~\cite{Tanabashi:2018oca} & 0.22343$\pm 0.00007 $
& $\ol{s}_\ell^2$~\cite{Tanabashi:2018oca} &0.23154$\pm 0.00003$   \\
\hline\hline
\end{tabular}
\end{table}

\begin{table}[p]
 \centering
\caption{\label{tab-obsB}
Central values and uncertainty of the observables for EW gauge bosons.
}
\begin{tabular}{c|c|c|c} \hline
Name                        & Exp.      & Unc.                & Remark   \\ \hline\hline
$\br{W^+}{e^+ \nu}$       & 0.1086 & 0.1 $\%$  & SM  \\
$\br{W^+}{\mu^+ \nu}$  & 0.1086 & 0.1 $\%$  & SM  \\
$\br{W^+}{\tau^+ \nu}$  & 0.1085 & 0.1 $\%$  & SM  \\
$\br{W^+}{\text{had}}$  & 0.6656 & 3.76 $\%$  & SM \\
$\br{W^+}{c \ol{s}}$       & 0.3239 & 10      $\%$  & SM \\
\hline
$\br{Z}{e^+ e^-}$           & 0.03333 & 0.187 $\%$  & SM  \\
$\br{Z}{\mu^+ \mu^-}$   & 0.03333 & 0.187 $\%$  & SM \\
$\br{Z}{\tau^+ \tau^-}$   & 0.03326 & 0.187 $\%$  & SM \\
$\br{Z}{\text{had}}$           & 0.6766 & 3.76 $\%$  & SM \\
$\br{Z}{(u\ol{u}+c\ol{c})}/2$   & 0.1157 & 3.76 $\%$  & SM \\
$\br{Z}{(d\ol{d}+s\ol{s}+b\ol{b})}/3$   & 0.1483 & 3.76 $\%$  & SM\\
$\br{Z}{c\ol{c}}$   & 0.1157 & 3.76 $\%$  & SM \\
$\br{Z}{b\ol{b}}$   & 0.1479 & 3.76 $\%$  & SM \\ \hline
$\br{Z}{e\mu}$       & 0.0 &3.8$\times 10^{-7}$   & Ref.~\cite{Tanabashi:2018oca} \\
$\br{Z}{e\tau}$       & 0.0 & 5.0$\times 10^{-6}$  & Ref.~\cite{Tanabashi:2018oca} \\
$\br{Z}{\mu\tau}$   & 0.0 & 6.1$\times 10^{-6}$ & Ref.~\cite{Tanabashi:2018oca} \\
\hline
$A_e$     &  0.1469 & 1 $\%$  & SM  \\
$A_\mu$ &  0.1469 & 10 $\%$   & SM  \\
$A_\tau$ & 0.1469 &  1 $\%$  & SM   \\
$A_s$      &  0.9406 & 10 $\%$ & SM   \\
$A_c$      &  0.6949 & 1 $\%$   & SM  \\
$A_b$      &  0.9406 & 1 $\%$   & SM    \\
\hline
$\mu_{\mu\mu}$                & 0.0    & 1.3  &     Ref.~\cite{Tanabashi:2018oca} \\
$\mu_{\tau\tau}$                & 1.12  &  0.23  &   Ref.~\cite{Tanabashi:2018oca} \\
$\mu_{bb}$                         &   0.95& 0.22  &   Ref.~\cite{Tanabashi:2018oca} \\
$\mu_{\gamma\gamma}$ & 1.16 & 0.18  &     Ref.~\cite{Tanabashi:2018oca} \\
$\br{h}{ee}$                             & 0.0    & 9.7$\times 10^{-4}$  &    Ref.~\cite{Tanabashi:2018oca} \\
$\br{h}{e\mu}$                         & 0.0    & 1.8$\times 10^{-4}$   &    Ref.~\cite{Tanabashi:2018oca} \\
$\br{h}{e\tau}$                         & 0.0    & 3.1 $\times 10^{-3}$ &    Ref.~\cite{Tanabashi:2018oca} \\
$\br{h}{\mu\tau}$                     & 0.0    & 1.3 $\times 10^{-3}$  & Ref.~\cite{Tanabashi:2018oca} \\
 \hline 
\end{tabular}
\end{table}

\subsubsection{\texorpdfstring{$\boldsymbol{W}$}{W} Boson Decays}
There is a right-handed charged current coupling to the $W$ boson which is absent in the SM.
Furthermore, the non-unitarity of the CKM and PMNS matrices can affect the $W$ boson couplings.
These will alter $W$ boson decays.

The branching fractions for $W$ boson decays are given by
\begin{align}
&\br{W^+}{e_i^+ \nu}
= \frac{m_W}{48\pi\Gamma_W} (1-x^W_{e_i})^2(2+x_{e_i}^W) \sum_{k=1}^3 \abs{\left[\hg_{\ell_L}^W \right]_{ki}}^2 ,
\\
&\br{W^+}{u_i \ol{d}_j}
=    \frac{m_W}{8\pi \Gamma_W} \lambda ( x^W_{u_i}, x^W_{ d_i})
\left[ \left(\abs{\left[\hg^W_{q_L}\right]_{ij}}^2+\abs{\left[\hg^W_{q_R}\right]_{ij}}^2\right) \right.  \notag \\
& \left. \hspace{1.7cm} \times
\left(1-\frac{x^W_{u_i}+x^W_{d_j}}{2}-\frac{(x^W_{u_i}-x^W_{d_j})^2}{2}\right)
+ 6\  \text{Re}\ \left( \left[\hg^W_{q_L}\right]_{ij} \left[\hg^W_{q_R}\right]_{ij}\right)
\sqrt{x^W_{u_i} x^W_{d_j}}
\right],
\end{align}
where $ x^W_{f_i}:={m_{f_i}^2}/{m_W^2}$ $(f=e,u,d)$.
The function $\lambda$ is defined in Eq.~\eqref{eq-lamfun}.

SM predictions for these decays are calculated by replacing
\begin{align}
\left[\hat{g}^W_{\ell_L}\right]_{ij} \to \frac{g}{\sqrt{2}} \left[V_\text{PMNS}\right]_{ij}  ,\quad
\left[\hat{g}^W_{q_L}\right]_{ij}    \to \frac{g}{\sqrt{2}} \left[V_\text{CKM}\right]_{ij},\quad
\hat{g}^W_{q_R} \to 0.
\end{align}
Here, experimental absolute values of the PMNS and CKM matrix elements are used.
For the leptonic decays, radiative corrections and experimental uncertainties are small.
We use a $0.1 \%$ relative uncertainty for these decays.
For the hadronic decay modes, QCD corrections may change the values by a 
factor proportional to $\alpha_s/\pi \sim 0.038$.
We use a relative uncertainty of $3.8\%$ for the total hadronic branching fraction,
while we use a $10\%$ relative uncertainty for $W\to cs$ because experimental 
uncertainties here are still large.

\subsubsection{\texorpdfstring{$\boldsymbol{Z}$}{Z} Decays and Asymmetry Parameters}
The $Z$ boson couplings are, in general, changed by the mixing effects.
This can affect the branching fractions and asymmetry parameters of $Z$ decays.
The $Z$ boson couplings depend on the weak mixing angle $\theta_W$.
For the lepton couplings,
we use the effective angle $\ol{s}_\ell = 0.23154$ including radiative corrections,
while the tree-level value $s_W = 0.22343$ is used for the quark couplings \cite{Tanabashi:2018oca}.
Using the effective angle is necessary to reproduce the observed values of $A_{\ell=e,\mu,\tau}$.

The branching fractions for flavor conserving decays are given by
\begin{align}
 \br{Z}{f_i \ol{f}_i} =
 \frac{N_c^{f} m_Z}{24\pi \Gamma_Z} \sqrt{1-4x^Z_{f_i}} &\    \left[
   \left( \abs{\left[\hg^Z_{f_L}\right]_{ii}}^2+\abs{\left[\hg^Z_{f_R}\right]_{ii}}^2 \right)(1-x^Z_{f_i}) \right. \\ \notag
   &\ \left. \quad\quad\quad\quad
          + 6 x^Z_{f_i}\ \text{Re}\left( \left[\hg^Z_{f_L}\right]^*_{ii} \left[\hg^Z_{f_R}\right]_{ii} \right)
\right],
\end{align}
where $N_c^{f}$ is a color factor and  $x^Z_{f_i} \define {m_{f_i}^2}/{m_Z^2}$ for a fermion $f_i$. 
Those for flavor violating decays are given by
\begin{align}
 \mathrm{BR}\left( Z \to f^\pm_i f^\mp_j\right) =&\
 \frac{N_c^{f} m_Z}{12\pi\Gamma_Z} \lambda(x^Z_{f_i},x^Z_{f_j})    \left[
 \left(\abs{\left[\hg^Z_{f_L}\right]_{ij}}^2+\abs{\left[\hg^Z_{f_R}\right]_{ij}}^2 \right)
   \right.  \\ \notag
 & \left.  \times \left(1-\frac{x^Z_{f_i}+x^Z_{f_j}}{2}-\frac{(x^Z_{f_i}-x^Z_{f_j})^2}{2}  \right)
 +6 \sqrt{x^Z_{f_i} x^Z_{f_j}}\ \mathrm{Re}\left( \left[\hg^Z_{f_L}\right]^*_{ij}
\left[\hg^Z_{f_R}\right]_{ij} \right)
\right].
\end{align}
For $\br{Z}{\mathrm{had}}$, 
the experimental value of a four-body decay $\br{Z}{b\ol{b}b\ol{b}} = 0.00036$~\cite{Tanabashi:2018oca}
is added to the sum of two-body decays to quarks. 
The $Z$-pole asymmetry parameters are defined as
\begin{align}
 A_{f_i} \define \frac{2 g_{V_{f_i}}  g_{A_{f_i}} }{g_{V_{f_i}}^2+g_{A_{f_i}}^2},
\end{align}
where $f_i = e,\mu,\tau,s,c,b$, and (axial-)vector couplings are obtained as 
\begin{align}
 g_{V_{f_i}} = \left[ \hg^Z_{f_L} + \hg^Z_{f_R} \right]_{ii},\quad
 g_{A_{f_i}} = \left[ \hg^Z_{f_L} - \hg^Z_{f_R} \right]_{ii}.
\end{align}

SM predictions for these observables  
are obtained by formally replacing $\Pfb,P_5 \to \id_3, 0_3$ in Eq.~\eqref{eq-Zcoup}.
The leading QED and QCD corrections to these decays 
are proportional to $3\alpha_e/4\pi\sim 0.0019$ and $\alpha_s/\pi\sim0.038$ for 
leptonic or hadronic decays, respectively. 
We, therefore, use a $0.19\%~(3.8\%)$ relative uncertainty on the leptonic (hadronic) decays.
The relative uncertainties for the asymmetry parameters are taken as $1\%$ for $A_e, A_\tau, A_c, A_b$,
and $10 \%$ for $A_\mu$ and $A_s$, consistent with their experimental uncertainties.
The uncertainties for flavor violating decays are determined from their experimental upper bounds.

Let us illustrate how the SM boson couplings, in general, are very close to their SM values.
We show this analytically in Appendix~\ref{sec-anal}. 
In general, one finds the mixing to be suppressed by
$\sim \eps_{f_\mathrm{SM}}^2 m_{f_\mathrm{SM}}^2/M_{F_\mathrm{VL}}^2$.   
Considering, for example, the left-handed $Z-\mu\tau$ coupling we find that it can be estimated as
\begin{align}
 \left[\hg^Z_{e_L}\right]_{23} \sim \frac{g}{2c_W} s_{\mu_L} \eps_{\tau_L} \frac{m_\mu m_\tau}{M_{E_L}^2}
                                             \sim   2.0 \times 10^{-9} \times
                                                       \left(\frac{s_{\mu_L}}{1/\sqrt{2}} \right)
                                                       \left(\frac{\eps_{\tau_L}}{0.01}\right)
                                                       \left(\frac{500\ \mathrm{GeV} }{M_{E_L}}\right)^2.
\end{align}
Corrections for lighter flavors are even more suppressed.

\subsubsection{Higgs Decays}
The Higgs boson couplings to SM fermions can, in general, depart from their SM predictions 
due to misalignment of the Yukawa couplings and mass matrices.
We have studied the signal strengths for the measured decay modes to $\mu\mu$, $\tau\tau$, $bb$, and $\gamma\gamma$ 
final states as well as the branching fractions for the unobserved decays $ee$, $e\mu$, $e\tau$, and $\mu\tau$.
Central values and uncertainties are set to their experimentally observed values.
Decay widths for flavor conserving decays are given by
\begin{align}
 \Gamma(h \to f_i \ol{f}_i)  =  \frac{m_h}{16\pi} \sqrt{1-4x^h_{f_i}}
\left[
 \abs{\left[\hat{Y}^h_f\right]_{ii}}^2
-4x^h_{f_i}  \left(\mathrm{Re} \left[\hat{Y}^h_f \right]_{ii}\right)^2
\right],
\end{align}
and those for flavor violating decays are
\begin{align}
\Gamma\left(h\to f_i^\pm f_j^\mp\right) =&\
 \frac{m_h}{16\pi} \lambda(x^h_{f_i},x^h_{f_j}) \\ \notag
& \hspace{-2.0cm} \times   \left[
 \left(
 \abs{\left[\hat{Y}^h_f \right]_{ij}}^2 +  \abs{\left[\hat{Y}^h_f \right]_{ji}}^2
\right) (1-x^h_{f_i}-x^h_{f_j})
-4\mathrm{Re}\left(\left[\hat{Y}^h_f \right]_{ij} \left[\hat{Y}^h_f \right]_{ji}\right) \sqrt{x^h_{f_i}x^h_{f_j}}
\right],
\end{align}
where $x^h_{f_i} \define m_{f_i}^2/m_h^2$.

In addition to the tree-level decays,
the VL families may significantly contribute to the loop-induced decay,
$h \to \gamma \gamma$ and $h\to gg$.
The decay width for $h\to\gamma\gamma$ is given by~\cite{Djouadi:2005gi}
\begin{align}
\label{eq-haa}
 \Gamma(h\to \gamma \gamma) = \frac{G_F \alpha_e^2 m_H^3}{128 \sqrt{2} \pi^3}
\abs{
A_1^H (\tau_W) +
\sum_{f_A}  N^f_c Q_f^2 \left(\frac{y_{f_A}  v_H}{m_{f_A}}\right) A^H_{1/2}(\tau_{f_A})
}^2,
\end{align}
where $\tau_I = m_H^2/(4m_I^2)$ with $I=f_A, W$.
Here, $f_A$ runs over all the fermions in this model and $A=1,\dots,5$ is a flavor index.
$N_c^f$ is the number of color degrees of freedom and $y_{f_A} \define [\hat{Y}^h_f ]_{AA}$ 
is a diagonal Yukawa coupling constant of the Higgs boson
to a fermion $f_A$.
The form factors are given by
\begin{align}
 A_{1/2}^H(\tau) = \frac{2}{\tau^2} \left[ \tau+(\tau-1)f(\tau)\right],\quad
 A_{1}^H(\tau)    = -\frac{1}{\tau^2} \left[ 2\tau^2 + 3\tau+3(2\tau-1)f(\tau)\right],
\end{align}
where
\begin{align}
 f(\tau) =
\begin{cases}
 \text{arcsin}^2\sqrt{\tau} & \tau \le 1, \\
 -\dfrac{1}{4} \left[\log{\dfrac{1+\sqrt{1-\tau^{-1}}}{1-\sqrt{1-\tau^{-1}}}} -i\pi \right]^2 & \tau>1.
\end{cases}
\end{align}
Similarly, the decay width of $h\to gg$ is given by~\cite{Djouadi:2005gi}
\begin{align}
 \Gamma(h\to g g) = \frac{G_F \alpha_s^2 m_H^3}{36\sqrt{2} \pi^3}
\abs{
\frac{3}{4}
\sum_{q_A}  \left(\frac{y_{q_A} v_H}{m_{q_A}}\right) A^H_{1/2}(\tau_{q_A})
}^2,
\end{align}
where  $q_A$ only runs over the quarks.

Naively, one expects the size of VL fermion contributions 
to these one-loop decays to be suppressed by the squared ratio of $\mathrm{SU}(2) \times \U1$ breaking mass to VL mass, 
i.e.\ by a factor $(\lambda_e^\prime v_H)^2/M_{F_{VL}}^2$.
This is exactly what we find. 
Using the result of Appendix~\ref{sec-anal},
contributions from VL fermions $F_L$, $F_R$ are given by 
\begin{align}
 \sum_{f = F_L, F_R} \left(\frac{y_f v_H}{m_f}\right) A^H_{1/2}(\tau_f)
 \sim&\  \left(\la^\prime_f v_H\right)^2 \frac{A^H_{1/2}(\tau_{F_L})-A^H_{1/2}(\tau_{F_R})}
                                                                   {\tM^2_{F_L}-\tM^2_{F_R} }
\sim   - \frac{7}{90}\frac{m_H^2 \left(\la^\prime_f v_H\right)^2}{M_{F_L}^2M_{F_R}^2}
\\ \notag
\sim&\  - 5.9 \times 10^{-4} \times\left(\frac{\la^\prime_f}{1.0}\right)^2
                                      \left(\frac{500\ \mathrm{GeV}}{\sqrt{M_{F_L}M_{F_R}}}\right)^4,
\end{align}
where $\tM_{F_{L(R)}}$ is the approximated mass of VL fermion $F_{L(R)}$ defined in Eq.~\eqref{eq-Mtilde}.
Here, $\la^\prime_f v_H \ll M_{F_{L,R}}$ and $\la_f \ll \la^\prime_f$ have been assumed.
For the last equality in the first line, 
we use the series expansion $A^H_{1/2}(\tau) \approx 4/3 + 14/45\ \tau$ 
around $\tau\approx0$.
A possible cancellation between the two VL fermions gives an extra suppression.
Altogether, we confirm that the VL fermions only give very small corrections to these decay rates.
Especially VL quarks will be heavy, and their effects therefore particularly suppressed.
Thus, there is no meaningful constraint arising from $h\to gg$ for our analysis, 
and also the Higgs boson production rate is unchanged with respect to the SM.

The signal strengths of the Higgs boson are defined as
\begin{align}
\mu_{XX} \define \frac{\sigma^\mathrm{prod}\cdot \br{h}{XX}}
                                 {\sigma^\mathrm{prod}_\mathrm{SM}\cdot{\br{h}{XX}}_\mathrm{SM}}
            \simeq  \frac{\br{h}{XX}}{{\br{h}{XX}}_\mathrm{SM}}
          , \quad
X = \mu, \tau, \gamma, b.
\end{align}

\subsection{Quarks}
We study the SM quark masses,
9 absolute values of the CKM matrix elements and 3 CP phases $\alpha, \beta, \gamma$.
The Wilson coefficients relevant to the $\bsll$ processes are fitted to explain the anomalies.
The new physics contributions will also affect neutral meson mixing,
namely $\MMbar{K}$, $\MMbar{B_d}$, $\MMbar{B_s}$ and $\MMbar{D}$ mixing,
(semi-)leptonic decays of B mesons and top quark decays.
Central values and uncertainties for quark masses and the CKM elements
are listed in Table~\ref{tab-mmQ}.
Values for the other observables are listed in Table~\ref{tab-obsQ}.
We do not assume unitarity of the CKM matrix
for our analysis and our parameters are fit directly to 
the values determined by experimental measurements.
\begin{table}[t]
 \centering
\caption{\label{tab-mmQ}
Central values and uncertainties of quark masses and CKM mixing parameters.}
\begin{tabular}{c|c|c|c} \hline
Name                        & Exp.      & Unc.    & Remark   \\ \hline\hline
$m_u(m_Z)$ [MeV] & 1.29  &   0.39    & Ref.~\cite{Antusch:2013jca}\\
$m_c(m_Z)$ [MeV] & 627  &  19.    & Ref.~\cite{Antusch:2013jca}    \\
$m_t(m_Z)$ [GeV]  & 171.7 &  1.5   & Ref.~\cite{Antusch:2013jca}  \\
$m_d(m_Z)$ [MeV] & 2.75  &   0.29    & Ref.~\cite{Antusch:2013jca} \\
$m_s(m_Z)$ [MeV] & 54.3  &  2.9    & Ref.~\cite{Antusch:2013jca}   \\
$m_b(m_Z)$ [GeV] & 2.853   & 0.026   & Ref.~\cite{Antusch:2013jca}  \\ \hline 
$\abs{V_{ud}}$      & 0.97420&0.00021 &   Ref.~\cite{Tanabashi:2018oca} \\
$\abs{V_{us}}$      & 0.2243  &0.0005      &   Ref.~\cite{Tanabashi:2018oca} \\
$\abs{V_{ub}}$      & 0.00394&0.00036 &   Ref.~\cite{Tanabashi:2018oca} \\
$\abs{V_{cd}}$      & 0.218    &0.004       &   Ref.~\cite{Tanabashi:2018oca} \\
$\abs{V_{cs}}$      & 0.997    & 0.017    &   Ref.~\cite{Tanabashi:2018oca} \\
$\abs{V_{cb}}$      & 0.0422 & 0.0008 &   Ref.~\cite{Tanabashi:2018oca} \\
$\abs{V_{td}}$      &  0.0081&0.0005 &   Ref.~\cite{Tanabashi:2018oca} \\
$\abs{V_{ts}}$      & 0.0394 &0.0023  &   Ref.~\cite{Tanabashi:2018oca} \\
$\abs{V_{tb}}$      & 1.019   &0.025    &   Ref.~\cite{Tanabashi:2018oca} \\  \hline
$\alpha $ [rad]       & 1.475   & 0.097    &   Ref.~\cite{Tanabashi:2018oca} \\
$\sin 2\beta $        & 0.691     & 0.017       &   Ref.~\cite{Tanabashi:2018oca} \\
$\gamma $ [rad]    &  1.283  & 0.081    &   Ref.~\cite{Tanabashi:2018oca} \\
 \hline
\end{tabular}
\end{table}
\begin{table}[th]
 \centering
\caption{\label{tab-obsQ}
Values of observables of the quark sector.
Theoretical uncertainties are included for $\Delta M_K$, $\Delta M_d$, $\Delta M_s$
and $\eps_K$.
Values for $C_{9,10}^{('), e,\mu}$ are discussed in the text.
}
\begin{tabular}{c|c|c|c} \hline
Name                        & Exp.      & Unc.    & Remark   \\ \hline\hline
$\Delta M_K$  [ps$^{-1}$] & 0.005293   & $0.0022$    &  Ref.~\cite{Tanabashi:2018oca} \\
$\abs{\eps_K}$ $\times 10^{3}$& 2.228  &  0.21 &  Ref.~\cite{Tanabashi:2018oca} \\
$\Delta M_d$ [ps$^{-1}$] & 0.5065 & 0.081 &  Ref.~\cite{Tanabashi:2018oca} \\
$S_{\psi K_S}$  & $\ 0.695 $ & $0.019$ &  Ref.~\cite{Amhis:2016xyh} \\ 
$\Delta M_s$ [ps$^{-1}$]  & $17.757 $& $2.5  $ & Ref.~\cite{Tanabashi:2018oca} \\
$S_{\psi\phi}$ & $0.021 $ & $0.031$ & Ref.~\cite{Amhis:2016xyh}  \\
$\abs{\Delta^\text{NP} x_D}$ [$\%$]& 0.0  & 0.5 &  Ref.~\cite{Tanabashi:2018oca}  \\
$R_K^{\nu\ol{\nu}}$      & 1.0 & 2.6 & Ref.~\cite{Lees:2013kla} \\
$R_{K^*}^{\nu\ol{\nu}}$ & 1.0 & 2.7 & Ref.~\cite{Lutz:2013ftz} \\  \hline
$R_{B_d\to \mu\mu}$  & 1.5 & 1.4 & Refs.~\cite{ Bobeth:2013uxa,Tanabashi:2018oca} \\
$R_{B_s\to \mu\mu}$  & 0.75 & 0.16 & Refs.~\cite{Altmannshofer:2017wqy, Tanabashi:2018oca} \\   \hline
$\Gamma_t$   [GeV]           & 1.41 & 0.17 & Ref.~\cite{Tanabashi:2018oca}  \\
$\br{t}{Zq}$                   & 0.0      &  2.6 $\times 10^{-4}$    &  Ref.~\cite{Tanabashi:2018oca}  \\
$\br{t}{hu}$                   & 0.0      &  9.7 $\times 10^{-4}$       &  Ref.~\cite{Tanabashi:2018oca}  \\
$\br{t}{hc}$                   & 0.0     &  8.2 $\times 10^{-4}$     &  Ref.~\cite{Tanabashi:2018oca}  \\  \hline
$\br{B_s}{K\tau\tau}$ & 0.0     & $1.4\times 10^{-3}$ & Ref.~\cite{TheBaBar:2016xwe} \\
 \hline
\end{tabular}
\end{table}

\subsubsection{\boldmath\texorpdfstring{$\bsll$}{}\unboldmath~Processes}
\label{sec-bsll}
The relevant effective Hamiltonian for $\bsll$ is given by~\cite{Buras:1994dj,Bobeth:1999mk}, 
\begin{align}
\Heff^{\ell} = -\frac{4G_F}{\sqrt{2}} \frac{\alpha}{4\pi} V_{tb}V^*_{ts} \sum_{a=9,10}
 \left(C_a^{\ell}\Ocal_a^{\ell}+C_a^{'\ell} \Ocal_a^{'\ell}  \right),
\end{align}
where
\begin{align}
 \Ocal_9^{\ell} =&\   \left[ \ol{s} \gam^\mu P_L b  \right]
                                                             \left[ \ol{\ell} \gam_\mu \ell \right],\quad
 \Ocal_{10}^{\ell} =  \left[ \ol{s} \gam^\mu P_L b  \right]
                                                             \left[ \ol{\ell} \gam_\mu \gamma_5 \ell \right], \\
 \Ocal_9^{'\ell} =&\   \left[ \ol{s} \gam^\mu P_R b  \right]
                                                             \left[ \ol{\ell} \gam_\mu \ell \right],\quad
 \Ocal_{10}^{'\ell} =   \left[ \ol{s} \gam^\mu P_R b  \right]
                                                             \left[ \ol{\ell} \gam_\mu \gamma_5 \ell \right].
\end{align}
Here, $\ell = e, \mu, \tau$.
The Wilson coefficients induced by $\Zp$ exchange are given by
\begin{align}
 C^{\ell}_9 =&\  -\frac{\sqrt{2}}{4 G_F} \frac{4\pi}{\alpha_e}\frac{1}{V_{tb}V_{ts}^*} \frac{1}{2m^2_{Z'}}
               \cdot \left[\hg^{\Zp}_{d_L} \right]_{23} \left[\hg^{\Zp}_{e_{R}}+\hg^{\Zp}_{e_L}  \right]_{ii}, \\
 C^{\ell}_{10} =&\ -\frac{\sqrt{2}}{4 G_F} \frac{4\pi}{\alpha_e}\frac{1}{V_{tb}V_{ts}^*} \frac{1}{2m^2_{\Zp}}
               \cdot \left[\hg^{\Zp}_{d_L} \right]_{23} \left[\hg^{\Zp}_{e_{R}}-\hg^{\Zp}_{e_L}  \right]_{ii}, \\
 C^{'\ell}_9 =&\  -\frac{\sqrt{2}}{4 G_F} \frac{4\pi}{\alpha_e}\frac{1}{V_{tb}V_{ts}^*} \frac{1}{2m^2_{\Zp}}
               \cdot \left[\hg^{\Zp}_{d_R} \right]_{23} \left[\hg^{\Zp}_{e_{R}}+\hg^{\Zp}_{e_L}  \right]_{ii}, \\
 C^{'\ell}_{10} =&\ -\frac{\sqrt{2}}{4 G_F} \frac{4\pi}{\alpha_e} \frac{1}{V_{tb}V_{ts}^*} \frac{1}{2m^2_{\Zp}}
               \cdot \left[\hg^{\Zp}_{d_R} \right]_{23} \left[\hg^{\Zp}_{e_{R}}-\hg^{\Zp}_{e_L}  \right]_{ii},
\end{align}
where $i = 1,2,3$ for $\ell = e,\mu,\tau$, respectively.

In Ref.~\cite{Aebischer:2019mlg}\footnote{%
See for the similar analyses before Moriond 2019~\cite{
Altmannshofer:2017fio,Altmannshofer:2017yso,Alok:2017sui,Capdevila:2017bsm,Ciuchini:2017mik,DAmico:2017mtc,Geng:2017svp,Ghosh:2017ber,Arbey:2018ics}
and after Moriond 2019~\cite{Aebischer:2019mlg,Alguero:2019ptt,Alok:2019ufo,Ciuchini:2019usw,Datta:2019zca,Kowalska:2019ley,Arbey:2019duh,Kumar:2019nfv}},
one or two of the Wilson coefficients are fitted
to the latest data for $R_{K^{(*)}}$ and $\bsll$ decay observables,
while all the other Wilson coefficients are assumed to vanish.
There are three scenarios in the one-dimensional analysis 
that have pulls larger than 5$\sig$ with respect to the SM prediction:
\begin{align}
(\mathrm{I})  &\quad\mathrm{Re}\,C_9^{\mu}     = -0.95 \pm 0.15, \label{eq:pattern1} \\
(\mathrm{II})  &\quad \mathrm{Re}\,C_{10}^{\mu} =  0.73 \pm 0.14, \\
(\mathrm{III}) &\quad \mathrm{Re}\,C_{9}^{\mu} = - \text{Re}\,C_{10}^\mu = -0.53 \pm 0.09.
\end{align}
For the two-dimensional analysis there are two patterns
that have pulls larger than $6\sig$ with respect to the SM prediction:
\begin{align}
 (\mathrm{IV}) &\quad \mathrm{Re}\,C_{9}^{\mu} = -0.7\pm 0.3, & & \text{Re}\,C_{10}^\mu = 0.4 \pm 0.25,  \label{eq:pattern4} \\
 (\mathrm{V})  &\quad \mathrm{Re}\,C_{9}^{\mu} = -1.04\pm 0.24,
                     && \text{Re}\,C_{9}^{\prime\mu} = 0.48 \pm 0.30.
\end{align}
In our analysis, we attempt to fit $C_{9,10}^{(\prime)\ell=e,\mu}$ to one of these 5 patterns.
The central values and uncertainties of the other coefficients are assumed to be $0.0\pm 0.1$.
We also include imaginary parts, and try to fit them to $0.0\pm 0.1$.
We remark here that non-zero imaginary parts are actually favored by the analysis of Ref.~\cite{Arbey:2018ics},
however, we do not consider this possibility in the present paper.

$C_9^\mu$ is sizable in most of the above preferred patterns of Wilson coefficients.
The $\Zp$ contribution to $C_9^\mu$ can be estimated as
\begin{align}
 C_9^\mu \sim - 1.0 \times
                                         \left(\frac{0.3}{\gp}\right)
                                         \left(\frac{1\ \mathrm{TeV}}{v_\Phi} \right)^2
                                         \left( \frac{s_{\mu_L}^2 + s_{\mu_R}^2}{1.0}\right)
                                         \left(\frac{\left[\hg^\Zp_{d_L}\right]_{23}}{0.001} \right),
\end{align}
where we note that $(s_{\mu_L}^2 + s_{\mu_R}^2)\sim1$ is required by a successful explanation of $\Delta a_\mu$.
Therefore, the $\bsll$ anomalies can be explained
with small, $\order{10^{-3}}$, couplings of the SM quarks to the $\Zp$ boson.

The Wilson coefficients with $\ell=\tau$ contribute to the semi-leptonic decays 
$B_s \to K^{(*)}\tau\ol{\tau}$ and $B_s \to \phi \tau \ol{\tau}$.
Branching fractions for these decays as a function of the Wilson coefficients 
are calculated in Ref.~\cite{Capdevila:2017iqn}.

\subsubsection{Neutral Meson Mixing}
\label{sec-nmm}
The neutral bosons, $\Zp$, $\chi$ and $\sig$, give contributions to neutral meson mixing.
We neglect contributions from the $Z$ boson and the Higgs boson,
since the flavor violating couplings are expected to be small, as discussed above and shown in Appendix~\ref{sec-anal}.
We study $\MMbar{K}$, $\MMbar[d]{B}$, $\MMbar[s]{B}$ and $\MMbar{D}$ mixing~\cite{
Buchalla:1995vs,Buras:1997fb,Buras:1998raa,
Buras:2012jb,Buras:2013rqa,DiLuzio:2017fdq}.

The relevant effective Hamiltonian is given by
\begin{align}
\Hcal^{\Delta F=2}_\text{eff}= \sum_{i,a} C_i^a  Q_i^a,
\end{align}
where $(i,a) = (1,\mathrm{VLL}),(1,\mathrm{VRR}), (1,\mathrm{LR}),(2,\mathrm{LR}),
                       (1,\mathrm{SLL}), (2,\mathrm{SLL}),(1,\mathrm{SRR}), (2,\mathrm{SRR})$.
The four-fermi operators are defined as
\begin{align}
 &Q_1^\text{VLL} = \left(\ol{F}^\alpha \gamma_\mu P_L f^\alpha \right)
                                 \left(\ol{F}^\beta \gamma^\mu P_L f^\beta \right),
&\quad &
Q_1^\text{VRR} = \left(\ol{F}^\alpha \gamma_\mu P_R f^\alpha \right)
                                 \left(\ol{F}^\beta \gamma^\mu P_R f^\beta \right), \\
&Q_1^\text{LR} = \left(\ol{F}^\alpha \gamma_\mu P_L f^\alpha\right)
                                 \left(\ol{F}^\beta \gamma^\mu P_R f^\beta \right),
 &\quad &
Q_2^\text{LR} = \left(\ol{F}^\alpha P_L f^\alpha \right)
                                 \left(\ol{F}^\beta  P_R f^\beta \right), \\
&Q_1^\text{SLL} = \left(\ol{F}^\alpha P_L f^\alpha \right) \left(\ol{F}^\beta  P_L f^\beta \right),
&\quad &
Q_1^\text{SRR} = \left(\ol{F}^\alpha P_R f^\alpha \right) \left(\ol{F}^\beta  P_R f^\beta \right), \\
&Q_2^\text{SLL} = \left(\ol{F}^\alpha \sigma_{\mu\nu} P_L f^\alpha \right)
                                 \left(\ol{F}^\beta \sigma^{\mu\nu} P_L f^\beta \right),
&\quad &
Q_2^\text{SRR} = \left(\ol{F}^\alpha \sigma_{\mu\nu} P_R f^\alpha \right)
                                 \left(\ol{F}^\beta \sigma^{\mu\nu} P_R f^\beta \right),
\end{align}
where $\alpha, \beta$ are the color indices and $\sigma_{\mu\nu} = [\gamma_\mu,\ \gamma_\nu]/2$.
Here, $(F,f) = (s,d),\ (b, d),\ (b, s),\ (c, u)$ for
$K$-$\ol{K}$, $B_d$-$\ol{B}_d$, $B_s$-$\ol{B}_s$ and $D$-$\ol{D}$ mixing, respectively.

The Wilson coefficients including $\mathcal{O}(\alpha_s)$ corrections are given by~\cite{Buras:2012fs}
\begin{align}
\label{eq-CVLL}
C_1^\mathrm{VLL}(\mu)=&\ \left[ 1+\frac{\alpha_s}{4\pi}\left(-2\log\frac{m_{\Zp}^2}{\mu^2}
                          + \frac{11}{3}\right) \right]    \frac{g^{\Zp}_L g^{\Zp}_L} {2m_\Zp^2},  \\
C_1^\mathrm{LR}(\mu)= &\ \left[1+\frac{\alpha_s}{4\pi}\left(-\log\frac{m_{\Zp}^2}{\mu^2} - \frac{1}{6}\right) \right]
                                  \frac{g^{\Zp}_L g^{\Zp}_R} {m_{Z'}^2}
                                  - \left( -\frac{3}{2}\frac{\alpha_s}{4\pi} \right)
                              \sum_{S=\chi, \sigma} \frac{y^{S}_L y^{S}_R} {2m_{S}^2},
                                                  \\
C_2^\mathrm{LR}(\mu)= &\ \frac{\alpha_s}{4\pi}\left(-6\log\frac{m_{\Zp}^2}{\mu^2} - 1 \right)
                              \frac{g^{\Zp}_L g^{\Zp}_R} {m_{\Zp}^2}
                              - \left(1-\frac{\alpha_s}{4\pi}\right)
                               \sum_{S=\chi, \sigma} \frac{y^{S}_L y^{S}_R} {2m_{S}^2},  \\
C_1^\mathrm{SLL}(\mu) =&\ - \sum_{S=\chi, \sigma}
                     \left[1+\frac{\alpha_s}{4\pi}\left(-3\log\frac{m_S^2}{\mu^2} + \frac{9}{2}\right)\right]
                                        \frac{y^{S}_L y^{S}_L} {4m_{S}^2},   \\
C_2^\mathrm{SLL}(\mu)=&\  - \sum_{S=\chi, \sigma}
                                  \frac{\alpha_s}{4\pi}\left(-\frac{1}{12}\log\frac{m_S^2}{\mu^2}+\frac{1}{8}\right)
                                      \frac{y^{S}_L y^{S}_L} {4m_{S}^2},
\label{eq-CSLL}
\end{align}
where $\mu$ is the $\overline{\mathrm{MS}}$-scheme renormalization scale.
The gauge couplings $g_{L,R}^\Zp$ and Yukawa couplings $y_{L,R}^S$ are given by
\begin{align}
 g^\Zp_L = \left[\hg^\Zp_{d_L}\right]_{ij},\quad
 g^\Zp_R = \left[\hg^\Zp_{d_R}\right]_{ij},\quad
 y^S_L = \left[\hY^S_{d}\right]_{ij},\quad
 y^S_R = \left[\hY^S_{d}\right]^*_{ji}.
\end{align}
where the flavor indices are $(i,j) = (2,1), (3,1), (3,2)$
for $\MMbar{K}$, $\MMbar[d]{B}$, or $\MMbar[s]{B}$ mixing, respectively.
For $\MMbar{D}$ mixing,
\begin{align}
 g^\Zp_L = \left[\hg^\Zp_{u_L}\right]_{21},\quad
 g^\Zp_R = \left[\hg^\Zp_{u_R}\right]_{21},\quad
 y^S_L = \left[\hY^S_{u}\right]_{21},\quad
 y^S_R = \left[\hY^S_{u}\right]^*_{12}.
\end{align}
The right-right Wilson coefficients ($C_1^\mathrm{VRR}$, $C_1^\mathrm{SRR}$, $C_2^\mathrm{SRR}$)
are obtained by formally replacing $L\to R$ in the above expressions.
The off-diagonal mixing matrix element in the respective meson's mass matrix is given by
\begin{align}
 \left(M_{12}(M)\right)^* = \left(M_{12}^{\mathrm{SM}}(M)\right)^* + \frac{1}{2m_{M}}\sum_{i,a}C_i^a  \bra{\ol{M}} Q^a_i\ket{M},
\quad\text{for}\quad M=K, B_d, B_s, D.
\end{align}
Here, $m_M$ is the mass of the meson $M$.
\begin{table}[t]
\centering
\caption{\label{tab-valOs}
Numerical values of the operators $O^a_i \define \bra{\ol{M}} Q^a_i \ket{M}/(2m_M) $
at $\mu_B = $ 1 TeV.
The corresponding right-right operators have the same values as the LL operators.
}
\begin{tabular}{c|ccccc} \hline
                          & $O^{VLL}_1(\mu_B)$ &$O^{LR}_1(\mu_B)$ &$O^{LR}_2(\mu_B)$ & $O^{SLL}_1(\mu_B)$ &  $O^{SLL}_2(\mu_B)$\\ \hline\hline
$K\text{-}\ol{K}$         &0.00159& -0.159& 0.261& -0.0761& -0.132  \\
$B_d\text{-}\ol{B}_d$ &0.0465  & -0.186 & 0.241& -0.0909& -0.167 \\
$B_s\text{-}\ol{B}_s$ & 0.0701  & -0.264 & 0.338& -0.136& -0.252 \\
$D\text{-}\ol{D}$        &0.0162   &  -0.157& 0.227& -0.0845& -0.152 \\
\hline
\end{tabular}
\end{table}
Values of the operators 
\begin{equation}
O_i^a\define  \bra{\ol{M}} Q^a_i(\mu_B)  \ket{M}/(2m_M) 
\end{equation}
at $\mu_B=$ 1 TeV according to our own evaluation are listed in Table~\ref{tab-valOs}.
For this we have used input values of meson masses, decay constants and quark masses 
which are listed in Table~\ref{tab-Cmesonmixing}.
Values of hadronic matrix elements are taken from the results of the respective lattice collaborations.
We refer to Ref.~\cite{Aoki:2019cca} for hadronic matrix elements of $\MMbar{K}$ and $f^2_{B_q} \hat{B}_{B_q}$.
Those for $f_{B_q}^2 B^{(2\text{-}5)}_{B_q}$ and $B_D^{(1\text{-}5)}$
are taken from Refs.~\cite{Carrasco:2013zta,Bazavov:2016nty}
and Ref.~\cite{Carrasco:2014uya}, respectively.
The QCD running between the respective lattice scales and $\mu = 1\,\mathrm{TeV}$
has been calculated based on the anomalous dimensions shown in Ref.~\cite{Buras:2001ra}.

SM contributions to $K\text{-}\ol{K}$, $B_q \text{-}\ol{B}_q$ $(q=d,s)$ mixing are given by
\begin{align}
\left(M_{12}^{\text{SM}}(K)\right)^* ~=~&\frac{G_F^2}{12\pi^2} m_W^2 m_{K} f^2_{K} \hat{B}_{K}  \notag \\ 
 &\ \times \left[ \left(\la^{K}_c\right)^2 S_0(x_c) \eta_1 + 
        \left(\la^{K}_t\right)^2 S_0(x_t) \eta_2 +
         2\la^{K}_c\la^{K}_t S_0(x_c,x_t) \eta_3 
 \right] ,  \\
 \ \left(M_{12}^{\text{SM}}(B_q)\right)^*~=~& \frac{G_F^2}{12\pi^2} m_W^2 \left(\la^{B_q}_t\right)^2 S_0(x_t) \eta_B m_{B_q} f^2_{B_q} \hat{B}_{B_q},
\end{align}
with $\la^K_q \define V^*_{qs} V_{qd}$, $\la^{B_q}_t \define V^*_{tb} V_{tq}$, as well as $x_t \define m_t^2/m_W^2$ and $x_c \define m_c^2/m_W^2$. 
Here, $V$ is the $3\times 3$ CKM matrix of the SM families.
The Inami-Lim functions are given by 
\begin{align}
&\ 
\label{eq-InamiS}
S_0(x) = \frac{4 x - 11 x^2 + x^3}{4(1-x)^2} - \frac{3x^3 \log{x}}{2(1-x)^3}, \\  
&\ S_0(x_c,x_t)  \simeq x_c\left[\log\frac{x_t}{x_c}-\frac{3x_t}{4(1-x_t)}-\frac{3x_t^2\log{x_t}}{4(1-x_t)^2} \right].
\end{align}
Short distance corrections are quantified by $\eta_{1,2,3}$ and $\eta_B$.
Values for all relevant factors and their respective references are listed in 
Table~\ref{tab-Cmesonmixing}.
\begin{table}[t]
 \centering
\caption{\label{tab-Cmesonmixing}
Values of constants used in our numerical analysis.
}
\begin{tabular}[t]{c|c|c|c}\hline\hline
$m_K$~\cite{Tanabashi:2018oca} & 497.611$\pm$0.013 MeV  &
$f_K$~\cite{Aoki:2013ldr}                &$156.3\pm 0.9$ MeV  \\
 $\hat{B}_K$~\cite{Aoki:2019cca}  & 0.733 $\pm 0.040$     &
$\eta_1$~\cite{Brod:2011ty}           & $1.87\pm0.76$           \\
$\eta_2$~\cite{Buras:1990fn}          & $0.5765\pm0.0065$  &
$\eta_3$~\cite{Brod:2010mj}            & $0.496\pm 0.047$     \\
$m_s(2\text{ GeV})$~\cite{Tanabashi:2018oca} & 93.8 $\pm 2.4$ MeV &
$m_d(2\text{ GeV})$~\cite{Tanabashi:2018oca} & 4.70$\pm$0.20 MeV  \\
$\kappa_\eps$~\cite{Buras:2008nn,Buras:2010pza}&  $ 0.94\pm0.02$               &
$\phi_\eps$~\cite{Buras:2008nn,Buras:2010pza}     & $(43.51\pm0.05)^{\circ}$ \\  \hline
$m_{B_d}$  ~\cite{Tanabashi:2018oca}& 5.27963$\pm$0.00015 GeV                 &
$m_{B_s} $ ~\cite{Tanabashi:2018oca} & 5.36689$\pm$ 0.00019 GeV               \\
$f_{B_d} \sqrt{\hat{B}_{B_d}}$~\cite{Aoki:2019cca} & 0.225$\pm$0.009 GeV   &
$f_{B_s} \sqrt{\hat{B}_{B_s}}$~\cite{Aoki:2019cca} & 0.274$\pm$ 0.008 GeV    \\
$\eta_B$~\cite{Buras:1990fn,Urban:1997gw} & 0.55$\pm$ 0.01                           &
$m_b(m_b)$~\cite{Tanabashi:2018oca} & 4.18 $\pm$ 0.03 GeV                          \\
$m_c(m_c)$~\cite{Tanabashi:2018oca} & 1.28 $\pm 0.025$ GeV                        &
$m_t(m_t)$~\cite{Marquard:2015qpa,Bazavov:2016nty} & 163.53 $\pm$ 0.83 GeV                         \\ \hline
$m_D$ ~\cite{Tanabashi:2018oca} & 1.86483$\pm$ 0.00005 GeV                       &
$f_D$ ~\cite{Aoki:2019cca}& $212.0 \pm 0.7$ MeV                                                \\
$\tau_D$ ~\cite{Tanabashi:2018oca}& 0.4101 $\pm$ 0.0015 ps                            &
$m_c(3\text{ GeV}) $~\cite{Aoki:2019cca} & 0.988 $\pm$ 0.007 GeV                  \\
$m_u(2\text{ GeV})$~\cite{Tanabashi:2018oca}& 2.15$\pm$0.15 MeV  &    &  \\ 
\hline\hline 
\end{tabular}
\end{table}
The relevant observables are defined as
 \begin{align}\label{eq-defmm}
\Delta M_K =&\ 2\ \text{Re}\left(M_{12}(K)\right),&  \epsilon_K =&\ \frac{\kappa_\eps e^{i \phi_\eps}}{\sqrt{2}\left(\Delta M_K\right)_\mathrm{exp} } \mathrm{Im}\left(M_{12}(K) \right),& \\
\Delta M_d =&\ 2 \abs{M_{12}(B_d)},& S_{\psi K_s} =&\ \sin\left(\text{Arg}\left(M_{12}(B_d)\right)\right),& \\
\Delta M_s =&\ 2 \abs{M_{12}(B_s)},& S_{\psi \phi} =&\ - \sin\left(\text{Arg}\left(M_{12}(B_s)\right)\right),& \\
x_{D} =&\  \abs{\frac{2 M_{12}(D)}{\Gamma_D}}.&
 \end{align}
Values of $\kappa_\eps$ and $\phi_\eps$ are stated in Table~\ref{tab-Cmesonmixing}. $\left(\Delta M_K \right)_\text{exp}$ is the experimentally determined
value of the Kaon mass splitting and $\Gamma_D$ is the experimentally determined decay width of the $D_0$ meson; both are taken from the PDG \cite{Tanabashi:2018oca}.

The experiments measure the mass differences and $\eps_K$ precisely.
On the other hand, there are large theoretical uncertainties to estimate the SM contributions
for these observables originating from the determination of the bag parameter, QCD factors and the CKM matrix elements.
For $K\text{-}\ol{K}$ mixing,
uncertainties come from $\eta_1, f_K, \hat{B}_K, \kappa_\eps$ and the CKM elements.
The uncertainty of $\Delta M_K$ is dominated by the NLO factor $\eta_1$,
while for $\eps_K$ it is dominated by the CKM elements.
With the Wolfenstein parametrization, $\eps_K$ is approximately proportional to $A^4$.
Hence, we include the uncertainty from $A=0.836 \pm 0.015$~\cite{Tanabashi:2018oca}
as the CKM uncertainty together with those from $f_K^2\hat{B}_K$ and $\kappa_\eps$.
The relative uncertainties are estimated as
$41 \%$ and $9.3 \%$ for $\Delta M_K$ and $\eps_K$, respectively.

For the mass differences $\Delta M_{B_q} \equiv \Delta M_{q}$,
we include the uncertainties originating from $\eta_B$, $f^2_{B_q} \hat{B}_{B_q}$
and the absolute values of the CKM matrix elements.
Note that unlike the analyses in Refs~\cite{DiLuzio:2017fdq,King:2019lal},
we cannot reduce the uncertainties by assuming exact unitarity of the CKM matrix,
because the unitarity of CKM matrix is not guaranteed in our model.
Altogether,
the relative uncertainties are estimated as $16 \%$ and $14 \%$ for $\Delta M_d$ and $\Delta M_s$,
respectively.
For the CP asymmetry parameters $S_{\psi K_S}$ and $S_{\psi \phi}$,
we require our model to fit them within their experimental uncertainties.

For $D\text{-}\ol{D}$ mixing there is a large theoretical uncertainty from long-distance effects. 
The observed value is $x_{{D}} = 0.32~(14) \%$~\cite{Amhis:2016xyh}.
We simply require that the new physics contribution to $x_{{D}}$ should be 
less or equal than the size of the observed value, that is $|\Delta^{\mathrm{NP}} x_{D}| \define|2M^{\mathrm{NP}}_{12}(D)/\Gamma_D| = 0.0 \pm 0.5  \%$.

It is convenient to express the size of new physics contributions relative to the SM,
\begin{align}
 R_{\Delta M_K} \define&\ 
 \frac{M_{12}(K)-M_{12}^{\mathrm{SM}}(K)}{\mathrm{Re}\left(M_{12}^{\mathrm{SM}}(K)\right)} ,\quad 
&\  R_{\eps_K} \define   &\ 
\frac{M_{12}(K)-M_{12}^{\mathrm{SM}}(K)}{\mathrm{Im}\left(M_{12}^{\mathrm{SM}}(K) \right)}  ,  \\
R_{\Delta M_d} \define&\ 
\frac{ M_{12}(B_d)-M_{12}^{\mathrm{SM}}(B_d)}{\abs{M_{12}^{\mathrm{SM}} (B_d)  } }, \quad 
&\ R_{\Delta M_s} \define &\  
\frac{ M_{12}(B_s)-M_{12}^{\mathrm{SM}}(B_s)}{\abs{M_{12}^{\mathrm{SM}}(B_s)  } },
\end{align}
and these are given by 
\begin{align}
\label{eq-R12}
R_{\Delta M_K}\times 10^{-3}  \approx&\  
     0.053 \cdot \left( r^K_{VLL}+r^K_{VRR} \right) - 10.6 \cdot r^K_{VLR}  \\ \nonumber
&\  + \sum_{S}\left( 1.27 \left(r^K_{SLL}+r^K_{SRR} \right) -8.70 \cdot r^K_{SLR}  \right), \\ 
R_{\eps_K}\times 10^{-3} \approx &\ 
     6.94 \cdot \left( r^K_{VLL}+r^K_{VRR} \right) - 1380 \cdot r^K_{VLR}  \\ \nonumber 
&\   + \sum_{S}\left( 166 \cdot \left(r^K_{SLL}+r^K_{SRR} \right) - 1140 \cdot r^K_{SLR}  \right), \\ 
R_{\Delta M_d}  \approx &\
     11.5 \cdot \left( r^{B_d}_{VLL}+r^{B_d}_{VRR} \right) - 91.5 \cdot r^{B_d}_{VLR}  \\ \nonumber
&\   + \sum_{S}\left( 11.2 \left(r^{B_d}_{SLL}+r^{B_d}_{SRR} \right) -59.2 \cdot  r^{B_d}_{SLR}  \right),  \\ 
R_{\Delta M_s} \approx &\
     0.538 \left( r^{B_s}_{VLL}+r^{B_s}_{VRR} \right) - 4.05 \cdot r^{B_s}_{VLR}  \\ \nonumber
&\   + \sum_{S}\left( 0.522 \left(r^{B_s}_{SLL}+r^{B_s}_{SRR} \right) -2.59 \cdot  r^{B_s}_{SLR}  \right).
\end{align}
Here, 
\begin{align}
\label{eq-rMXY}
 r_{VXY}^M
\define&\  \left[ \hg^{\Zp}_{d_X} \right]_{Ff} \left[\hg^{\Zp}_{d_Y} \right]_{Ff}
                        \left( \frac{10^5\ \mathrm{GeV}}{m_\Zp}\right)^2 , \\
 r_{SXY}^M
\define&\ \left[ \hY^{S}_{d_X} \right]_{Ff} \left[ \hY^{S}_{d_Y} \right]_{Ff}
                        \left(\frac{10^5\ \mathrm{GeV}}{m_S}\right)^2,
\end{align}
with $X,Y=L,R$ and $(F,f) = (2,1), (3,1), (3,2)$ for $M=K, B_d, B_s$, respectively,
and we identify $\hY_{d_L} \equiv \hY_d$ and $\hY_{d_R} \equiv \hY_d^\dag$ (and similarly for the up sector).
The numerical coefficients are obtained by
using the values listed in Table~\ref{tab-valOs} and
neglecting the $\order{\alpha_s}$ corrections in Eqs.~\eqref{eq-CVLL}-\eqref{eq-CSLL}.
The SM contributions are calculated with the unitary CKM matrix fitted
to the experimental values~\cite{Tanabashi:2018oca}.
The coefficients for the lighter mesons tend to be larger because the SM contributions are smaller.
Left-right contributions are enhanced, especially $r^K_{VLR}$,
by the large hadronic matrix element itself and the enhancement by the running effects~\cite{Buras:2001ra},  
see Table~\ref{tab-valOs}.

Similarly, $\Delta^{\mathrm{NP}} x_{D}$ is given by
\begin{align}
\Delta^{\mathrm{NP}} x_{D} = \abs{1.01\ (r^D_{VLL}+r^D_{VRR}) - 19.5\ r^D_{VLR}
                                        + 2.63\ (r^D_{SLL}+r^D_{SRR})  - 14.1\ r^D_{SLR} },
\end{align}
with
\begin{align}
\label{eq-rDXY}
 r_{VXY}^D
\define&\  \left[ \hg^{\Zp}_{u_X} \right]_{21} \left[\hg^{\Zp}_{u_Y} \right]_{21}
                        \left( \frac{10^5\ \mathrm{GeV}}{m_\Zp}\right)^2 , \\
 r_{SXY}^D
\define&\ \left[ \hY^{S}_{u_X} \right]_{21} \left[ \hY^{S}_{u_Y} \right]_{21}
                        \left(\frac{10^5\ \mathrm{GeV}}{m_S}\right)^2.
\end{align}

We now comment on the box-diagram contributions involving $W$ bosons and up-type quarks  
which are the dominant contribution in the SM. 
In general, 
the unitarity of the CKM matrix is violated by the mixing with the $\mathrm{SU(2)_L}$ singlet VL quark.
The GIM mechanism may, in principle, become invalid in our model. 
The mass independent contributions is proportional to a sum over the five internal quarks, 
\begin{align}
 \sum_{A=1}^5 \left[\hat{V}_\mathrm{CKM}^\dagger \right]_{iA} \left[\hat{V}_\mathrm{CKM} \right]_{Aj}
 = \left[ \mhc{U^d_L} \Pfb U^u_L   \cdot \mhc{U^u_L} \Pfb U^d_L  \right]_{ij}
 = \left[ \mhc{U^d_L} \Pfb U^d_L \right]_{ij}.
\end{align}
This has the same structure as the weak-isospin part of the $Z$ boson couplings.
Using the analytical expressions of Appendix~\ref{sec-anal}, 
the size of flavor violating contribution is estimated as
\begin{align}\notag
 \left[ \mhc{U^d_L} \Pfb U^d_L \right]_{ij} -\delta_{ij}\sim&\ 
           \eps_{D_i}\eps_{D_j} \frac{m_{d_i}m_{d_j}}{M_{D_\mathrm{VL}}^2} \\
                                                        \lesssim&\ 1.6 \times 10^{-11}
                                                        \left(\frac{\sqrt{\eps_{D_i}\eps_{D_j}}}{10^{-2}}\right)^2
                                                        \left(\frac{\sqrt{m_{d_i} m_{d_j}} }{0.6\ \mathrm{GeV} }\right)^2
                                                        \left(\frac{1.5\ \mathrm{TeV}}{M_{D_\mathrm{VL}}}\right)^2,
\label{eq-exCKMunit}
\end{align}
where $\eps_{D_i}$ is a mixing angle between the singlet VL quark $D_R$ and the SM down quark $d_i$, 
and $M_{D_\mathrm{VL}}$ is a typical VL down quark mass. 
In addition, there can be an, in principle, important contribution which is enhanced by the 
heavy VL quark mass.  
Using Eq.~\eqref{eq-V5j}, the dominant contribution is estimated as 
\begin{align}
\label{eq-CKMS}
\sum_{b=4,5} \left(\left[\hat{V}_\mathrm{CKM}^\dagger \right]_{ib} \left[\hat{V}_\mathrm{CKM} \right]_{bj} \right)^2 
      S_0\left( \frac{m_{u_b}^2}{m_W^2} \right) 
\sim&\  \frac{M_{U_R}^2}{4m_W^2} \left(\eps_{U_3}^2 \frac{m_t^2}{M_{U_R}^2} V_{ti}^* V_{tj}  \right)^2  \notag \\
\sim&\  2.5 \times 10^{-9}\times \left(\frac{\eps_{U_3}}{0.1}\right)^4 \left(\frac{V_{ti}^* V_{tj}}{0.04} \right)^2                     
                                         \left(\frac{1.5\ \mathrm{TeV} }{M_{U_R}} \right)^2,   
\end{align}
where $\eps_{U_i}$ is defined in the same way as $\eps_{D_i}$. 
For $B_s$-$\ol{B}_s$ mixing, $(i,j)=(3,2)$,  
this should be compared with the top loop contribution 
$\left(\la^{B_s}_t \right)^2 S_0(x_t) \sim 0.004$ and is, thus, much smaller than the SM contribution. 
The same suppression happens for the other meson mixing. 
Thus, the violation of the GIM mechanism is extremely small.

\subsubsection{\boldmath\texorpdfstring{$B_{d,s}\to\mu^+\mu^-$}{}\unboldmath}
The new bosons, in general, induce new physics contributions to $B_{q} \to \mu^+\mu^-$ ($q=d,s$).
We refer to Ref.~\cite{Altmannshofer:2017wqy}.
The relevant effective interactions are 
\begin{align}
 -\lag_{B_q\to\mu\mu} =C^V_{AA}
              \left(\ol{q}\gamma^\mu \gamma_5 b \right) \left(\ol{\mu} \gamma_\mu  \gamma_5 \mu \right)
         + C^S_{PS}
              \left(\ol{q} \gamma_5 b \right) \left(\ol{\mu} \mu \right)
         + C^S_{PP}
              \left(\ol{q} \gamma_5 b \right) \left(\ol{\mu} \gamma_5 \mu \right).
\end{align}
In this model, the coefficients are given by 
\begin{align}
 C_{AA}^V =&\  C_\text{SM}(B_q)
                             + \frac{1}{m_{Z'}^2} \left(\left[\hg^{Z'}_{e_R} \right]_{22}-\left[\hg^{Z'}_{e_L} \right]_{22}\right)
                                \left(\left[\hg^{Z'}_{d_L} \right]_{i3}-\left[\hg^{Z'}_{d_R} \right]_{i3} \right), \\
 C_{PS}^S =&\ \sum_{S=\chi, \sigma}
                          \frac{1}{m_S^2} \cdot \text{Re} \left(\left[\hat{\la}^{S}_{e} \right]_{22}\right)
                                \left(\left[\hat{\la}^{S}_{d} \right]_{i3}-\left[\hat{\la}^{S}_{d} \right]^*_{3i} \right), \\
 C_{PP}^S =&\  - \sum_{S=\chi, \sigma}
                          \frac{i }{m_{S}^2}\cdot \text{Im}\left(\left[\hat{\la}^{S}_{e} \right]_{22} \right)
                                \left(\left[\hat{\la}^{S}_{d} \right]_{i3}-\left[\hat{\la}^{S}_{d} \right]^*_{3i} \right),
\end{align}
where $i=1,2$ for $q = d,s$, respectively. 
The SM contribution in $C^V_{AA}$ is given by
\begin{align}
  C_\text{SM} (B_q)=  
  4 \frac{G_F}{\sqrt{2}} \frac{\alpha}{2\pi \sin^2 \theta_W}
  V_{tq}^*V_{tb}\, \eta_Y\, Y_0(x_t),
\end{align}
where $ \eta_Y = 1.012$ quantifies QCD corrections~\cite{Buchalla:1998ba,Misiak:1999yg} 
and the loop function is given by
\begin{align}
 Y_0(x_t)=&\ \frac{x_t}{8}\left[\frac{x_t-4}{x_t-1}+\frac{3x_t}{(x_t-1)^2}\log{x_t}\right]
, \quad
x_t = \frac{m_t^2}{m_W^2}.
\end{align}
The decay width of $B_q \to \mu^+ \mu^- $ is proportional to
\begin{align}
\abs{P_q}^2 + \abs{S_q}^2 \define&\ 
                    \left| \frac{C^V_{AA}}{C_\mathrm{SM}(B_q) } 
                         -\frac{m_{B_q}^2}{2 m_\mu (m_q+m_b)} \frac{C^S_{PP}}{C_\mathrm{SM}(B_q) }\right|^2 \\ 
          \notag &\ \quad         + \left| \sqrt{1-\frac{4m_\mu^2}{m_{B_q}^2}}\frac{m_{B_q}^2}{2 m_\mu (m_q+m_b)} 
                      \frac{C^S_{PS}}{C_\mathrm{SM}(B_q)}
                          \right|^2.  
\end{align}
We define the ratios of branching fractions of our model to the SM,
\begin{align}
R_{B_d\to \mu\mu}^\mathrm{th} \define&\ \frac{\br{B_d}{\mu^+\mu^-}}{\br{B_d}{\mu^+\mu^-}_\text{SM}}
                                          =  \abs{P_d}^2+\abs{S_d}^2, \\
R_{B_s\to \mu\mu}^\mathrm{th} \define&\  \frac{\ol{\text{BR}}\left(B_s\to\mu^+\mu^-\right)}
                                                      {\ol{\text{BR}}(B_s\to\mu^+\mu^-)_\text{SM}}
                                           =  \frac{1+A_{\Delta\Gamma} y_s}{1+y_s} \left(\abs{P_s}^2+\abs{S_s}^2\right). 
\label{eq-RBsmm}
\end{align}
Mind the bars: In the $B_s$-$\ol{B}_s$ system, the measured width difference between light and heavy mass eigenstates, $y_s \define \Delta \Gamma_{B_s}/ \left(2 \Gamma_{B_s}\right) = 0.065\pm0.005$~\cite{Amhis:2016xyh},
is not negligible \AT{\cite{DeBruyn:2012wj,DeBruyn:2012wk}}. 
The experimentally determined value for the branching ratio, therefore, corresponds to the 
time-integrated value 
\begin{align}
 \ol{\text{BR}}\left(B_s\to \mu^+\mu^-\right)
= \frac{1+A_{\Delta \Gamma}\, y_s}{1-y_s^2}\cdot  \br{B_s}{\mu^+\mu^-},
\end{align}
where the mass-eigenstate rate asymmetry $A_{\Delta \Gamma}$
is given by~\cite{Buras:2013uqa,Fleischer:2008uj}
\begin{align}
 A_{\Delta \Gamma} =
\frac{\abs{P_s}^2 \cos\left(2\phi_P-\phi_s^\text{NP}\right) -\abs{S_s}^2 \cos\left(2\phi_S-\phi_s^\text{NP}\right)}
 {\abs{S_s}^2+\abs{P_s}^2}.
\end{align}
Here, $P_s = \abs{P_s} e^{i \phi_P}$, $S_s = \abs{S_s} e^{i \phi_S}$ and
$\phi_s^\text{NP}$ relates to $S_{\psi\phi}$, defined in Eq.~(\ref{eq-defmm}), as
\begin{align}
S_{\psi \phi}  = \sin(2 \beta_s - \phi_s^\text{NP} ),\quad \text{where}\quad
 V_{ts} = - \abs{V_{ts}} e^{i\beta_s},
\end{align}
in the standard phase convention of the CKM matrix. $A_{\Delta\Gamma} = 1$ in the SM.

The SM predictions are~\cite{Altmannshofer:2017wqy, Bobeth:2013uxa},   
\begin{align}
\text{BR} \left(B_d\to\mu^+\mu^-\right)_\text{SM} =&\ (1.06\pm0.09)\times 10^{-10}, \\
\ol{\text{BR}} \left(B_s\to\mu^+\mu^-\right)_\text{SM} =&\ (3.60\pm0.18)\times 10^{-9}.
\end{align}
The experimental values are~\cite{Tanabashi:2018oca},
\begin{align}
\text{BR} \left(B_d\to\mu^+\mu^-\right)_\mathrm{exp} =&\ \left(1.6 \pm 1.5    \right)\times 10^{-10},\\
\ol{\text{BR}} \left(B_s\to\mu^+\mu^-\right)_\mathrm{exp} =&\ \left(2.7 \pm 0.55 \right)\times 10^{-9}.
\end{align}
Altogether, the values of the ratios are given by
\begin{align}
R_{B_d\to \mu\mu}^\mathrm{exp} \define&\ \frac{\br{B_d}{\mu^+\mu^-}_\mathrm{exp}}{\br{B_d}{\mu^+\mu^-}_\text{SM}} 
                                               = 1.5 \pm 1.4, \\
R_{B_s\to \mu\mu}^\mathrm{exp} \define&\  \frac{\ol{\text{BR}}\left(B_s\to\mu^+\mu^-\right)_\mathrm{exp}}{\ol{\text{BR}}(B_s\to\mu^+\mu^-)_\text{SM}}
= 0.75\pm 0.16.
\end{align}
The current data for $\ol{\mathrm{BR}}\left({B_s}\to{\mu^+\mu^-}\right)$ has a slight tension with the SM prediction.
We note that $\br{B_s}{\mu^+\mu^-}$ is included in the analysis of Ref.~\cite{Aebischer:2019mlg},
where due to the tension a larger $C_{10}^\mu$ is favored.
Nonetheless, 
we additionally include $\br{B_s}{\mu^+\mu^-}$ individually in our $\chi^2$ analysis in order 
to take into account scalar contributions which were not included in~\cite{Aebischer:2019mlg}.

\subsubsection{\boldmath\texorpdfstring{$B \rightarrow K^{(*)} \nu\ol{\nu}$}{}\unboldmath}

The $\Zp$ boson typically affects $B \to K^{(*)} \nu\ol{\nu}$. 
We consider the observables given by~\cite{Buras:2014fpa,Altmannshofer:2009ma,Calibbi:2015kma},
\begin{align}
 R_K^{\nu\ol{\nu}}     \define \frac{1}{3} \sum_{i,j=1,2,3} \left[1-2\eta_{ij} \right]\eps_{ij}^2,\quad
R_{K^*}^{\nu\ol{\nu}} \define \frac{1}{3} \sum_{i,j=1,2,3}  \left[1+1.31\eta_{ij} \right]\eps_{ij}^2,
\end{align}
where
\begin{align}
 \eps_{ij}^2 \define \frac{ \abs{X^{ij}_L(B_s)}^2+\abs{X^{ij}_R(B_s)}^2}{\abs{\eta_X X_0(x_t)}^2 },
\quad
\eta_{ij} \define - \frac{\text{Re}\left(X^{ij}_L(B_s) X^{ij*}_R(B_s)\right)}{\abs{X^{ij}_L(B_s)}^2+\abs{X^{ij}_R(B_s)}^2}.
\end{align}
Here, $i,j = 1,2,3$ run over the three neutrino flavor.
In this model, $X^{ij}_L, X^{ij}_R$ are given by
\begin{align}
 X^{ij}_L(B_s)=  \eta_X X_0(x_t) \delta_{ij}
+ \frac{ \left[\hg^{Z'}_{\nu_L}\right]_{ij} }{g_\text{SM}^2 m^2_{Z'}}
     \frac{\left[\hg^{Z'}_{d_L}\right]_{23}}{V_{ts}^*V_{tb} },\quad
 X^{ij}_R(B_s) =
 \frac{\left[\hg^{Z'}_{\nu_L}\right]_{ij}}{g_\text{SM}^2 m^2_{Z'}}
 \frac{\left[\hg^{Z'}_{d_R}\right]_{23}}{V_{ts}^*V_{tb} }.
\end{align}
The first term in $X_L^{ij}(B_s)$ is the SM contribution.
The loop function is defined as
\begin{align}
X_0(x_t)=&\ \frac{x_t}{8}\left[\frac{x_t+2}{x_t-1}+\frac{3x_t-6}{(x_t-1)^2}\log{x_t}\right], \quad
x_t = \frac{m_t^2}{m_W^2}.
\end{align}
$\eta_X = 0.994 $ is the QCD factor~\cite{Buchalla:1998ba,Misiak:1999yg}. 
The experimental limits are~\cite{Lees:2013kla,Lutz:2013ftz},
\begin{align}
 R_{K}^{\nu\ol{\nu}} < 4.3,\quad
 R_{K^*}^{\nu\ol{\nu}} < 4.4 
\end{align}
at 90\%\ C.L~\cite{Buras:2014fpa}.

\subsubsection{Top Quark Decays}
The mixing with the VL quarks may affect top quark decays.
We study the dominant top quark decay $t\to W^+b$
and the flavor violating decays $t\to Z q$ and $t\to h q$ ($q = u, c$).
The partial decay width and the branching fractions for the flavor violating decays are,
\begin{align}
\Gamma(t \to W^+ b) =&\ \frac{m_t}{32\pi} \la\left(y_b^t, z_W^t\right)
  \left[   \left( \abs{\left[\hg^W_{u_L}\right]_{33}}^2+\abs{\left[\hg^W_{u_R}\right]_{33}}^2  \right)  \right. \\ \notag
&\left.  \quad\quad  \times         \left(  \frac{\left(1-y^t_b\right)^2}{z^t_W} + 1+ y^t_b - 2 z^t_W \right)
                              - 3 \sqrt{y^t_b}\
           \text{Re}\left( \left[\hg^W_{u_L}\right]_{33}  \cdot \left[\hg^W_{u_R}\right]_{33}  \right)
            \right],  \\
\br{t}{Z q_i} =&\ \frac{m_t}{32\pi \Gamma_t} \frac{1+2 z^t_Z }{z^t_Z}
                          \left(1- z^t_Z \right)^2
                                    \left( \abs{\left[\hg^Z_{u_L}\right]_{i3}}^2+\abs{\left[\hg^Z_{u_R}\right]_{i3}}^2  \right),  \\
\br{t}{h q_i} =&\ \frac{m_t}{64\pi \Gamma_t}\left(1- z^t_h \right)^2
                                    \left( \abs{\left[\hat{Y}_u\right]_{i3}}^2+\abs{\left[\hat{Y}_u\right]_{3i}}^2  \right) ,
\end{align}
where $x^t_B \define m_B^2/m_t^2$ ($B=h, Z, W$) and $y^t_b \define m_b^2/m_t^2 $.
Here, the light quark masses are neglected.
$\Gamma(t\to W^+ b)$ is compared with the total decay width of the top quark.
The other modes, CKM suppressed and flavor violating decays,
are neglected to calculate the total decay width of top quark, i.e we use the approximation $\Gamma_t \approx \Gamma(t\to W^+ b)$.
Uncertainties for the flavor violating decays are determined
from the experimental upper limits~\cite{Tanabashi:2018oca}.

\subsection{\boldmath\texorpdfstring{$\Zp$}{}\unboldmath~Physics}
We now study potential signals of the $\Zp$ gauge boson at the LHC, 
in gauge kinetic mixing, and in neutrino trident production.
In general, there are exclusion regions from all these observables.
Note that we do not include these observables in our $\chi^2$ analysis, but 
only check a posteriori whether the respective constraints are fulfilled.

\subsubsection{Dimuon Signals at the LHC}
In the present model, the $\Zp$ gauge boson should be lighter 
than about $800\,\mathrm{GeV}$ to explain $\Delta a_\mu$.
The most relevant $\Zp$-related process at the LHC is resonant dimuon production,
\begin{align}
  p p \to Z' \to \mu^+\mu^-.
 \end{align}
$\Delta a_\mu$ requires sizable couplings to muons,
while small couplings to the SM quarks are enough to explain $\bsll$ anomalies.
Hence, the $\Zp$ boson will dominantly decay to muons and muon neutrinos,
and its production cross section will be suppressed by the small couplings to the SM quarks.
General LHC limits on $\Zp$ bosons responsible for $\bsll$ anomalies are studied
in Refs.~\cite{Kohda:2018xbc,Allanach:2019mfl}.
Exclusion bounds are given in Ref.~\cite{Aad:2019fac} based on $139\,\mathrm{fb}^{-1}$ of data.
We have calculated the fiducial cross section, using the definition and cuts of Ref.~\cite{Aad:2019fac},
with \texttt{MadGraph5$\_$2$\_$6$\_$5}~\cite{Alwall:2014hca}
based on an \texttt{UFO}~\cite{Degrande:2011ua} model file generated
with \texttt{FeynRules$\_$2$\_$3$\_$32}~\cite{Christensen:2008py,Alwall:2014hca}.

\subsubsection{Gauge Kinetic Mixing}
We assume that the gauge kinetic mixing between the
$\mathrm{\U1_Y}$ and $\mathrm{\U1^\prime}$ gauge boson
is absent at tree-level.
At the one-loop level mixing is unavoidable and the corresponding 
$Z$-$\Zp$ mixing parameter $\eps$ is estimated as
\begin{align}
 \eps \simeq \frac{g_Y g'}{6\pi^2}
         \log{ \left(\frac{m_E^2}{m_L^2}\,\frac{m_Q^2 m_D^2}{m_U^4}\right) }\;,
\end{align}
where $m_F \sim \la_V^F v_\phi$ for $F=E,L,Q,U,D$ and $g_Y$ is the 
$\mathrm{\U1_{\mathrm{Y}}}$ gauge coupling constant.
Current experimental limits are summarized in Ref.~\cite{Hook:2010tw}.
Values of $\eps\sim 0.05$ are not excluded provided that the $\Zp$
is heavier than a few $100\,\mathrm{GeV}$.

\subsubsection{Neutrino Trident Production}
The $\Zp$ contributes to muon-neutrino induced muon pair production off a nucleus $\nu_\mu N \to \nu_\mu \mu^+ \mu^- N$, 
the so-called neutrino trident process~\cite{Altmannshofer:2014cfa,Altmannshofer:2014pba,Magill:2016hgc,Ge:2017poy,Ballett:2018uuc,Altmannshofer:2019zhy}.
The cross section for this process at the CCFR experiment is estimated as~\cite{Altmannshofer:2019zhy}
(see also \cite{Zhou:2019vxt} for a complete SM computation)
\begin{align}
R_\mathrm{CCFR} \define \frac{\sigma_\text{CCFR}}{\sigma^\text{SM}_\text{CCFR}}
\simeq \frac{(1+4s_W^2 + \Delta g^V_{\mu\mu\mu\mu})^2 + 1.13 (1-\Delta g^A_{\mu\mu\mu\mu})^2 }
          {(1+4s_W^2 )^2 + 1.13 }.
\end{align}
The experimentally observed rate is $\sigma_\text{CCFR}/\sigma^\text{SM}_\text{CCFR} = 0.82\pm 0.28$ at 95\% C.L.
The relevant effective interactions are
\begin{align}
 \mathcal{H}_\text{eff} = \frac{G_F}{\sqrt{2}}  \left[
g^V_{\mu\mu\mu\mu} \left(\ol{\nu}_\mu \gamma_\alpha P_L \nu_\mu  \right)
 \left(\ol{\mu} \gamma^\alpha \mu \right)  +
g^A_{\mu\mu\mu\mu}
 \left(\ol{\nu}_\mu \gamma_\beta P_L \nu_\mu  \right)\left(\ol{\mu} \gamma^\beta\gamma_5 \mu \right)
\right],
\end{align}
where the neutrinos are taken as flavor states.
In our model, the coupling constants are given by
\begin{align}
 g^V_{\mu\mu\mu\mu} =  1 + 4 s_W^2 + \Delta g^V_{\mu\mu\mu\mu} \quad \text{and}\quad
 g^A_{\mu\mu\mu\mu} = -1 + \Delta g^A_{\mu\mu\mu\mu},
\end{align}
with $\Zp$ boson contributions given by
\begin{align}
 \Delta g^{V,A}_{\mu\mu\mu\mu} =&\ \frac{\sqrt{2}}{G_F\cdot 2 m_{Z'}^2} \left[ \hat{g}^{Z'}_{e_L}\right] _{\nu_\mu\nu_\mu}
                                                 \left(\left[\hg^{Z'}_{e_R}\right] _{22}\pm\left[\hg^{Z'}_{e_L}\right] _{22}  \right)\;. 
\end{align}
Here, $\left[ \hat{g}^{Z'}_{e_L}\right] _{\nu_\mu\nu_\mu}$ is given by 
\begin{align}
 \left[ \hat{g}^{Z'}_{e_L}\right]_{\nu_\mu\nu_\mu}=g' \left[ U_{e_L}^\dagger\, Q'_{n_L}  U_{e_L}\right]_{22},
\end{align}
and we have used $s_W^2 = 0.23129$ specifically for this process as in Ref.~\cite{Altmannshofer:2019zhy}.
This constraint is relevant only for light $Z'$'s and becomes unimportant for 
$m_\Zp\gtrsim 200\,\mathrm{GeV}$.

\section{Results}
\label{results}
\subsection{\boldmath\texorpdfstring{$\chi^2$}{}\unboldmath~Fitting}
We search for parameters that can explain both $\Delta a_\mu$ and $\bsll$ anomalies
consistently with the other observables.
For this, we attempt to minimize the $\chi^2$ function,
\begin{align}
 \chi^2 (x) \define \sum_{I} \frac{(y_I(x)-y_I^{0})^2}{\sigma_I^2},
\end{align}
where $x$ is a parameter space point, $y_I(x)$ is the value of an observable $I$ 
with central value $y_I^0$ and uncertainty $\sigma_I$.
Altogether, we include $98$ observables with central values
and uncertainties listed in Tables~\ref{tab-obsL}, \ref{tab-obsB}, \ref{tab-mmQ} and \ref{tab-obsQ}.
Values for $C_{9,10}^{(\prime)e,\mu}$ have been stated in Section~\ref{sec-bsll}.
We use exact numerical evaluation to compute the observables, not the analytic expressions that we have
only used to illustrate the general features of the model. 
In our analysis there are $65$ free model parameters. Five of these 
are in the bosonic sector, namely
\begin{align}
 m_{\Zp},\ v_\phi,\ \gp,\ \la_\chi,\ \la_\sigma,
\end{align}
which are the $\Zp$ mass, the VEV of $\phi$, the $\U1^\prime$ gauge coupling constant,
and the effective quartic couplings of the scalars $\Phi$ and $\phi$.
All other parameters are Yukawa coupling constants
appearing in Eqs.~\eqref{eq-yukSM}-\eqref{eq-yukPhi}.
Generally, we assume that the Yukawa coupling constants are real, 
except for the couplings $y^{u,d}_{13}, y^{u,d}_{31}$ which are taken to be complex
in order to explain the complex phases in the CKM matrix. 
The Yukawa couplings involving the right-handed neutrinos with heavy Majorana masses,
that is $\la^N_{i}$ and $y^{n}_{ij}$, are not included in our $\chi^2$ analysis
as none of our $98$ observables is sensitive to them.  
As discussed in Section~\ref{sec-RGE},
$g^\prime < 0.35$ is imposed, such that the gauge coupling stays perturbative up to $\sim 10^{16}$ GeV.
We restrict all Yukawa and effective quartic coupling constants to be smaller than unity and impose $v_\phi \le 5.0$ TeV.

\subsection{Best Fit Points}
\begin{table}[t]
 \centering
\caption{\label{tab-bestinfo}
Values of $\chi^2$, selected input parameters 
and observables for $\Zp$ physics at the best fit points A, B, C and  D.
The degree of freedom in our analysis
is $N_\mathrm{obs}-N_\mathrm{inp} = 98-65 = 33$.
}
\begin{tabular}[t]{c|cccc}\hline
 Parameters                         & Point A            & Point B    & Point C & Point D      \\   \hline\hline
$\chi^2$                            & $22.6$               & $25.0$    &  $23.3$ & $23.8$       \\ 
$g'$                                   & $0.250$             & $0.340$  & $0.323$& $0.349$     \\
$(v_\Phi,\ v_\phi)\,[\mathrm{TeV}]$
                                          & $(0.785,\ 4.08)$   & $(1.11,\ 3.12)$ &$(1.07,\ 4.98)$& $(1.54,\ 4.82)$\\ \hline
$m_\Zp$ [GeV]                                                          &  277.6 & 535.3 & 486.7 & 758.7     \\
$\sig_\mathrm{fid}(pp\to\Zp\to\mu^+\mu^-)$ [fb] & 0.618 & 0.245 & 0.126 & 0.069      \\
$\eps_{Z\text{-}\Zp} \times 10^3$                           & -1.33   & 3.15   & 1.62   & -0.365    \\
$R_\mathrm{CCFR}$                                               & 1.019 & 1.010 & 1.028 & 1.008      \\
\hline
\end{tabular}
\end{table}
\begin{table}[h]
\caption{\label{tab-selobs}
Values of selected observables at the best fit points A, B, C and D. 
The last column shows experimental central values and their uncertainties.  
The upper limits on the LFV decays are $90\%$ C.L. limits.
}
\small 
 \begin{tabular}{c|cccc|c} \hline 
 Observables                                   & Point A            & Point B  &Point C & Point D          & Exp. \\   \hline\hline
 $\Delta a_\mu \times 10^9$                  & $2.62$ & $2.52$ & $2.52$    & $2.45$           & $2.68\pm0.76$ \\
 $\br{\mu}{e\gamma}\times 10^{13}$  & $0.147$& $1.597$& $0.061$ & $0.822$          &$<4.2$ \\
 $\br{\tau}{\mu\gamma}\times10^{8}$
                                        &$\val{3.34}{-4}$&$\val{3.62}{-4}$&$\val{3.27}{-6}$&$\val{8.45}{-7}$&$<4.4$  \\
 $\br{\tau}{\mu\mu\mu}\times10^{8}$
                                        &$\val{6.96}{-3}$&$\val{4.77}{-4}$&$\val{6.55}{-5}$ &$\val{4.36}{-7}$ &$<2.1$ \\ \hline
 $\mathrm{Re}\,C_9^\mu$               & $-0.548$               & $-0.806$& $-0.838$& $-0.808$ & $-0.7\pm0.3$  \\
 $\mathrm{Re}\,C_{10}^\mu$          & $0.370$                & $0.252$&$0.347$     & $0.322$  & $0.4\pm0.2$  \\
 $\Delta M_d\,[\mathrm{ps}^{-1}]$   & $0.561$               & $0.610$ & $0.598$    & $0.590$ & $0.506\pm 0.081$   \\
 $\Delta M_s\,[\mathrm{ps}^{-1}]$   & $19.6$                & $19.8$    & $19.4$      & $20.0  $  & $17.76\pm 2.5$  \\
 $S_{\psi K_s}$                     & $0.697$               & $0.696$         &$0.692$     & $0.695$   &$0.695\pm0.019$  \\
$S_{\psi \phi}$                     & $0.0366$               & $0.0374$      &$0.0373$    & $0.0379$&$0.021\pm0.031$  \\
$R^\mathrm{th}_{{B_s}\to{\mu\mu}}$    &  0.841 & 0.890 &0.850  & 0.861 & $0.75\pm0.16$ \\ \hline
 \end{tabular}
\end{table}

We find a landscape of good fit points in similar phenomenological regions. 
We will focus our discussion on the four best fit points 
A, B, C and D with $\chi^2 = 22.6,\ 25.0,\ 23.3$ and $23.8$ (for $N_\mathrm{d.o.f.}=98-65=33$ degrees of freedom),
respectively.
All four best fit  points are selected from points with the charged VL lepton heavier than $250$ GeV
and the fiducial cross section $\sigma_\mathrm{fid}\left(p p \to \Zp \to \mu^+\mu^-\right)$
smaller than the latest experimental limit.
Point A is the global best fit point under these conditions.
The point B is the best fit point of points with $m_{E_1} > 1.2$ TeV.
This point has slightly larger $\chi^2$ value than the other three best fit points (see Table~\ref{tab-selobs}), 
mainly because $R^\mathrm{th}_{B_s\to\mu\mu} \sim 0.9$ due to the smaller $\mathrm{Re}C_{10}^\mu$.
The points C  and D are the best fit points under the conditions $m_\chi > 750$ GeV and $m_{\Zp} > 750$ GeV,
respectively.

The values of selected input parameters and observables are listed 
in Tables~\ref{tab-bestinfo} and~\ref{tab-selobs}.
All input parameters are shown in Appendix~\ref{sec-inputs}
and complete lists of all observables at the best fit points are listed in Appendix~\ref{sec-fullobs}.
Masses and predicted dominant decay modes of new particles are summarized in 
Tables~\ref{tab-massdecayA},~\ref{tab-massdecayB},~\ref{tab-massdecayC} and~\ref{tab-massdecayD}.
The decay widths are calculated based on the formulae in Appendix~\ref{sec-twobody}.

\subsection{Phenomenology }
\begin{figure}[t]
 \centering
\includegraphics[height=100mm]{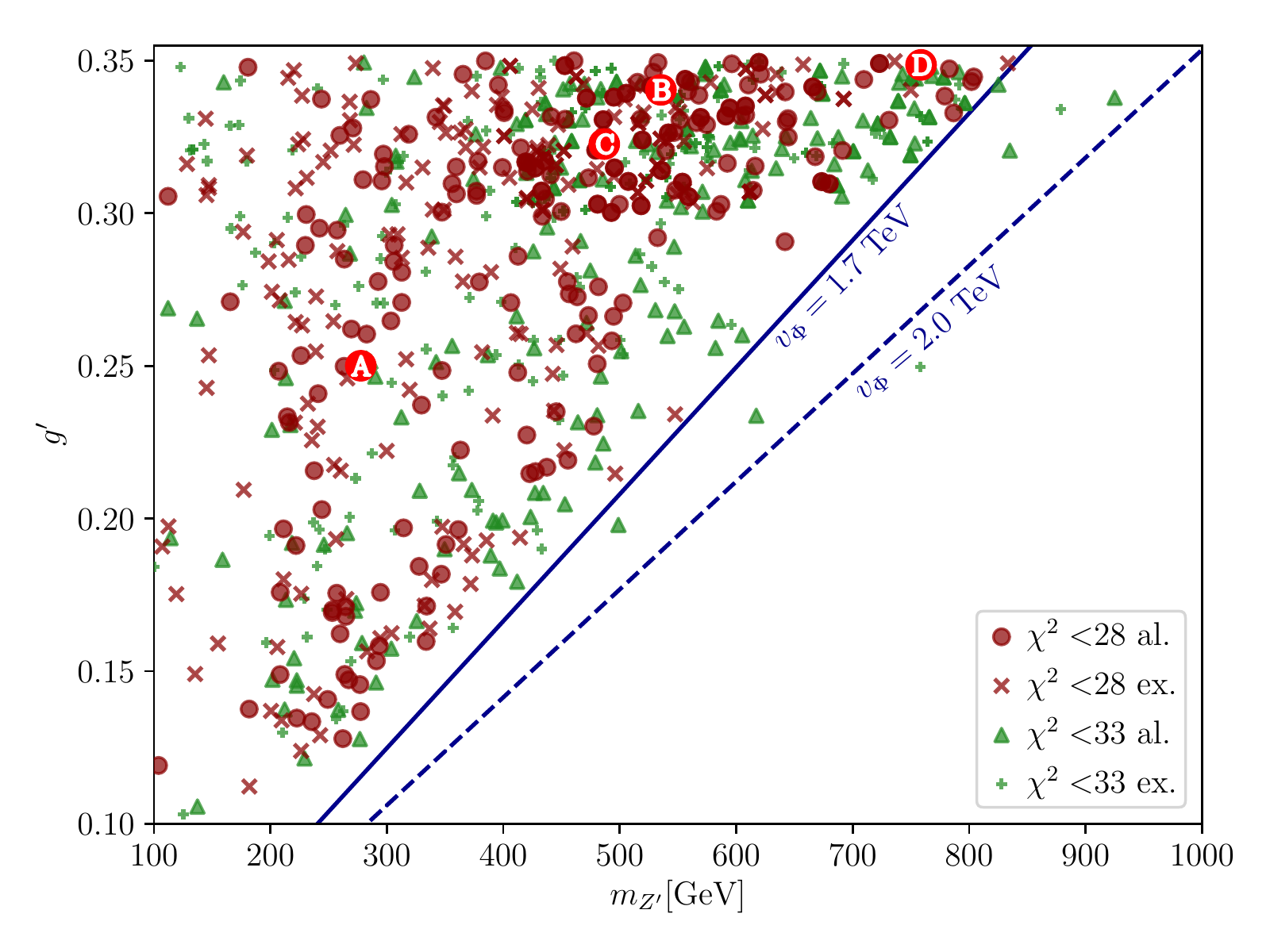}
\caption{
\label{fig-scatmZpgp}
Good fit points in the ($m_\Zp$, $\gp$) plane.
The red~(green) dots represent $\chi^2 < 28~(33)$, where 
points which are already excluded by direct $Z'$ searches or neutrino trident production
are shown with crosses or pluses. 
The blue solid (dashed) line shows the maximally allowed values of $v_\Phi$ to
give an explanation of $\Delta a_\mu$.}
\end{figure}
We now discuss some global features of our model.
Figure~\ref{fig-scatmZpgp} shows fit points with $\chi^2 < 33 = N_\mathrm{d.o.f.}$.
The red circles (green triangles) have $\chi^2 < 28~(33)$. 
Points which are excluded by $\Zp$ physics, namely LHC searches and/or neutrino trident production,
are denoted by red crosses (green pluses) with the same color coding as above.
The $Z$-$\Zp$ mixing parameter $\epsilon$ is always less than or equal $\order{10^{-3}}$
and, thus, much smaller than the experimental limit.
All subsequent plots show \textit{the same} model parameter points as Figure~\ref{fig-scatmZpgp}.

The blue solid (dashed) line in Figure~\ref{fig-scatmZpgp} corresponds to $v_\Phi = 1.7~(2.0)$ TeV.
Consistent with our analytical analysis of $\Delta a_\mu$ in Section~\ref{sec-AMM}, c.f.\ especially Eq.~\eqref{eq:amu_upperbound},
there is no point with $\chi^2 < 28~(33)$ whenever $v_\Phi > 1.7~(2.0)$ TeV.
This results in an upper bound on the $\Zp$ mass: $m_{\Zp} \lesssim 840$ GeV for $\gp < 0.35$.
We note that in Fig.~\ref{fig-scatmZpgp} allowed and excluded points co-exist for 
similar values of $m_\Zp$ and $\gp$. This is because in this plane one does not resolve the different 
textures for Yukawa couplings, which can lead to vastly different phenomenology of $\Zp$ physics.
For example, $\Delta C_9^\mu$ requires $\left[\hg^\Zp_{d_L}\right]_{23} \sim 0.001$,
but this would not exclude $\order{1}$ values of $\left[\hg^\Zp_{d_L}\right]_{22}$ 
or  $\left[\hg^\Zp_{d_L}\right]_{33}$ which have dramatic consequences for $\Zp$ direct production
as we will discuss now.

\subsubsection{\boldmath\texorpdfstring{$\Zp$}{}\unboldmath~Physics}
\begin{figure}[t]
\centering
\includegraphics[height=100mm]{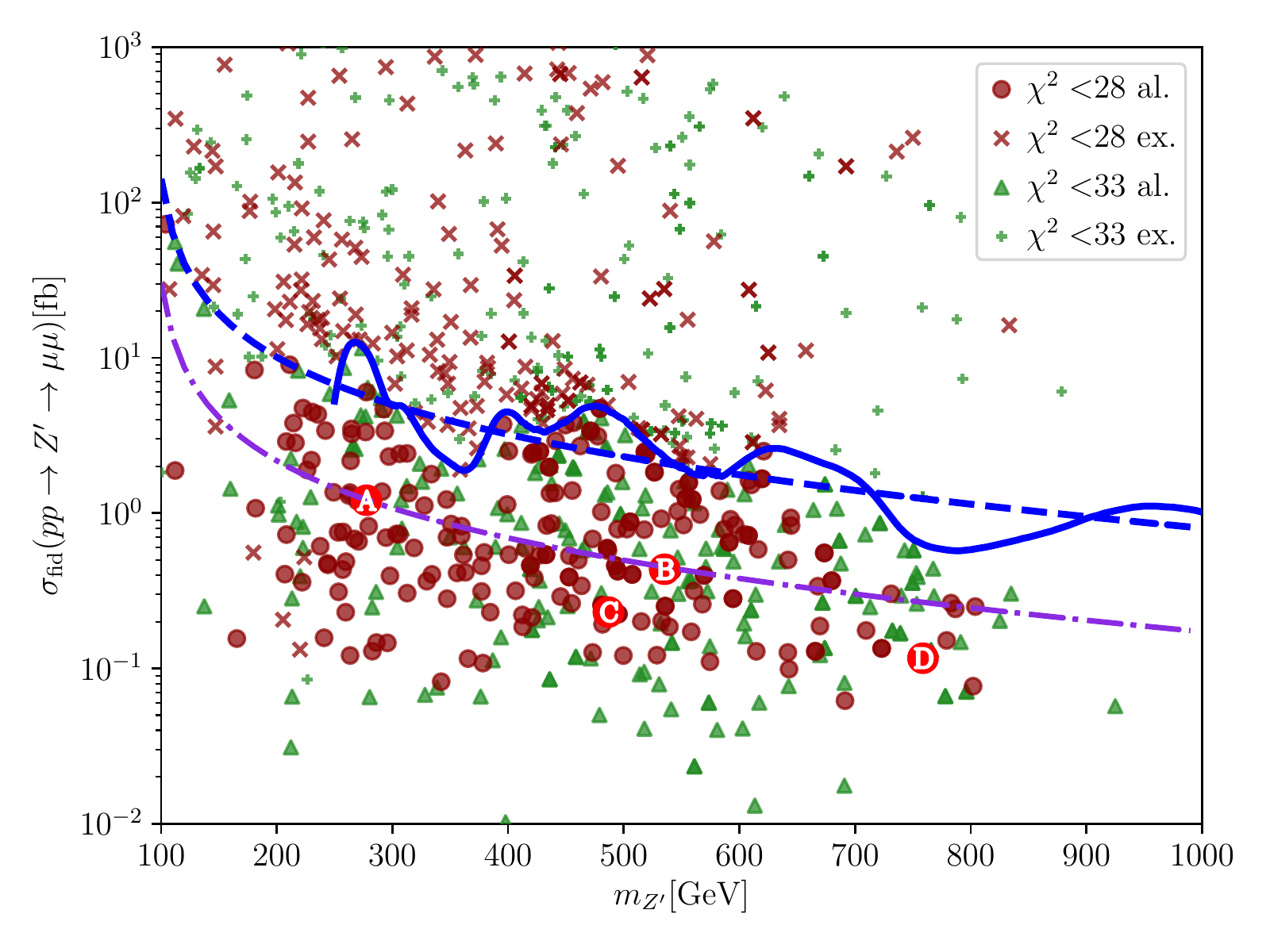}
\caption{\label{fig-scatfid}
Good fit points in the ($m_\Zp$, $\sigma_\mathrm{fid}\left(pp\to\Zp\to\mu\mu\right)$) plane.
The color coding is the same as in Fig.~\ref{fig-scatmZpgp}. 
The solid blue line is the 95\% C.L.\ exclusion limit from ATLAS~\cite{Aad:2019fac}.
The blue dashed line extrapolates this line to test the points with $m_\Zp < 250$ GeV.
As purple dot-dashed line we show a rough estimate of the future sensitivity to be expected after HL-LHC.
Excluded points below the blue line are not excluded by LHC direct searches but by neutrino trident production.
}
\end{figure}
Figure~\ref{fig-scatfid} shows the good fit points in the ($m_\Zp$, $\sigma_\mathrm{fid}(pp\to\Zp\to\mu\mu))$ plane,
where we use the definition and cuts for the fiducial cross section $\sigma_\mathrm{fid}$ of Ref.~\cite{Aad:2019fac}.
The blue solid line is the 95\% C.L.\ limit from the ATLAS analysis~\cite{Aad:2019fac}.
Since the limit is given only for $m_\Zp > 250$ GeV, we use an extrapolation down to 
lower masses shown by the dashed blue line.
As a rough estimate for the sensitivity to be expected at the HL-LHC we 
can scale the limit on the cross section by $\sqrt{139/3000}$, the square root of the expected 
ratio of integrated luminosities. This sensitivity is shown as a purple, dot-dashed line in Fig.~\ref{fig-scatfid}.

A small flavor violating coupling to $\Zp$, $\left[\hg^\Zp_{d_L}\right]_{23}\sim 10^{-3}$
is enough to explain the $\bsll$ anomalies.
A diagonal coupling of $\Zp$ to bottom quarks or to the light quarks could 
be sizable without changing other flavor violating observables.
However, fitting the observed CKM matrix sets limits on the size of such couplings.
Therefore, a good fit prefers small diagonal couplings to quarks.
In agreement with that, our best fit points predict fiducial cross section roughly about an order of magnitude 
smaller than the current limits. 
We stress that the LHC limits were not part of the fit and only checked subsequently on good fit points.

Since $\Delta a_\mu$ requires sizable $\Zp$ coupling to muons,
a sizable muon neutrino coupling is also predicted.
Our model, therefore, is sensitive to the neutrino trident process if $v_\Phi \lesssim 350$ GeV.
Focusing on this mass range in Fig.~\ref{fig-scatfid}, we see that there are a handful of 
points which are excluded exclusively by the trident constraints and not by LHC searches.
On a different note, the one-loop induced gauge kinetic mixing for all points is $\order{10^{-3}}$ or less,
much smaller than the current limits.

\subsubsection{\boldmath\texorpdfstring{$\bsll$}{}\unboldmath}
\begin{figure}[t]
\begin{minipage}[c]{0.5\hsize}
 \centering
\includegraphics[height=75mm]{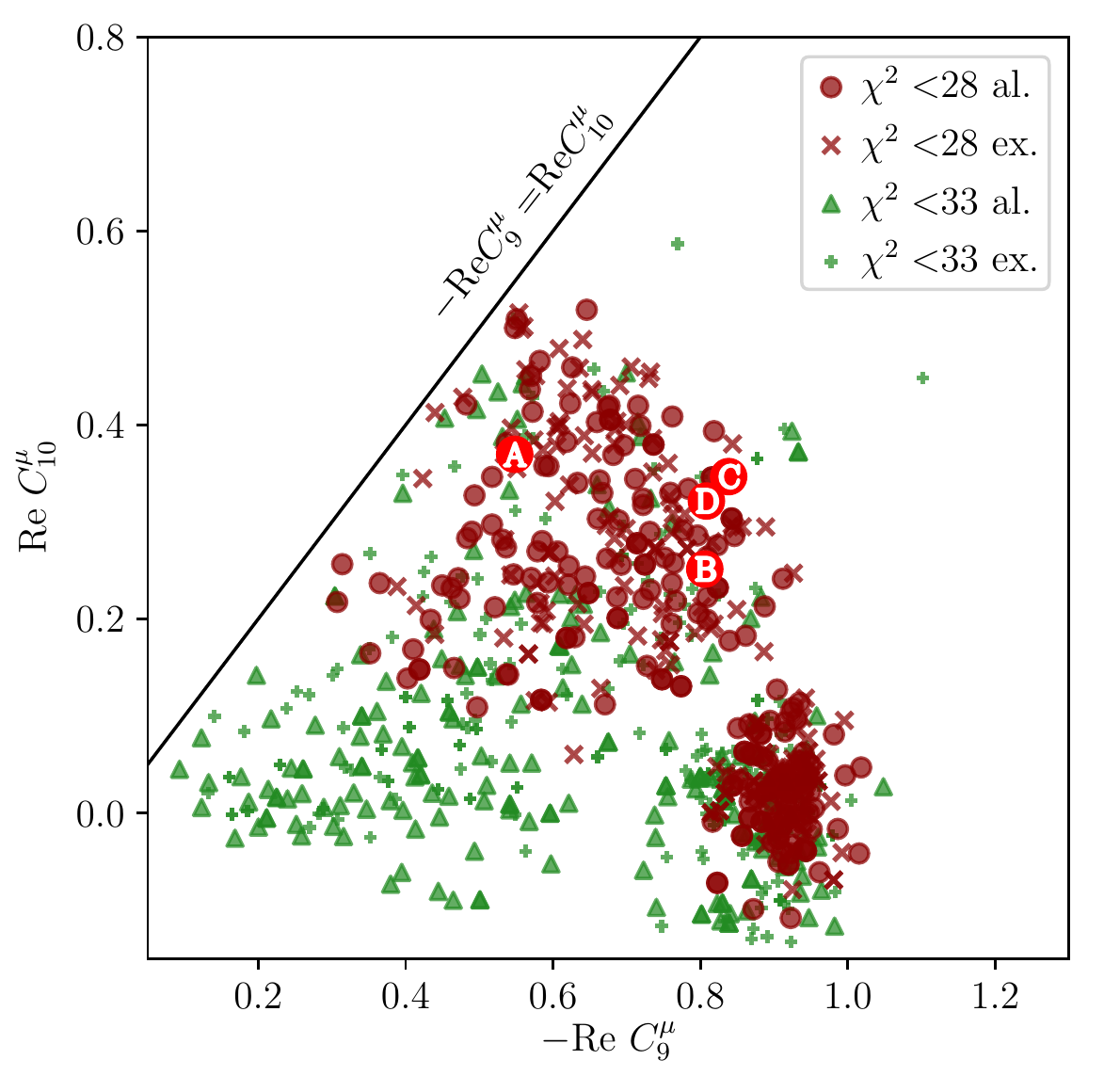}
\end{minipage}
\begin{minipage}[c]{0.5\hsize}
 \centering
\includegraphics[height=75mm]{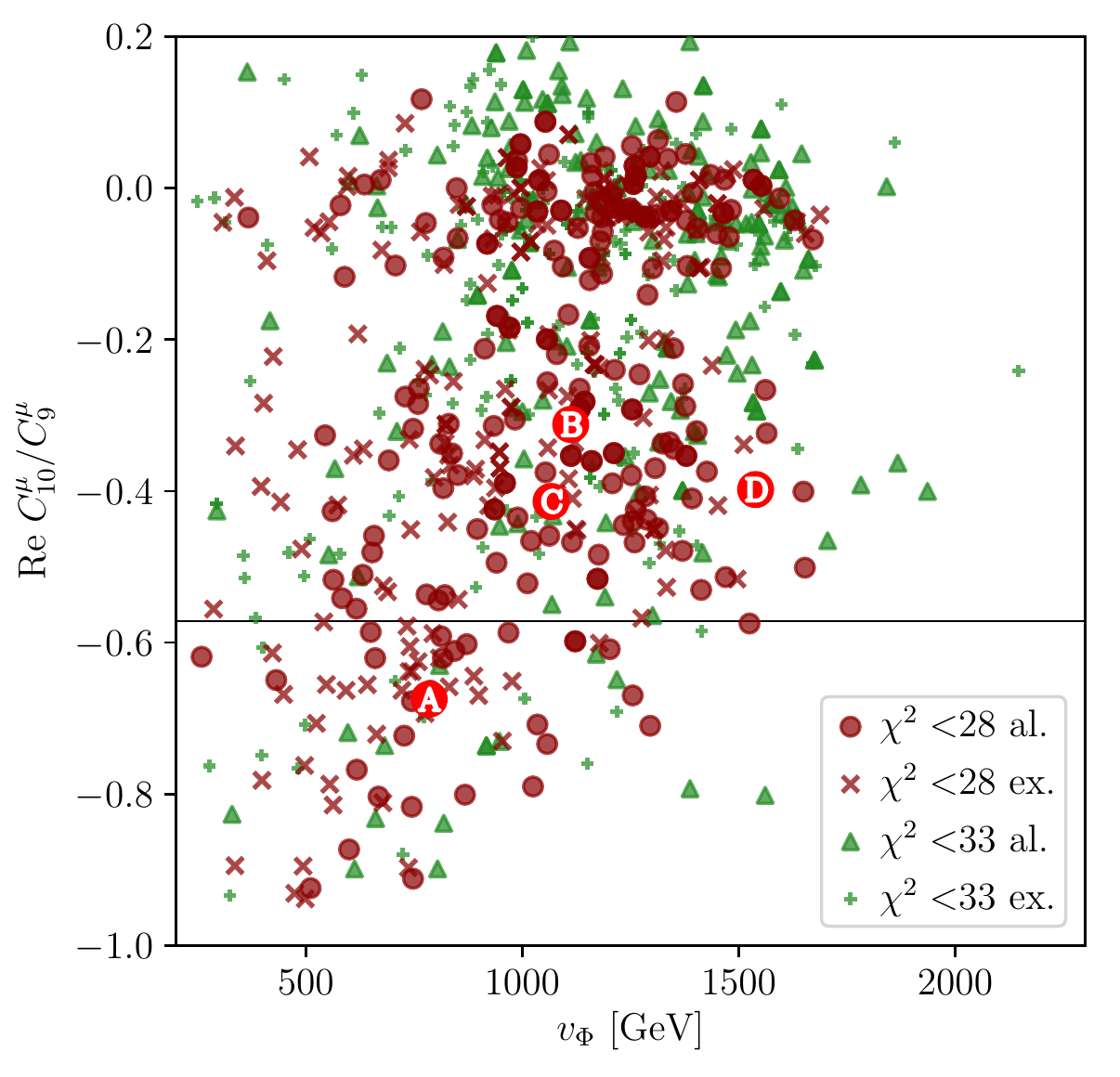}
\end{minipage}
\caption{\label{fig-scatvPC0C10}
Good fit points in the ($-\mathrm{Re}\,C_9^\mu$, $\mathrm{Re}\,C_{10}^\mu$) 
and ($v_\Phi$, $\mathrm{Re}(C_{10}^\mu/C_{9}^\mu)$) planes.
The color coding is the same as in Fig.~\ref{fig-scatmZpgp}.
The black line in the right plot correspond to pattern~(IV) of Eq.~\eqref{eq:pattern4}.}
\end{figure}
All the best fit points A-D are fitted to pattern (IV) (``$C_{9}$ and $C_{10}$'', cf.\ eq.~\eqref{eq:pattern4}).
There are also a lot of points which are fitted to pattern (I) (``$C_{9}$ only'', eq.~\eqref{eq:pattern1}).
We show our good fit points the ($\mathrm{Re}\,C_9^{\mu}$, $\mathrm{Re}\,C_{10}^\mu$) plane in 
the left panel of Fig.~\ref{fig-scatvPC0C10}. 
Points with pattern (IV) tend to have smaller $\chi^2$ 
because of the tension in $R_{B_s\to \mu\mu}$ which favors non zero $C_{10}$.  
The other patterns (II) (``$C_{10}$ only''), (III) (``$C_{9}=-C_{10}$''), and (V) (``$C_{9}$ and $C'_{9}$''),
are hardly compatible with other observables and we will now discuss this in some detail.
Making use of the analytic discussion in Appendix~\ref{sec-anal}, 
the $\Zp$ couplings to muons can be expressed as
\begin{align}
 \left[\hg^\Zp_{e_L}\right]_{22} \sim - \gp s_{\mu_L}^2,\quad
 \left[\hg^\Zp_{e_R}\right]_{22} \sim - \gp s_{\mu_R}^2.
\end{align}
Hence, the ratio $C_{10}^\mu/C_{9}^\mu$ is given by
\begin{align}
 \frac{C_{10}^\mu}{C_{9}^\mu} \sim \frac{s_{\mu_R}^2-s_{\mu_L}^2}{s_{\mu_R}^2+s_{\mu_L}^2}.
\end{align}
This indicates $\abs{C_{10}^\mu/C_{9}^\mu} \le 1$, and that pattern (II) (``$C_{10}$ only'') can never be realized.
pattern (III) is $C_{9}^\mu\approx-C_{10}^\mu$, implying $s^2_{\mu_R} \ll 1$.
However, as $\Delta a_\mu \propto s_{\mu_L}s_{\mu_R}/v_\Phi^2$ (cf.\ Eqs.~(\ref{eq:aMu_anal}), (\ref{eq:aMu_anal2})) it would be 
suppressed in this case unless the suppression is compensated by a small $v_\Phi\lesssim 500\,\mathrm{GeV}$.
We show this on the right panel of Fig.~\ref{fig-scatvPC0C10}, 
where one can clearly see that there are no good points with $\mathrm{Re}(C_{10}^\mu/C_{9}^\mu)\lesssim -0.8$ for $v_\Phi \gtrsim 1.0$ TeV.  
Finally, pattern (V) is incompatible with neutral meson mixing:
As can be seen from Eqs.~\eqref{eq-R12}, mixed LR contributions of 
$\Zp$ exchange are enhanced by large negative coefficients.
Since pattern (V) requires that $\left[\hg^\Zp_{d_L} \right]_{23}$ and $\left[\hg^\Zp_{d_R} \right]_{23}$
have opposite signs, their LR contribution to meson mixing adds constructively with the SM.
As the SM prediction for $\Delta M_s$ is already larger than the experimentally measured value, 
$\Zp$ couplings compatible with pattern (V) would only ever increase the tension with experiment. 
This could be overcome if there were sizable negative contributions from the scalar exchange,
however, the scalar couplings in our model are always suppressed as shown in Appendix~\ref{sec-anal}.

\subsubsection{Standard Model Quark Sector}
\label{sec-QFV}
\begin{figure}[!t!]
 \centering
\includegraphics[height=120mm]{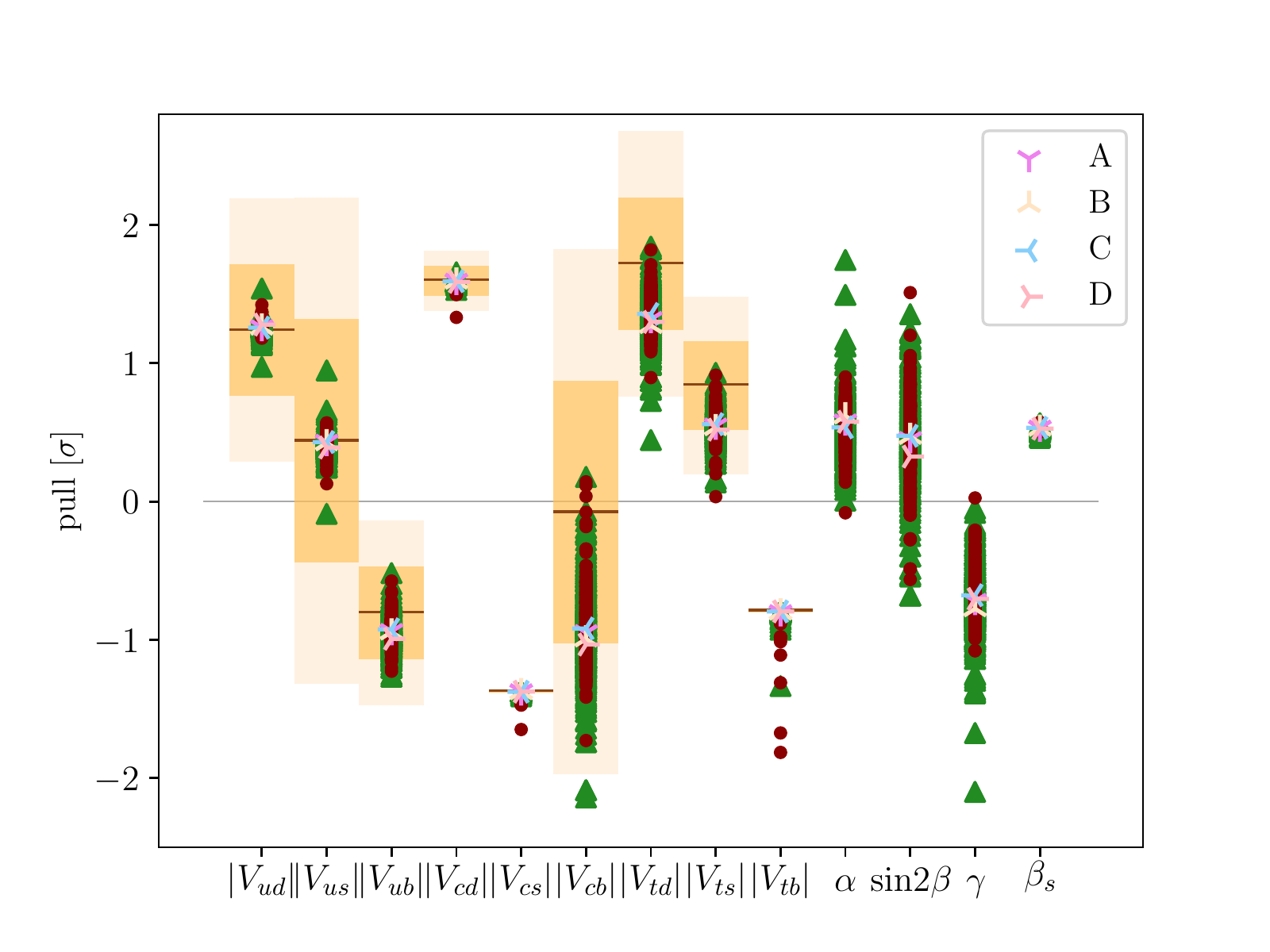}
\caption{\label{fig-CKM}
Pulls of good fit points of our model with respect to experimental determinations of absolute values and physical relative phases of CKM matrix elements. 
The color coding for the red and green points is the same as in Fig.~\ref{fig-scatmZpgp}. 
The brown lines and dark~(lighter) yellow bands are the central values and $1\sigma$~($2\sigma$) ranges of CKM elements as determined in a global SM fit taken from~\cite{Tanabashi:2018oca}. 
}
\end{figure}
We fit our model parameters to match the quark masses, 
absolute values of the CKM matrix elements, 
and relative physical phases $\alpha$, $\beta$ and $\gamma$.
We do not assume unitarity of the CKM matrix and fit our parameters directly to 
the experimentally determined absolute values and angles.
In addition, we require our model to fit the Wilson coefficients of $\bsll$ processes 
such that the anomalies are matched. 
Furthermore, we fit to CP-even and CP-odd 
observables in $\MMbar{K}$, $\MMbar{B_d}$, $\MMbar{B_s}$ and $\MMbar{D}$ mixing 
as well as $R_{K^{(*)}}^{\nu\ol{\nu}}$, $R_{B_{d(s)}\to\mu\mu}$, 
$\br{B_s}{K\tau\tau}$ and top quark decays.
Of course, to some extent this approach consists of a ``double fitting'' 
as CKM angles and phases are themselves extracted also from some of these observables under the assumption of the SM. 
However, 
our approach should be valid here as NP contributions to the relevant observables are 
typically less than $\mathcal{O}(10\%)$.

Our best fit values for CKM matrix elements and angles, 
relative to the SM extraction, are shown in Fig.~\ref{fig-CKM}. 
The brown lines and yellow bands show 
central values and their uncertainties as obtained in a global fit to the SM~\cite{Tanabashi:2018oca}. 
It is an important non-trivial crosscheck of our fitting procedure that we reproduce the SM 
best-fit values for most elements.
In general, our results are consistent with the SM as most of the values 
agree within their $1\sigma$ region.
However, some elements, namely $\abs{V_{cb}}$, $\abs{V_{td}}$ and $\abs{V_{ts}}$
show consistent deviations from the SM extraction.
Perhaps these deviations could be tested by future experiments, which would be especially 
interesting if they are correlated with other observables. 
\begin{figure}[t]
\begin{minipage}[c]{0.5\hsize}
\centering
\includegraphics[height=75mm]{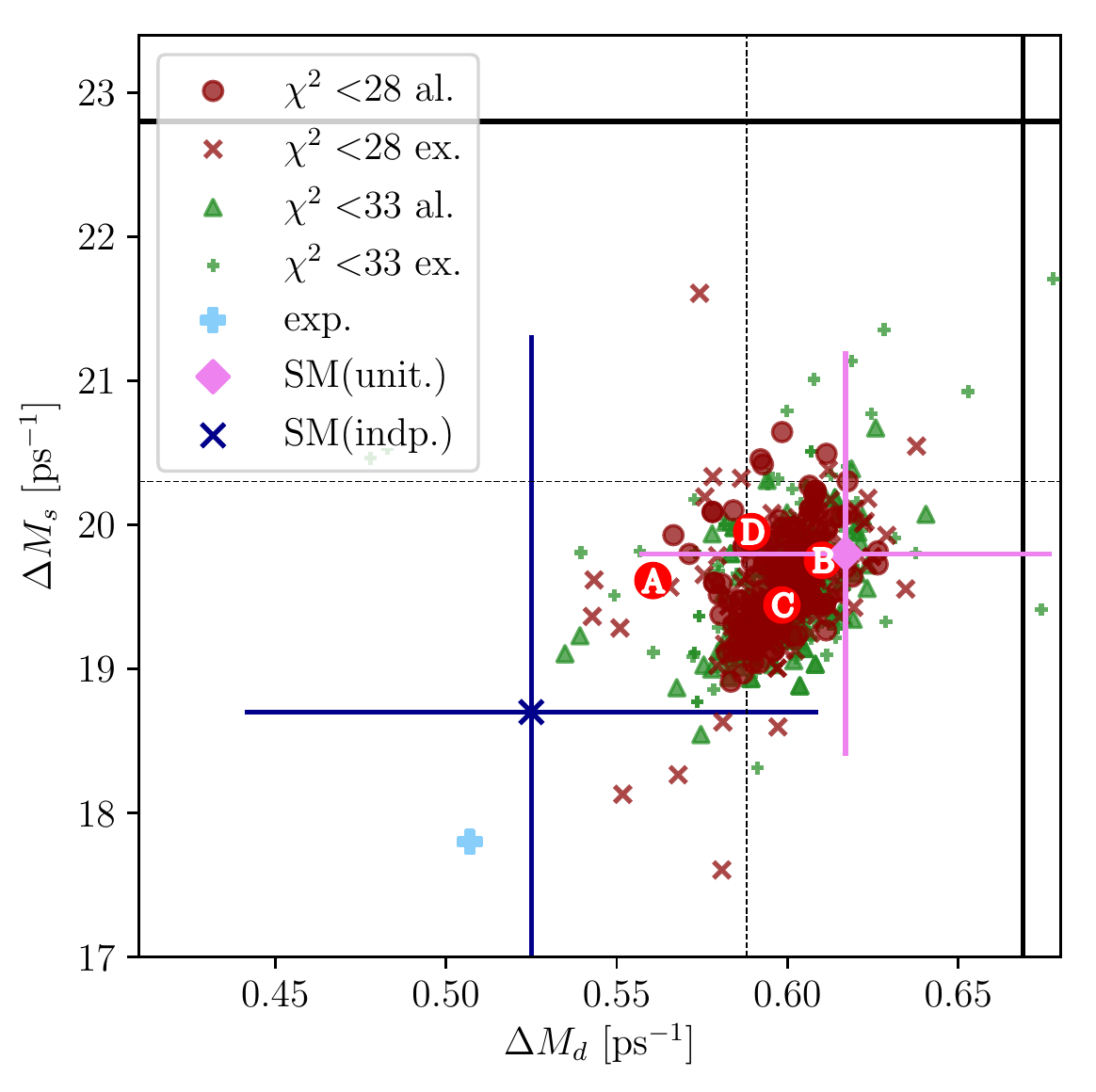}
\end{minipage}
\begin{minipage}[c]{0.5\hsize}
 \centering
\includegraphics[height=75mm]{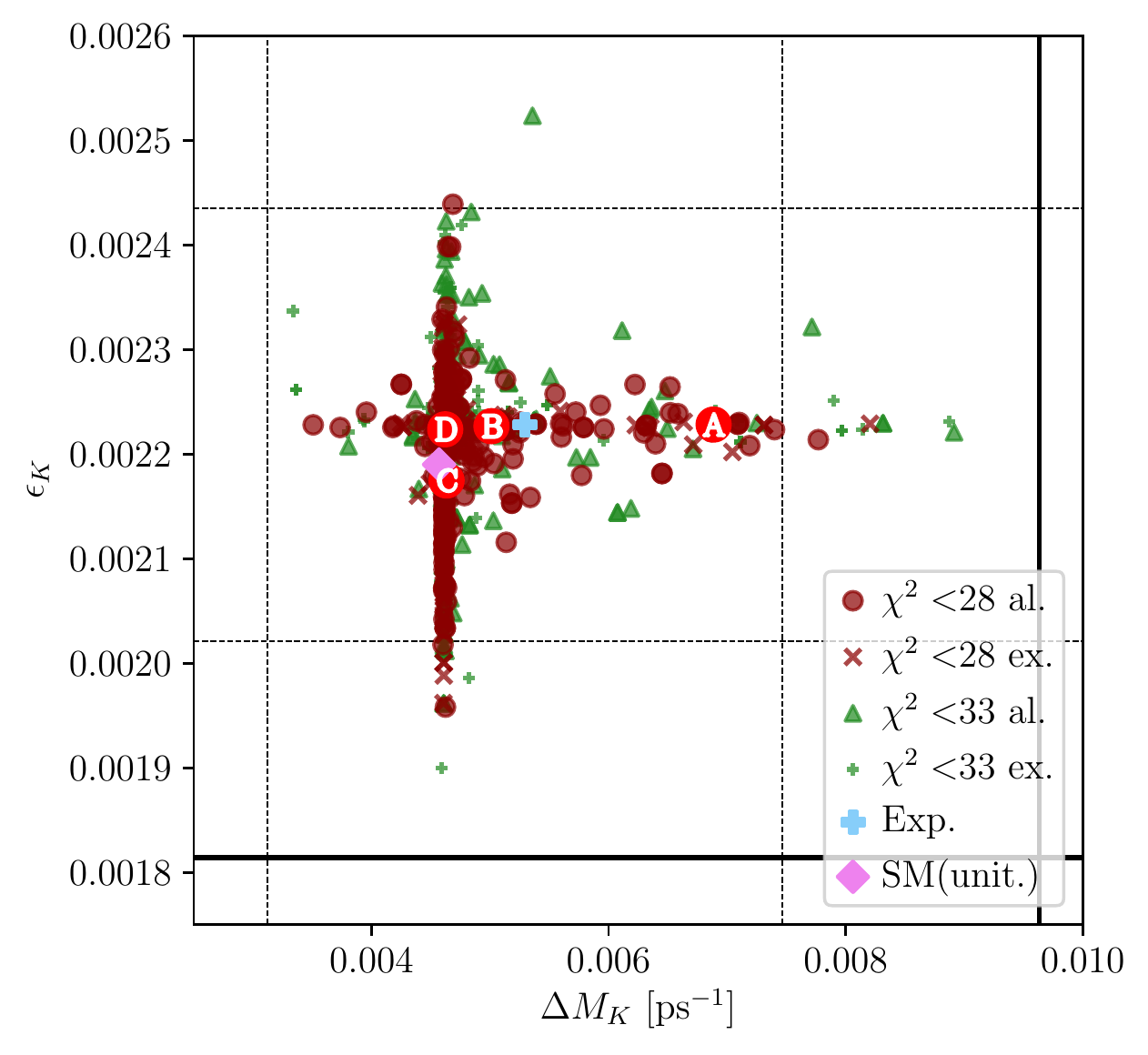}
\end{minipage}
 \caption{
\label{fig-scatNMM}
Good fit points in the ($\Delta M_d$, $\Delta M_s$) (left)
and ($\Delta M_K$,$\eps_K$) plane (right).
The color coding is the same as in Fig.~\ref{fig-scatmZpgp}.
Black thin-dashed (thick) lines show $1\sigma$ ($2\sigma$) deviations from 
the experimental values.  
The experimental central values are shown as light-blue plus.
SM predictions with (without) assuming CKM unitarity are shown
as a purple diamond (dark-blue $\textbf{x}$ in the left panel). 
}
\end{figure}
In Fig.~\ref{fig-scatNMM} we show the good fit points 
in the ($\Delta M_d$, $\Delta M_s$) plane compared to experimental measurements 
and SM prediction with and without the assumption of CKM unitarity.
Uncertainties of the SM predictions are shown in the figure.  
As discussed Section~\ref{sec-nmm},
the relative uncertainties for $\Delta M_d$ and $\Delta M_s$ are estimated as $16\,\%$ and $14\,\%$ 
without assuming unitarity of the CKM matrix. These uncertainties 
reduce to $9.8\,\%$ and $7.1\,\%$ if CKM unitarity is assumed~\cite{Tanabashi:2018oca}.

Although there are sizable $\Zp$ contributions 
these are still small compared with the uncertainties originating from the CKM elements without assuming unitarity.
The CP asymmetry parameters $S_{\psi K_s}$ and $S_{\psi \phi}$ are well fit to the experimental values,
cf.\ their values at the best fit points in Table~\ref{tab-selobs} and experimental values in Table~\ref{tab-obsQ}.

The right panel of Fig.~\ref{fig-scatNMM} shows the good fit points in the ($\Delta M_K$, $\eps_K$) plane.
Similar to $B$-$\ol{B}$ mixing, the $\Zp$ contributions are smaller than the uncertainties
from the CKM values and QCD corrections.

There may be a sizable contribution from $\Zp$ exchange to $D$-$\ol{D}$ mixing as well.
However, theoretical errors here are too large to hope for a discrimination
of SM and NP effects. Also the effects in top quark decays, 
including flavor violating ones, are too small to be tested by experiment.

Potentially, there are also contributions from the new scalar fields.
However, as shown in Appendix~\ref{sec-anal}, the Yukawa coupling of the new scalars to the SM fermions
first arise at the second order of the small mixing angles.
Therefore, scalar contributions are very suppressed for $B_s \to \mu\mu$.
The ratio $R_{B_s\to\mu\mu}$ is predominantly determined by $C_{10}^\mu$
where a larger $C_{10}^\mu$ relaxes the tension between measurements and the SM prediction, see Table~\ref{tab-selobs}.
Finally, $\br{B_s}{K\tau\tau}$ is generally not much affected as the $\Zp$ coupling to $\tau$'s can be suppressed.
At all the best fit points, $R_{K^{(*)}}^{\nu\nu} \lesssim 1.06 $ which is a deviation 
much smaller than the experimental sensitivities.

\subsubsection{Charged Lepton Flavor Violation}
\label{sec-LFV}
\begin{figure}[t]
\begin{minipage}[c]{0.5\hsize}
 \centering
\includegraphics[height=75mm]{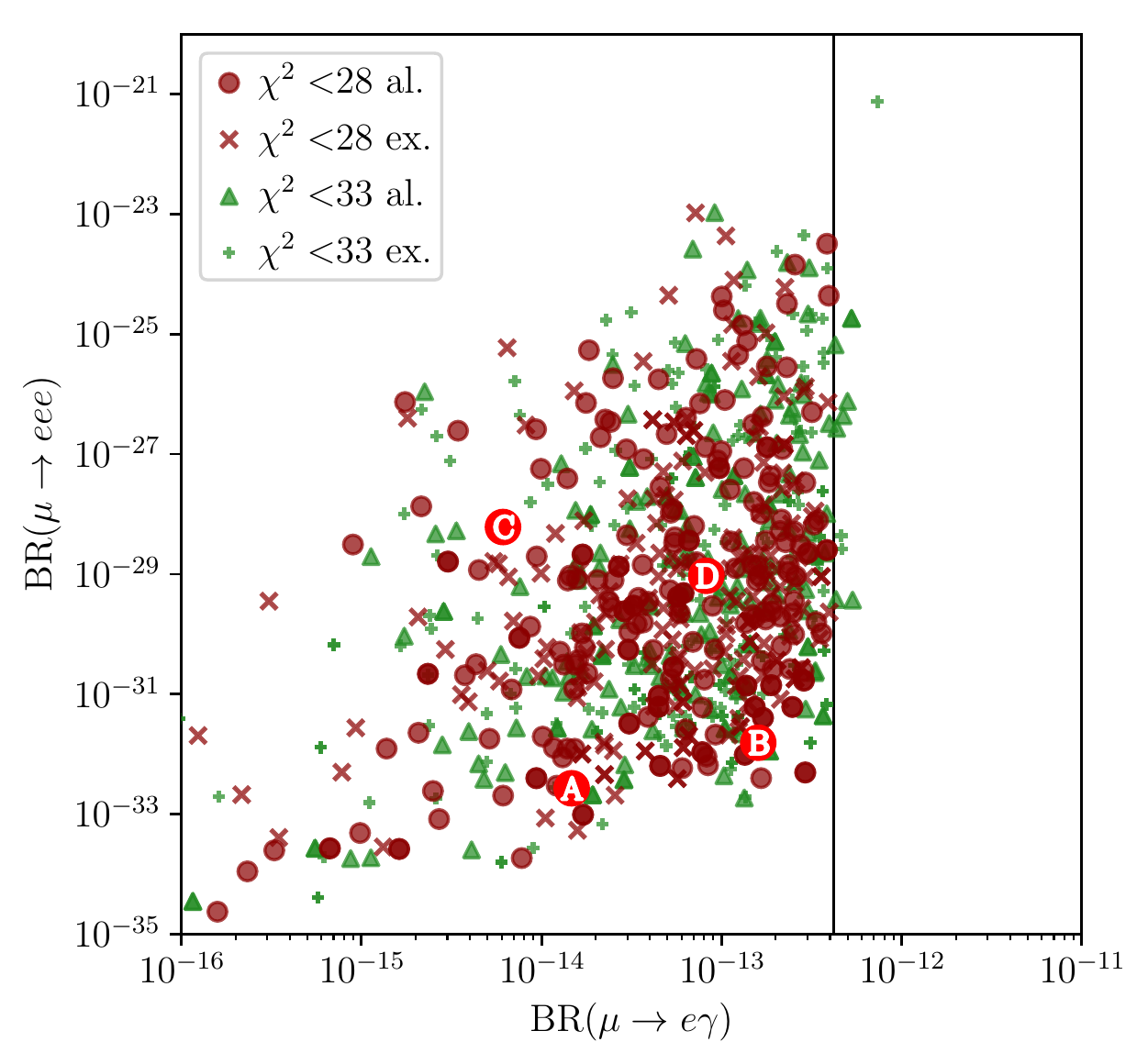}
\end{minipage}
\begin{minipage}[c]{0.5\hsize}
 \centering
\includegraphics[height=75mm]{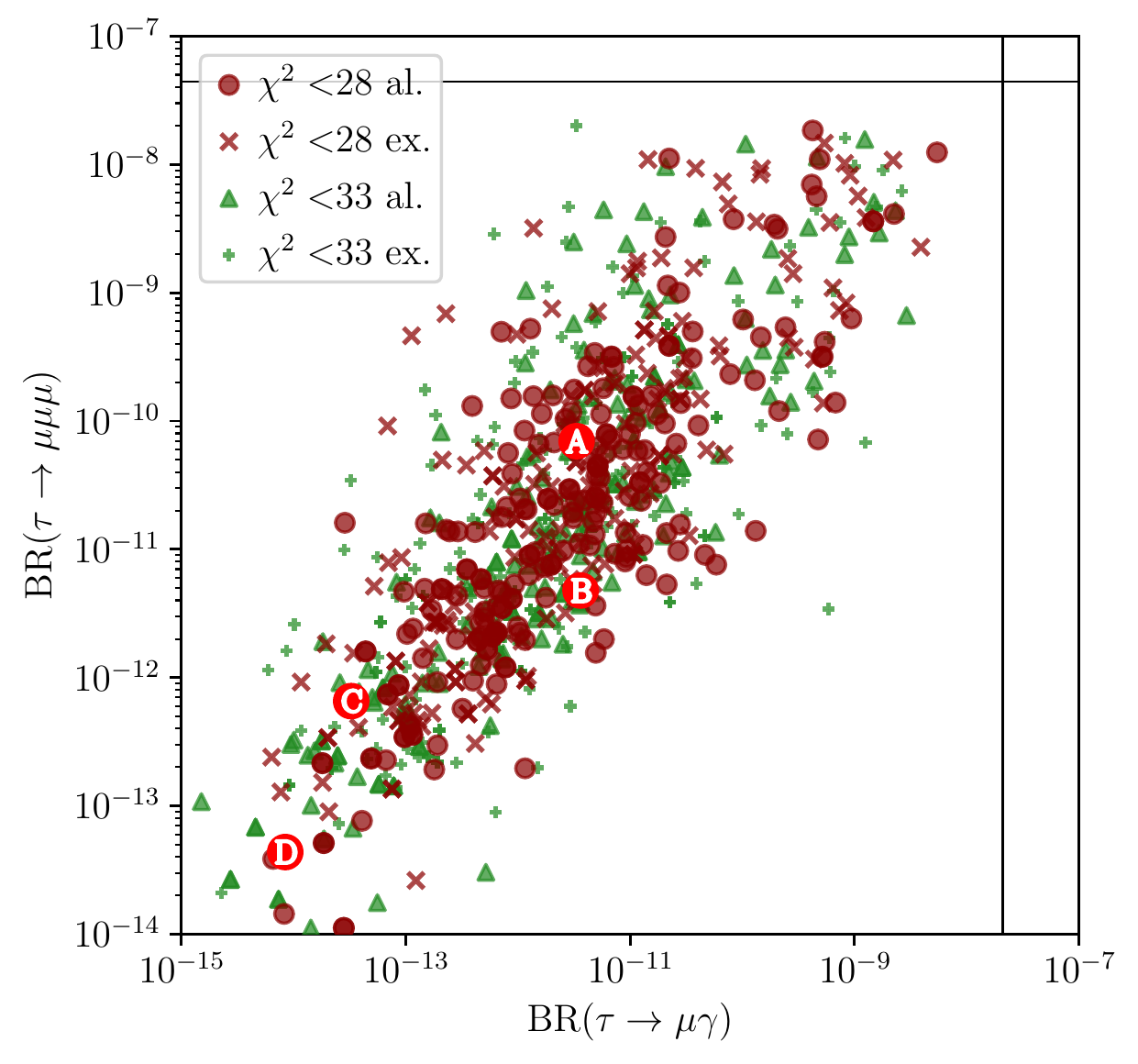}
\end{minipage}
 \caption{
\label{fig-scatLFV}
Predictions of the best fit points for $\br{\mu}{e\gamma}$, $\br{\mu}{eee}$, 
$\br{\tau}{\mu\gamma}$, as well as $\br{\tau}{\mu\mu\mu}$ and their correlations. 
The color coding is the same as in Fig.~\ref{fig-scatmZpgp}.
The black lines are the current experimental $90\%$ C.L.\ exclusion limits. 
}
\end{figure}
We now discuss predictions for charged LFV in this model.
While we have included the experimental upper bounds on charged LFV in the fit, 
it still is interesting to see the size and spread of LFV obtained in this model. 
Figure~\ref{fig-scatLFV} shows the best fit points in the ($\br{\mu}{e\gamma}$, $\br{\mu}{eee}$)
and ($\br{\tau}{\mu\gamma}$, $\br{\tau}{\mu\mu\mu}$) planes, respectively.
For LFV muon decays, $\br{\mu}{e\gamma}$ is much larger than $\br{\mu}{eee}$.
As already discussed in Section~\ref{sec-LFV-lgl} and \ref{sec-LFV-l3l} this can be understood analytically. 
While the former decay is quadratically proportional to the tiny mixing angle $\eps_e$ between electron and VL leptons, 
the latter decay scales with $\eps_e^6$. Thus, our model predicts $\br{\mu}{e\gamma} \gg \br{\mu}{eee}$.

For LFV $\tau$ decays, in contrast, $\br{\tau}{\mu\mu\mu}$ is roughly of the same order of magnitude as
$\br{\tau}{\mu\gamma}$. This can be understood because both of them are scaling as $\eps_\tau^2$,
where $\eps_\tau$ is the small mixing angle between $\tau$ and the VL leptons.
All the other LFV tau decays are suppressed by additional powers of $\eps_\tau$ and/or $\eps_e$.

We see that especially for $\mu\to e\gamma$ there are many best fit points close to the current
upper limit. 
This limit will be significantly improved to $<6\times10^{-14}$ by the upcoming MEG II experiment \cite{Baldini:2018nnn}.
Similarly an improvement on the upper limit of $\br{\tau}{\mu\gamma}$ by up to two orders of magnitude is expected from Belle II \cite{Kou:2018nap}.
Nonetheless, we find good fit points that extend into regions which will not be probed by upcoming experiments.
We therefore conclude that while a future excess in the charged LFV channels $\mu\to e\gamma$ and/or $\tau\to \mu\gamma$ could consistently be explained in our model,
those observables will not be the first to exclude this model.

Regarding charged LFV decays of SM bosons at tree level, 
we find that the respective branching fractions are more 
than several orders of magnitude smaller than the current bounds.
In fact, the couplings of SM bosons to SM fermions are very close to their SM values
which we have already discussed in Section~\ref{sec-EWBoson} based on our analytical treatment 
shown in Appendix~\ref{sec-anal}.

\subsubsection{Signals of Vector-Like Leptons}
\label{sec:VLfermions}
\begin{figure}[p]
\begin{minipage}[c]{0.5\hsize}
 \centering
\includegraphics[height=75mm]{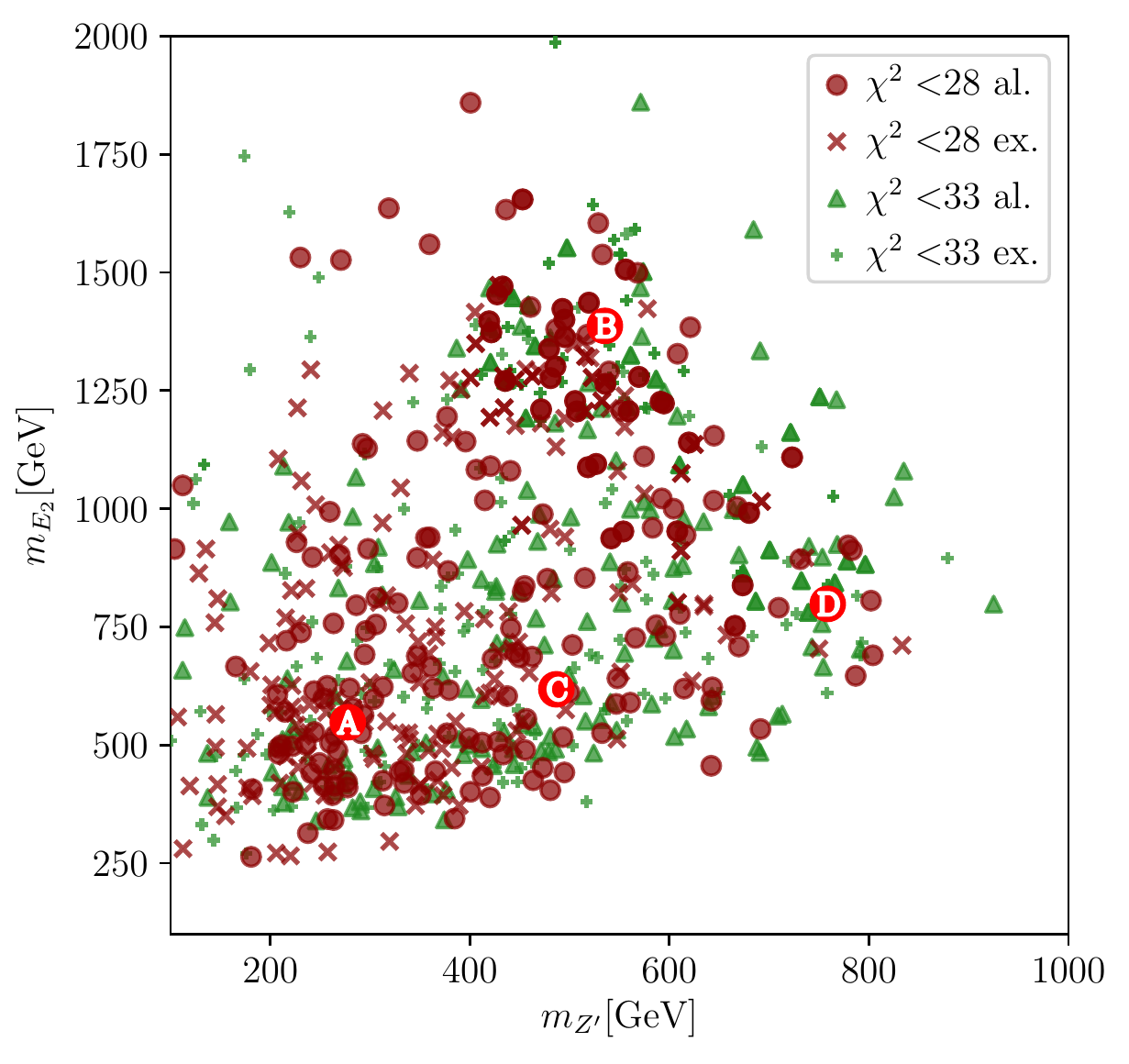}
\end{minipage}
\begin{minipage}[c]{0.5\hsize}
 \centering
\includegraphics[height=75mm]{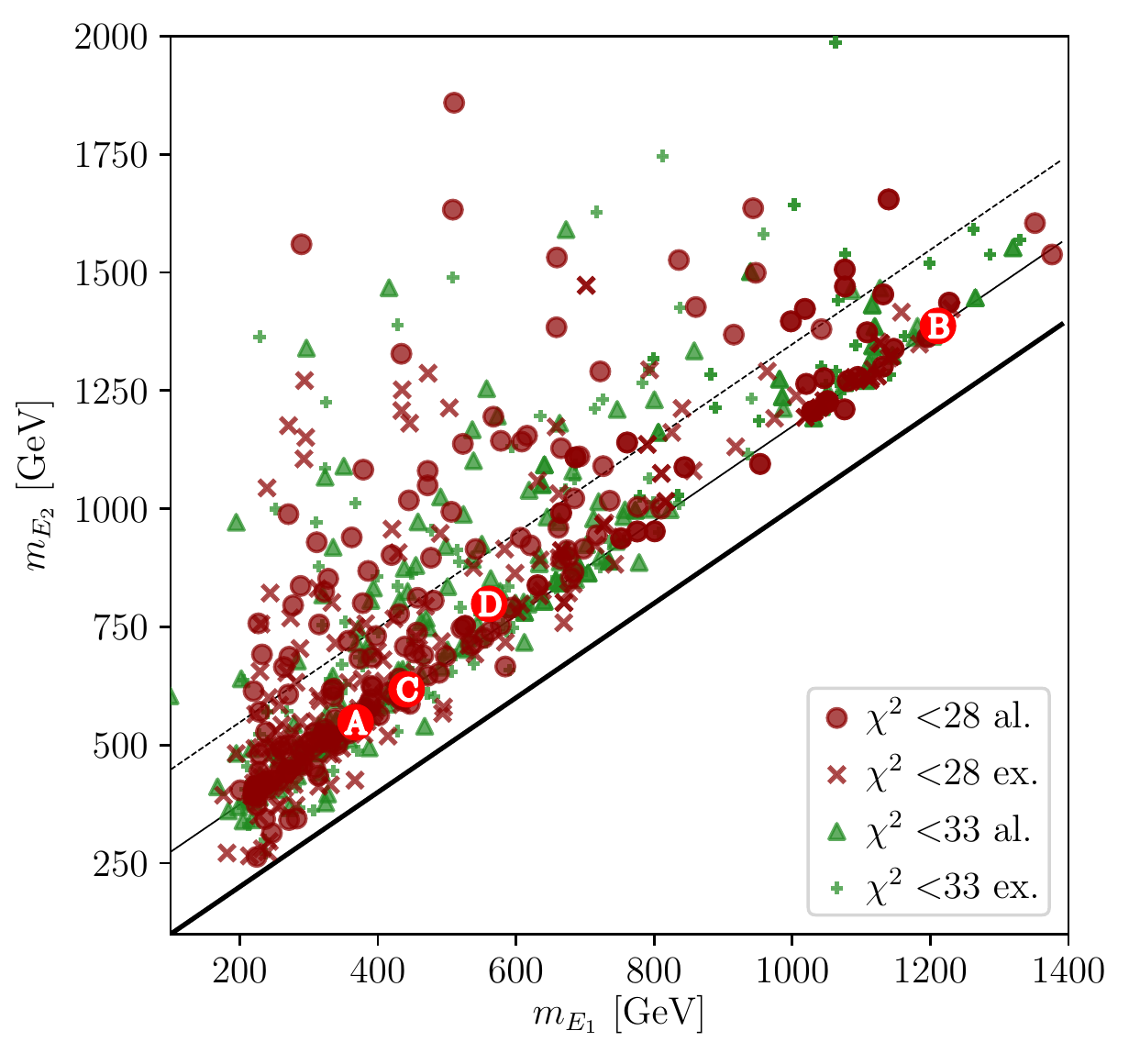}
\end{minipage}
\begin{minipage}[c]{0.50\hsize}
 \centering
\includegraphics[height=75mm]{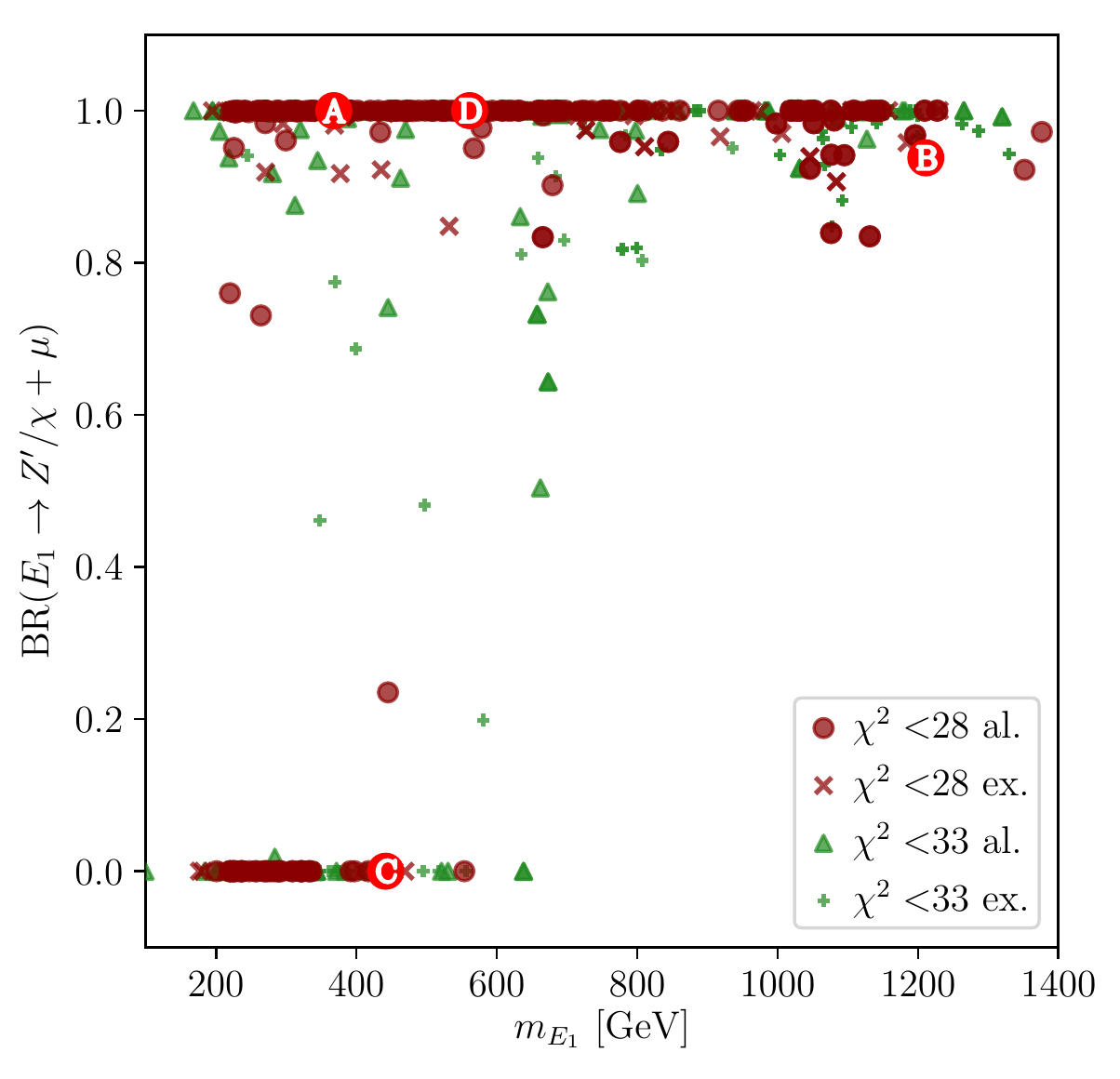}
\end{minipage}
\begin{minipage}[c]{0.50\hsize}
 \centering
\includegraphics[height=75mm]{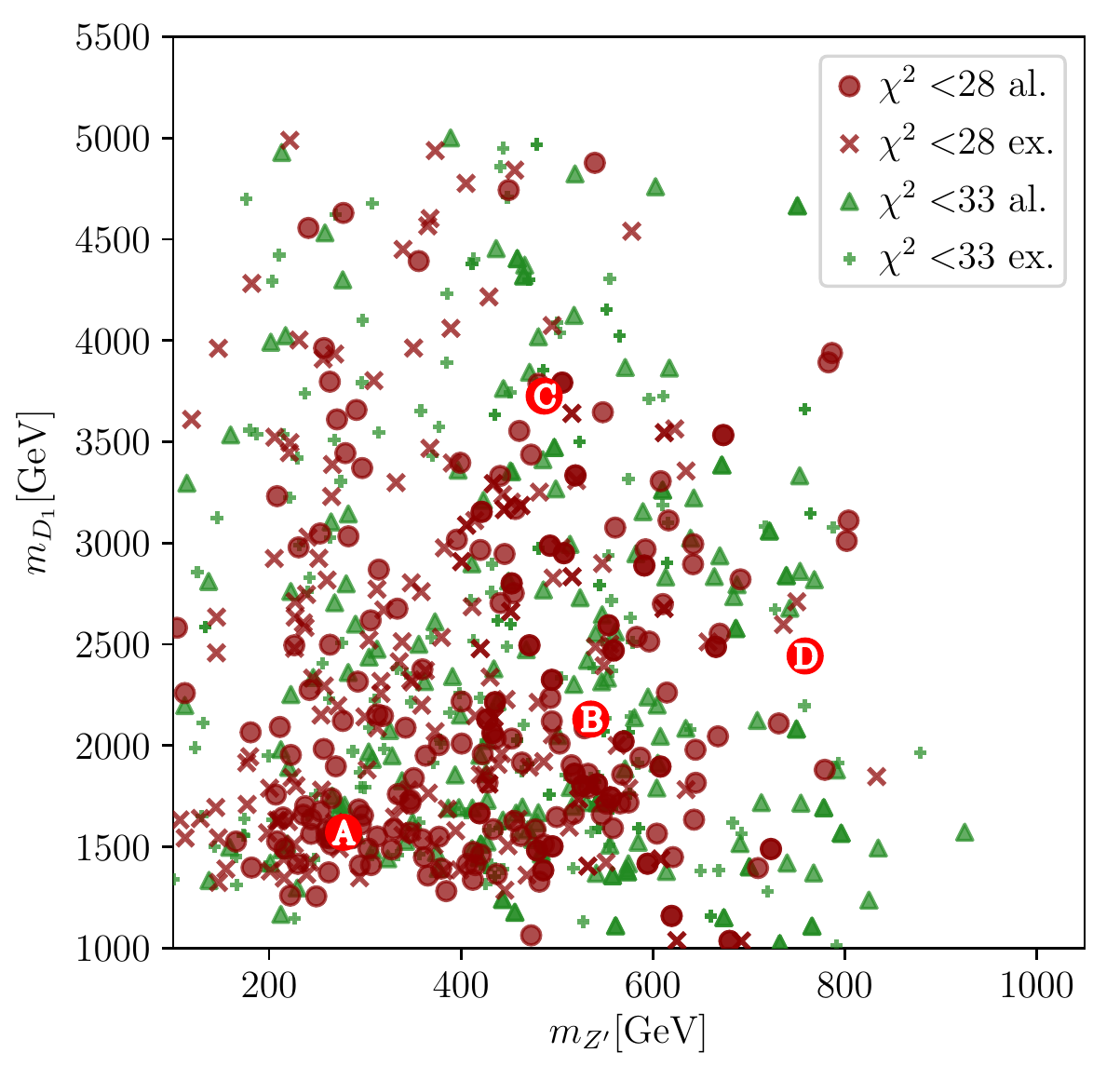}
\end{minipage}
 \caption{
\label{fig-scatVLL}
Good fit points and their predicted values of the observables, $m_\Zp$, $m_{E_1}$, $m_{E_2}$, $m_{D_1}$, and $\br{E_1}{\Zp/\chi + \mu}$.
The color coding is the same as in Fig.~\ref{fig-scatmZpgp}.
}
\end{figure}
Finally, let us investigate collider signatures of VL fermions. 
As discussed in Section~\ref{sec-AMM}, the VL lepton mass is constrained to explain the muon anomalous magnetic moment,
and $\Delta a_\mu = 2.68\times 10^{-9}$ can be realized only within the parameter space shown in Fig.~\ref{fig-E1bound}.
In the same figure we show the masses of the lightest VL lepton $m_{E_1}$ and $m_\Zp$
for our best fit points. 
Most points have $m_{E_1} \lesssim 800$ GeV
and the points with $m_{E_1} > 800$ GeV are found only where $m_\Zp \sim 500$ GeV
as expected from our analytical discussion illustrated by the contours in Fig.~\ref{fig-E1bound}.
In the upper panels of Fig.~\ref{fig-scatVLL} we show the distribution of 
the heavier VL lepton masses $m_{E_2}$ with respect to $m_\Zp$ (left) and 
with respect to the lighter VL lepton $m_{E_1}$ (right).
The black thick, thin, and dashed lines show mass splittings $\Delta m_E \define m_{E_2}-m_{E_1} = 0, 174$, and $2\cdot 174$ GeV, respectively.
The mass splitting is bounded by $\sim \la^\prime_e  v_H$, and it is typically not very large, since
the loop function $C_\Zp$ contributing to $\Delta a_\mu$, defined in Eq.~\eqref{eq-CLR}, 
is maximized when the masses are degenerate.
Consequently, the heavier VL lepton $E_2$ is typically not much heavier than about $1.5\,\mathrm{TeV}$.
According to Ref.~\cite{Bhattiprolu:2019vdu},
the production cross section of a doublet VL lepton with mass $1.5\,\mathrm{TeV}$ is about $\order{0.1}$ fb
which corresponds to about $30~(300)$ total events at the end of LHC (HL-LHC).
Therefore, the HL-LHC could exclude the whole parameter space compatible
with $\Delta a_\mu$ if the signals for VL leptons are very clean.

In the present model, high-multiplicity muon signals are expected from the production and decay of VL leptons.
The decay modes crucially depend on the masses of the $\Zp$ and $\chi$ bosons.
We show the dominant two-body decay modes and their branching fractions 
at our best fit points in Tables~\ref{tab-massdecayA}-\ref{tab-massdecayD} in Appendix~\ref{sec-fullobs}. 
If either of the following final states is kinematically allowed, the lightest VL lepton decays dominantly to 
\begin{align}
\label{eq:NP_VL_decay}
 E_1 \to \Zp + \mu,\quad \mathrm{and/or} \quad E_1 \to \chi + \mu.
\end{align}
For illustration, the lower left panel in Fig.~\ref{fig-scatVLL} shows 
the sum of $\br{E_1}{\Zp\mu} + \br{E_1}{\chi\mu}$ in dependence of $m_{E_1}$
for our good fit points, cf.\ also Tables~\ref{tab-massdecayA}-\ref{tab-massdecayD}.
Often either $\chi$ or $\Zp$ is lighter than the VL leptons, as is the case for the best fit points A, B, and D.
This comes about because a light scalar $\chi$ is favored to suppress its destructive contribution to $\Delta a_\mu$,
while the $\Zp$ mass is controlled by the overall scale $v_\Phi$ which is rarely above $\sim 1$ TeV.

On the contrary, if $m_{E_1} < m_{\Zp}, m_\chi$ (as in our best fit point C), 
$E_1$ decays predominantly to a SM boson and a muon or neutrino,
\begin{align}\label{eq:SM_VL_decay}
 E_1 \to h/Z + \mu,\quad \mathrm{and/or} \quad E_1 \to W + \nu_\mu.
\end{align}
The detailed ratio of these branching fractions depends on the
mixing between the singlet-like and doublet-like VL states.

The final states in Eq.~\eqref{eq:SM_VL_decay} have been studied as a signature of VL leptons in several papers~\cite{Aad:2015dha,
Dermisek:2014qca,Sirunyan:2019ofn,Bhattiprolu:2019vdu,Kumar:2015tna,Falkowski:2013jya,Ellis:2014dza}.
However, there is no study by LHC experiments of VL leptons decaying to a muon based on LHC Run-2 data.
A dedicated analysis of the experimental data shows that a singlet-like VL lepton at the point C with mass above 200 GeV
can not be excluded~\cite{Dermisek:2014qca}.

The final states in Eq.~\eqref{eq:NP_VL_decay} have not been considered so far; 
they give rise to characteristic multi-lepton final states.
This comes about because the $\Zp$ boson predominantly decays to a pair of muons or muon neutrinos, 
cf.\ Tables~\ref{tab-massdecayA}-\ref{tab-massdecayD}.
The scalar $\chi$ couples to SM fermions through the tiny Yukawa couplings induced by mixing effects,
$\sim \eps_{f_\mathrm{SM}} m_{f_\mathrm{SM}}/{M_{F_\mathrm{VL}}}$.
However, at most points $\chi$ has a large coupling to muons because of the large mixing.
In addition, couplings to third generation quarks could also be large due to an enhancement by their heavy masses.
This is the case at our best fit point D, cf.\ Table~\ref{tab-massdecayD}.
The sizable branching fractions of the exotic boson to pairs of muons provide clean resonance signals,
\begin{align}
 \Zp \to \mu^+ \mu^- \quad \mathrm{and} \quad \chi \to \mu^+ \mu^-.
\end{align}
Therefore, processes with dramatic multi-resonant multi-lepton final states, such as
\begin{align}
 p p \to E^+_1 E^-_1 \to \mu^+ \left(\mu^+ \mu^- \right)_B + \mu^-  \left(\mu^+ \mu^- \right)_{B},\quad
B=\chi, \Zp
\end{align}
are expected from the VL lepton production.
Here, $\left(\mu^+\mu^-\right)_B$ is a pair of muons with a resonant feature at the invariant mass $m_{\mu\mu}^2 \sim m_B^2$.
At point D, the doublet-like VL neutrino also decays to the exotic scalar and signals such as
\begin{align}
 p p \to E_1 N_1 \to \mu \left(\mu^+ \mu^- \right)_B + \nu \left(\mu^+ \mu^- \right)_{B},
\end{align}
are expected. This signal is expected if the lightest VL lepton is doublet-like.

The heavier VL lepton decays in a more complicated way.
It will predominantly decay to the lighter VL lepton under the emission of a SM boson, 
since there is sizable mixing between the VL leptons in order to enhance the left-right effects 
to $\Delta a_\mu$. The most dramatic case may be
\begin{align}
 E_2 \to E_1 Z \to \mu (\mu^+\mu^-)_\Zp + (\ell^+ \ell^-)_Z,
\end{align}
which results in five SM leptons from one VL lepton.
A pair of VL leptons, thus, could produce up to ten leptons per event.
Although it may be difficult to reconstruct all of them,
such high-multiplicity lepton signals provide a very distinctive event topology.

\subsubsection{Signals of Vector-Like Quarks}
The lower right panel of Fig.~\ref{fig-scatVLL} shows the good fit points in the ($m_\Zp$, $m_{D_1}$) plane.
Unlike for the VL leptons, there is no stringent upper limit on the VL quark masses.
This is because small $\Zp$ couplings to the SM quarks are enough to explain the $\bsll$ anomalies.
Moreover, the mixing itself is given by $\la^Q_i v_\Phi/\la^Q_V v_\phi$ and is independent of the Higgs VEV.
The VL quarks can be within the reach of current and future collider experiments if they are light.
For instance, point A has a singlet-like VL quark with mass $\sim 1.6$ TeV.
Since $v_\Phi < 1.7$ TeV is required,
VL quark decays to $\Zp$ or $\chi$ is always kinematically allowed.
As for the VL leptons, dramatic signals involving $\Zp$ or $\chi$, e.g.
\begin{align}
  p p  \to Q Q  \to \mathrm{jet}\ (\mu^+\mu^-)_{\Zp} +  \mathrm{jet}\ (\mu^+\mu^-)_{\Zp},
\end{align}
are expected.
These high-multiplicity leptons with resonant features in association with jets
provide another distinctive signal of this model.

\section{Summary}
\label{summary}
We have presented an extension of the Standard Model with the addition of a vector-like family of quarks and leptons which also
carry a new $\U1^\prime$ charge.  The model is constructed to address the known experimental anomalies associated with muons, i.e.
the anomalous magnetic moment of the muon and the decays $b \rightarrow s \ell^+ \ell^-$.  
SM quarks and leptons feel the new $Z^\prime$ gauge interactions only via mass mixing with the VL family.  
The model contains two additional SM singlet scalars, one that is used to model the masses of the VL family and another one that mixes the VL family 
with the SM states and obtains a VEV that spontaneously breaks the $\U1^\prime$ symmetry.  
We performed a global $\chi^2$ analysis of
the data, with 65 arbitrary model parameters and taking into account 98 observables.
We have found many model points which satisfy $\chi^2/N_{\mathrm{d.o.f.}}\leq 1$. 
We cannot simultaneously fit the anomalous magnetic moment of the electron and muon, 
because the theory is severely constrained by the experimental upper bound on $\br{\mu}{e \gamma}$.\footnote{%
During the completion of our work, Ref.~\cite{CarcamoHernandez:2019ydc} appeared on the arXiv which reaches 
the same conclusion for a model with $\Zp$ and VL leptons (see also the somewhat related discussion in \cite{Crivellin:2018qmi}).} 
We, therefore, decided to only fit $\Delta a_\mu$.  
All good fit points have $\mathrm{BR}(\mu\to e\gamma)>10^{-16}$ and $\mathrm{BR}(\tau\to \mu\gamma)>10^{-15}$ with the latter being
strongly correlated with $\mathrm{BR}(\tau\to 3\mu)$.  
Roughly half of our best fit points have $\mathrm{BR}(\mu\to e\gamma)$ in a range that is accessible by upcoming experiments.
However, note that this is not a necessity of the model, i.e.\ $\br{\mu}{e\gamma}$ could always be suppressed by tuning
$\eps_{e_{L,R}}$ to zero without affecting the explanation of the anomalies or SM predictions.

With regards to $b \rightarrow s \ell^+ \ell^-$ decay processes, 
we fit the Wilson coefficients for new physics contributions as discussed in Ref.~\cite{Aebischer:2019mlg}. 
Only two of the five possible good fit points of this analysis can be fit in our model.
The flavor violating decays of SM bosons, i.e.\ Higgs, $W$ and $Z$ are severely suppressed in our model. 
The vector-like quark induced coupling to $\Zp$ also gives sizable contributions to neutral meson mixing, 
particularly $K$-$\bar K$, $B_{d,s}$-$\bar B_{d,s}$, and $D$-$\bar D$ mixing.
The best-fit values for many CKM elements in our model consistently deviate from their experimental 
central values at the level of 1-2$\sigma$ (as they do also in the Standard Model).
Hence, more accurate constraints of CKM non-unitarity and more precise measurements of CKM elements
would be very welcome to further test the model. 

In order to understand the predictions for new physics we have presented four ``best fit points'' 
- A, B, C, and D with the 
masses of the new particles and their decay rates 
given in Tables \ref{tab-massdecayA}, \ref{tab-massdecayB}, \ref{tab-massdecayC} and \ref{tab-massdecayD}, respectively.  
Many more details are given in the Appendices.  
The fit values for some selected observables are given in Table \ref{tab-selobs}.  
Although the $Z^\prime$ mass is typically significantly less than a TeV and it decays with a
significant branching fraction to $\mu^+ \mu^-$,  
we find many points not excluded by recent ATLAS searches for a dimuon resonance. 
We are also constrained by neutrino trident processes. 
The VL leptons are typically light, while the VL quarks are significantly heavier with mass of order a few TeV.
Since the lightest VL leptons at best fit points, A, C and D, have mass between 300 - 600 GeV, 
these states may be accessible even at the LHC, and even more so at the HL-LHC. 
They typically result in multi-muon production channels as discussed in Section \ref{sec:VLfermions}.

This model is a prototype which highlights that fixing anomalies with consistent models, while maintaining the successful 
Standard Model predictions, comes at a price:
The model appears eminently testable and, therefore, can be excluded in many complementary ways.

\section*{Acknowledgments}
We thank R.\ Dermisek for useful discussions.
The work of J.K.\ and S.R.\ is supported in part by the Department of Energy (DOE) under Award No.\ DE-SC0011726.
The work of J.K.\ is supported in part by the Grant-in-Aid for Scientific Research from the
Ministry of Education, Science, Sports and Culture (MEXT), Japan No.\ 18K13534.
The work of A.T.\ was partly supported by a postdoc fellowship of the German Academic Exchange Service~(DAAD).
A.T.\ is grateful to the Physics Department of Ohio State University and
Centro de F\'isica Te\'orica de Part\'iculas (CFTP) at Instituto Superior T\'ecnico, Lisbon
for hospitality during parts of this work.

\appendix
\section{Decay Width Formulas}
\label{sec-twobody}
Widths of two-body decays with both left-handed and right-handed interactions
are summarized in this Appendix.
\subsection{Scalar Decays}
With the Yukawa interactions of a real scalar field $\phi$ and two fermions $f_1, f_2$,
\begin{align}
\label{eq-Lsdec}
\Lcal_{\phi\to f_1 \ol{f}_2} = - \phi \left(y_L \ol{f}_1 P_L f_2 + y_R \ol{f}_1 P_R f_2 \right),
\end{align}
the partial width of $\phi$ is given by
\begin{align}
 \Gamma\left(\phi\to f_1 \ol{f}_2\right)
= \frac{m_\phi}{16\pi} \la(x_1,x_2) \left[
      \left(\abs{y_L^2}+\abs{y_R^2}\right)\left(1-x_1-x_2\right)
     -\mathrm{ Re} \left(y_L y_R^* \right) 4 \sqrt{x_1 x_2}
       \right],
\end{align}
where $x_i = m_{f_i}/m_\phi^2$ and
\begin{align}
\label{eq-lamfun}
 \la(x_1, x_2) \define \sqrt{1-2(x_1+x_2)+(x_1-x_2)^2}.
\end{align}

\subsection{Gauge Boson Decays}
Gauge interactions of a vector field $V$, two fermions $f_1, f_2$,
\begin{align}
\label{eq-Lvdec}
\Lcal_{V\to f_1 \ol{f}_2} = V_\mu
        \left(g_L \ol{f}_1 \gamma^\mu P_L f_2+g_R \ol{f}_1\gamma^\mu P_R f_2 \right).
\end{align}
give the partial width,
\begin{align}
 \Gamma\left(V \to f_1 \ol{f}_2\right)
= \frac{m_V}{24\pi} \la(x_1,x_2) &\ \Bigg[ 
      \left(\abs{g_L^2}+\abs{g_R^2} \right) 
        \left(1-\frac{x_1+x_2}{2}-\frac{(x_1-x_2 )^2}{2} \right)  \\ \notag 
     &\quad   + \mathrm{ Re} \left(g_L g_R^* \right) 6 \sqrt{x_1 x_2}
       \Bigg],
\end{align}
where $x_i = m_{f_i}/m_V^2$.

\subsection{Fermion Decays}
If $m_{f_2} > m_{f_1}+m_\phi$,
a fermion $f_2$ can decay as $f_2 \to f_1 \phi$ through the Yukawa interaction
in Eq.~\eqref{eq-Lsdec}.
The partial width is given by
\begin{align}
 \Gamma\left(f_2 \to f_1 \phi \right)
= \frac{m_{f_2}}{32\pi} \la(y, z) \left[
      \left(\abs{y_L^2}+\abs{y_R^2} \right)\left(1-y-z\right)
       + \mathrm{ Re} \left(y_L y_R^* \right) 4 \sqrt{y}
       \right],
\end{align}
where $y=m_{f_1}^2/m_{f_2}^2$ and $z=m_{\phi}^2/m_{f_2}^2$.

The gauge interactions in Eq.~\eqref{eq-Lvdec}
induce a decay $f_2 \to f_1 V$, if $m_{f_2} > m_{f_1}+m_V$.
The partial width is given by
\begin{align}
 \Gamma\left(f_2 \to f_1 V \right)
= \frac{m_{f_2}^3}{32\pi m_V^2} \la(y, z) &\ \Bigg[ 
      \left(\abs{g_L^2}+\abs{g_R^2} \right)
      \left\{ (1-y)^2+z(1+y)-2z^2 \right\}   \notag \\ 
    &\quad   - \mathrm{ Re} \left(g_L g_R^* \right) 3 z \sqrt{y}
       \Bigg],
\end{align}
where $y=m_{f_1}^2/m_{f_2}^2$ and $z=m_{V}^2/m_{f_2}^2$.

\section{Analytical Analysis}
\label{sec-anal}
Many analytical formulae used in the main text are derived in this Appendix.
\subsection{Diagonalization of the Dirac Mass Matrices}
\label{subsec-diag}
The $5\times 5$ Dirac mass matrices are given by
\begin{align}
 \ol{f}_\mathrm{R}
 \mcal^{f}
 f_\mathrm{L}
=
\begin{pmatrix}
 {\ol{f}_\mathrm{R}}_i & \ol{F}_\mathrm{R} & \ol{F}'_\mathrm{R}
\end{pmatrix}
\begin{pmatrix}
 y^f_{ij} v_H   &  0    & \la^R_i v_\Phi         \\
 0    &    \la_f v_H    &  \la_V^R  v_\phi  \\
 \la^L_j v_\Phi   & \la_V^L v_\phi & \la_f^\prime v_H
\end{pmatrix}
\begin{pmatrix}
 {f_\mathrm{L}}_j \\ F'_\mathrm{L} \\ F_\mathrm{L}
\end{pmatrix},
\end{align}
where $(f, F, L, R) = (e,E,L,E), (u, U,Q, U), (d, D,Q, D)$
for charged leptons, up or down quarks, respectively, see Eqs.~\eqref{eq:Me}-\eqref{eq:Md}.
We are interested in the case $v_H \ll \la^{L,R}v_\phi$,
in which case the VL fermions are heavy enough to be consistent with LHC limits.
We diagonalize all the Dirac mass matrices perturbatively in $v_H/v_\phi$.
At leading order, i.e.\ $v_H$, the mass matrices are block diagonalized by the unitary matrices,
\begin{align}
\label{eq-ULR0}
 U^0_{f_L} =
\begin{pmatrix}
\vz_{L_j}^f & \vn_L^f & \bzero \\
 0_j     & 0     & 1
\end{pmatrix},
\quad
 U^0_{f_R} =
\begin{pmatrix}
\vz_{R_j}^f & \nv_R^f & \bzero \\
 0_j     & 0     & 1
\end{pmatrix}.
\end{align}
Here, the four-component vectors obey the following conditions,
\begin{align}
\label{eq-fvcond}
&\ \vn_L^f = \frac{1}{M_{F_L}}
\begin{pmatrix}
 \la^L_{i} v_\Phi \\ \la_V^L v_\phi
\end{pmatrix}^*,
\quad
 \vn_R^f =
\frac{1}{M_{F_R}}
\begin{pmatrix}
 \la^R_{i} v_\Phi \\ \la_V^R v_\phi
\end{pmatrix}, \\
&
\left(\vz^f_{L_i}\right)^\dagger \vn^f_L = \left(\vz^f_{R_i}\right)^\dagger \vn^f_R = 0,\quad
\left(\vz^f_{L_i}\right) ^\dagger \zv^f_{L_j} = \left(\zv^f_{R_i}\right) ^\dagger \zv^f_{R_j} =  \delta_{ij}, 
\label{eq-zorthn}
\end{align}
where
\begin{align}
 M_{F_L} = \sqrt{\sum_{i=1}^3\abs{\la^L_i v_\Phi}^2+\abs{\la^L_V v_\phi}^2 },
\quad
 M_{F_R} = \sqrt{\sum_{i=1}^3\abs{\la^R_i v_\Phi}^2+\abs{\la^R_V v_\phi}^2 }.
\end{align}
The vectors $\zv_{L_i}$, $\zv_{R_i}$ can be any four-component vectors which satisfy Eq.~\eqref{eq-zorthn}.  
Another set of 
 $\zv^\prime_{L_i} =[u_L]_{ij} \zv_{L_j}$, $\zv_{R_i}^\prime =  [u_R]_{ij} \zv_{R_j}$,
with arbitrary unitary matrices $u_L, u_R$ also satisfy the conditions Eq.~\eqref{eq-zorthn}.   
We define these vectors such that the upper-left $3\times3$ block is diagonalized 
by using this degree of freedom.

Rotating the mass matrix by these unitary matrices while keeping $\order{v_H}$ elements we obtain
\begin{align}\label{eq:m0}
 \tilde{\mathcal{M}}^f \define&\ \mhc{U_{f_R}^{0}} \mathcal{M}^f U_{f_L}^0   \\ \notag 
=&\ 
\begin{pmatrix}
\left(\vz_{R_i}^f\right)^\dagger \hat{m}_f\,\vz^f_{L_j}  &
\left(\vz_{R_i}^f\right)^\dagger \hat{m}_f \,\vn^f_L  & 0_i   \\
\left(\vn_{R}^f \right)^\dagger \hat{m}_f\,\vz^f_{L_j}  &
\left(\vn_{R}^f \right)^\dagger \hat{m}_f\,\vz^f_{L_j} &  M_{F_R} \\
0_j & M_{F_L} & \la^\prime_f v_H
 \end{pmatrix}
\define 
\begin{pmatrix}
\ty^f_{i}\delta_{ij} v_H  & \ty^f_{R_i} v_H & 0_i \\
 \ty^f_{L_j} v_H & \tla_f v_H & M_{F_R} \\
  0_j  & M_{F_L} & \la_f^\prime v_H \\
\end{pmatrix},  
\end{align}
where the $4\times 4$ matrix $\hat{m}_f$ is defined as
\begin{align}
 \hat{m}_f \define
 \begin{pmatrix}
  y^f_{ij} & 0_i \\
 0_j      & \la_f
 \end{pmatrix}
v_H.
 \end{align}
The matrix $\tilde{\mathcal{M}}^f$ is $3\oplus2$ block diagonal, except $\ty^f_{L_i,R_j} v_H$.
The mixing effects induced by these elements are $\mathcal{O}()\ty^f_{L,R} v_H/M_{F_{L,R}})$, 
suppressed by Yukawa couplings and VL fermion masses.

Next, we diagonalize the lower-right block.
We are interested in parameters where $M_{E_{L,R}} \gtrsim 250$ GeV 
and $M_{Q_{L,R}} \gtrsim 1.5$ TeV to be consistent with LHC searches.
The VL quarks are substantially heavier than $v_H$, while the VL leptons 
are at a few hundred GeV with a Yukawa coupling $\la^{\prime}_e\sim\order{1}$ as required in order 
to explain $\Delta a_\mu$. Fortunately, the other Yukawa couplings are small enough due to the small 
charged leptons masses. Keeping $\la^\prime_e v_H$, the next order of unitary matrices is parametrized as 
 \begin{align}
 U_{f_R}^1 =
\begin{pmatrix}
 \delta_{ij} & 0_i & 0_i \\
 0_j & s_{\theta_R}  & c_{\theta_R}   \\
 0_j & c_{\theta_R} & -s_{\theta_R}  \\
\end{pmatrix},
\quad
 U_{f_L}^1 =
\begin{pmatrix}
 \delta_{ij} & 0_i & 0_i \\
 0_j & c_{\theta_L} & - s_{\theta_L} \\
 0_j & s_{\theta_L} & c_{\theta_L}  \\
\end{pmatrix},
\end{align}
with angles $\theta_{L,R}$ that satisfy
\begin{align}
\label{eq-Mtilde}
\begin{pmatrix}
 s_{\theta_R}  & c_{\theta_R}   \\
 c_{\theta_R} & -s_{\theta_R}  \\
\end{pmatrix}
\begin{pmatrix}
 \tla_f v_H & M_{F_R} \\
M_{F_L} & \la^\prime_f v_H
\end{pmatrix}
\begin{pmatrix}
 c_{\theta_L} & - s_{\theta_L} \\
 s_{\theta_L} & c_{\theta_L}  \\
\end{pmatrix}
=: \mathrm{diag}\left(\tM_{F_L}, \tM_{F_R} \right). 
\end{align}
The rotated mass matrix is
\begin{align}
\label{eq-M1}
 \left(U_{f_R}^1 \right)^\dag \tilde{\Mcal}^f U_{f_L}^1 =&\
\begin{pmatrix}
 m_{f_i} \delta_{ij} & c_{\theta_L} \ty^f_{R_i} v_H & - s_{\theta_L}  \ty^f_{R_i} v_H  \\
s_{\theta_R} \ty^f_{L_j} v_H &  \tM_{F_L} & 0 \\
c_{\theta_R} \ty^f_{L_j} v_H & 0 & \tM_{F_R} 
\end{pmatrix}. 
\end{align}
We now give approximate forms for angles $\theta_{L,R}$ and masses $\tilde{M}_{F_{L,R}}$. 
Here, we neglect $\la_f v_H$.  
If $\la^\prime_f v_H \ll\abs{ M_{F_{L}} - M_{F_R} }$, 
the mixing angles and masses are approximately given by 
\begin{align}\notag
 c_{\theta_{L,R}} &=  1 + 
 \order{\delta_{f_{L,R}}^2 },& 
 \tilde{M}_{F_{L,R}} &= M_{F_{L,R}} 
                                                  + \order{\delta_{f_{L,R}} \la^\prime_f v_H },& \\ \label{eq-thetaLR} 
 s_{\theta_{L,R}} &=  \delta_{f_{L,R}} +  \order{\delta_{f_{L,R}}^2 },&  
\end{align} 
with an expansion parameter defined as 
\begin{align}
\label{eq-delta}
 \delta_{f_{L,R}} = \frac{ \la_f^\prime v_H M_{F_{L,R}}}{M_{F_L}^2-M_{F_R}^2}.
 \end{align}
Clearly, this approximation becomes inaccurate 
if the VL fermions are nearly mass-degenerate. 
For the nearly mass-degenerate case one can proceed as follows. 
We introduce three mass parameters, 
\begin{align}
 M_F \define \frac{M_{F_L}+M_{F_R}}{2},\quad 
 \Delta_F \define \frac{M_{F_L}-M_{F_R}}{2}, \quad 
\mu_F \define \sqrt{\Delta_F^2 + \frac{\left(\la^\prime v_H\right)^2}{4}}.  
\end{align}
If $\abs{\la^\prime_f v_H}, \abs{\Delta_F} \ll M_{F}$, 
the masses are given by 
\begin{align}
\label{eq-masscombi}
 \tilde{M}_{F_L} = M_F + \mu_F + \order{\frac{ (\la^\prime v_H)^2}{M_F}},
\quad 
 \tilde{M}_{F_R} = M_F - \mu_F + \order{\frac{ (\la^\prime v_H)^2}{M_F}}. 
\end{align}
The mixing angles are given by 
\begin{align}
\label{eq-theta2}
 c_{\theta_L} =&\ \frac{1}{\sqrt{2}} \left( \alpha_F 
                                         - \frac{\la^\prime_f v_H}{4M_F} \beta_F  \right)
, \quad  
 s_{\theta_L} = \frac{1}{\sqrt{2}} \left( \beta_F 
                                    + \frac{\la^\prime_f v_H}{4M_F} \alpha_F  \right)
, \\ 
 c_{\theta_R} =&\ \frac{1}{\sqrt{2}} \left( \alpha_F 
                                         + \frac{\la^\prime_f v_H}{4M_F} \beta_F  \right)
, \quad  
 s_{\theta_R} = \frac{1}{\sqrt{2}} \left( \beta_F 
                                         - \frac{\la^\prime_f v_H}{4M_F} \alpha_F  \right)
,  
\end{align}
where 
\begin{align}
 \alpha_F \define  \frac{1}{\sqrt{2}} \left( 
     \sqrt{1+\dfrac{\la^\prime_f v_H}{2\mu_F}}+
          \sqrt{1-\dfrac{\la^\prime_f v_H}{2\mu_F}} 
\right), \quad 
 \beta_F \define  \frac{1}{\sqrt{2}} \left( 
     \sqrt{1+\dfrac{\la^\prime_f v_H}{2\mu_F}}- 
          \sqrt{1-\dfrac{\la^\prime_f v_H}{2\mu_F}} 
\right). 
\end{align}
Here, higher orders of $\la^\prime v_H/M_F$ and $\Delta_F/M_F$ are neglected.

We now proceed to further diagonalize Eq.~\eqref{eq-M1}. 
At the first order in  $\ty^f_{L,R} v_H/\tM_{F_{L,R}}$, Eq.~\eqref{eq-M1} is diagonalized by unitary matrices,
\begin{align}
\label{eq-U2fL}
U_{f_R}^2 =&\
\begin{pmatrix}
 \delta_{ij} &
c_{\theta_L}  \ty^{f}_{R_i} v_H/\tM_{F_L} &-s_{\theta_L}\ty^{f}_{R_i} v_H / \tM_{F_R}  \\
-c_{\theta_L} \ty^{f*}_{R_j} v_H / \tM_{F_L}& 1 & 0 \\
 s_{\theta_L} \ty^{f*}_{R_j} v_H / \tM_{F_R} & 0 & 1
\end{pmatrix},
\\
 U_{f_L}^2 = &\
\begin{pmatrix}
 \delta_{ij} &
s_{\theta_R} \ty^{f}_{L_i} v_H / \tM_{F_L} & c_{\theta_R} \ty^{f}_{L_i} v_H / \tM_{F_R}  \\
- s_{\theta_R} \ty^{f*}_{L_j} v_H / \tM_{F_L} & 1 & 0 \\
- c_{\theta_R} \ty^{f*}_{L_j} v_H / \tM_{F_R}  & 0 & 1
\end{pmatrix}. 
\end{align}
Multiplying these unitary matrices one finds
\begin{align}
\label{eq-diag2}
\left(U_{f_R}^2 \right)^\dag  \left(U_{f_R}^1 \right)^\dag \tilde{\Mcal}^f
U_{f_L}^1 U_{f_L}^2=&\ \mathrm{diag}\left(
 m_{f_i} \delta_{ij} ,\ \tM_{F_L},\ \tM_{F_R}
\right)+ \order{\frac{\left(\ty^f_{L,R} v_H\right)^2}{\tM_{F_{L,R}}}}.
\end{align}
The second-order corrections to the upper-left block are given by
\begin{align}
\label{eq-delm}
\Delta m^f_{ij}\define \ty^f_{R_i} \ty^f_{L_j} \frac{ v_H^2}{\tM_{F_L} \tM_{F_R}}
\left(\tM_{F_L} c_{\theta_R}s_{\theta_L}  - \tM_{F_R} c_{\theta_L}s_{\theta_R}  \right)
\sim   \ty^f_{R_i} \ty^f_{L_j} \frac{\la_f^\prime v_H^3}{\tM_{F_L} \tM_{F_R}}.  
\end{align}
Here, it does not matter whether Eq.~\eqref{eq-thetaLR} or Eq.~\eqref{eq-theta2} is used in the second equality;
both give the same result at this accuracy. 
These corrections are negligibly small in the relevant parameter space.
Altogether, the approximate mass basis $\hat{f}_L, \hat{f}_R$ is defined as
\begin{align}\label{eq-appmass}
 \hat{f}_L \define \left(U_{f_L} \right)^\dag f_L
           \define \left(U_{f_L}^2 \right)^\dag\left(U_{f_L}^1 \right)^\dag
                    \left(U_{f_L}^0 \right)^\dag  f_L , \\ 
 \hat{f}_R \define \left(U_{f_R} \right)^\dag f_R
           \define \left(U_{f_R}^2 \right)^\dag\left(U_{f_R}^1 \right)^\dag
                    \left(U_{f_R}^0 \right)^\dag  f_R.
\end{align}

\subsection{EW Boson Couplings}
Couplings of the fermions to SM bosons are completely SM-like at the leading order.
The leading order unitary matrices of Eq.~\eqref{eq-ULR0}
transforms the $\mathrm{SU(2)_L}$ gauge couplings as
\begin{align}
\mhc{U^1_{f_L}}\mhc{U^0_{f_L}} \Pfb U^0_{f^\prime_L} U^0_{f^\prime_L}
=&\ 
V_{ij}\oplus
\begin{pmatrix}
 c_{\theta_L} c_{\theta^\prime_L} &  - c_{\theta_L} s_{\theta^\prime_L} \\
- s_{\theta_L} c_{\theta^\prime_L} &   s_{\theta_L} s_{\theta^\prime_L} \\
\end{pmatrix}\;,
\\
\mhc{U^1_{f_R}} \mhc{U^0_{f_R}} P_5 U^0_{f^\prime_R} U^1_{f^\prime_R}
=&\  
0_{3\times3}\oplus
\begin{pmatrix}
 c_{\theta_R} c_{\theta^\prime_R} &  - c_{\theta_R} s_{\theta^\prime_R} \\
- s_{\theta_R} c_{\theta^\prime_R} &   s_{\theta_R} s_{\theta^\prime_R} \\
\end{pmatrix}.
\end{align}
Here, $V_{ij}$ is a $3\times 3$ unitary matrix,
which is an identity matrix for the $Z$ boson couplings where $f=f^\prime$.
Since $U^1_{f_{L,R}}$ do not affect SM fermion couplings,
only the mixing via $U^2_{f_{L,R}}$ induces flavor violating couplings of SM generations.
Their size is estimated as
\begin{align}
\left[\mhc{U_{f_L}} \Pfb U_{f^\prime_L}\right]_{ij} =&\ V_{ij} + \order{ \frac{\ty_{L_i} \ty_{L_j}\,v_H^2}{M_{F_\mathrm{VL}}^2}},&
\left[\mhc{U_{f_L}} \Pfb U_{f^\prime_L}\right]_{ib} =&\  \order{  \frac{\ty_{L_i}\,v_H}{M_{F_\mathrm{VL} }}},&\\
\left[\mhc{U_{f_R}} P_5 U_{f^\prime_R}\right]_{ij} =&\  \order{  \frac{\ty_{R_i} \ty_{R_j}\,v_H^2}{M_{F_\mathrm{VL}}^2}},& 
\left[\mhc{U_{f_R}} P_5 U_{f^\prime_R}\right]_{ib} =&\  \order{ \frac{\ty_{R_i} \,v_H}{M_{F_\mathrm{VL}}}},&
\end{align}
where $\ty_{L(R)_i} \sim \mathrm{max}\left(\ty^f_{L(R)_i}, \ty^{f^\prime}_{L(R)_i} \right)$ and $M_{F_\mathrm{VL}}$ 
is the typical scale of VL fermions.

The Higgs boson couplings are aligned with the mass matrix by the rotation via $U_{f_{L,R}}^0$,
\begin{align}
\mhc{U^0_{f_R}} Y_f^h  U^0_{f_L} =
\begin{pmatrix}
 \ty^f_{i} \delta_{ij} & \ty_{L_i}^f & 0_i \\
\ty^f_{R_j}& \tilde{\la}_f & 0 \\
0_j &  0 & \la^\prime_f
\end{pmatrix}
= v_H^{-1} \left.\left[\tilde{\Mcal}^f \right]_{ij}\right|_{M_{F_L}=M_{F_R}=0}.
\end{align}
Hence, the Yukawa couplings to SM fermions are diagonal in the mass basis 
if $\ty^f_{L,R}$ are neglected.
The mixing $U^2_{f_{L,R}}$ induces flavor violating couplings of size
\begin{align}
 \left[ \mhc{U_{f_R}} {Y}^h_f U_{f_L} \right]_{ij} = &\  \ty^f_{i} \delta_{ij} +
\order{  \frac{\ty^f_{L_i}\ty^f_{L_j}v_H^2}{M_{F_\mathrm{VL}}^2},\ 
\frac{\ty^f_{R_j}\ty^f_{R_j}  v_H^2}{M_{F_\mathrm{VL}}^2}}, \\
 \left[ \mhc{U_{f_R}} {Y}^h_f U_{f_L} \right]_{ib} = &\  \order{\ty^f_{L_i}},
\quad
 \left[ \mhc{U_{f_R}} {Y}^h_f U_{f_L} \right]_{aj} =  \order{\ty^f_{R_j}}.
\end{align}

\subsection{Charged Leptons}
\label{subsec-diagcl}
For charged leptons, let us start in a basis in which the upper-left $3\times3$ block is diagonal,
\begin{align}
\label{eq-matlepd}
 \Mcal^{e} =
\begin{pmatrix}
  y_{i}^e \delta_{ij} v_H  & 0_i &  \la^E_i v_\Phi \\
  0_j     & \la_e v_H & \la^E_V v_\phi \\
 \la^L_j v_\Phi& \la^L_V v_\phi & \la_e^\prime v_H \\
\end{pmatrix},
\end{align}
such that SM-LFV effects are induced only by $\la^{L,E}_i$.
We can achieve this form by redefining $e_{L_i}, e_{R_i}$.
Such a redefinition does not change the $Z$ and $\Zp$ couplings.
The $W$ boson couplings are changed, but this can be absorbed by a redefinition of the neutrinos.
The Yukawa couplings to the scalars, namely Higgs boson and $\chi$,
are still aligned with the mass matrix.
In our analysis, we assume that all parameters in the charged lepton sector are real.

There should be sizable mixing between the muon and VL leptons to explain $\Delta a_\mu$,
while mixing involving $e$ or $\tau$ can be small to avoid LFV processes.
We introduce mixing parameters involving muon,
\begin{align}
c_{\mu_L}\define  \frac{\la_V^L v_\phi}{M_{E_L}},\quad
s_{\mu_L}\define  \frac{\la_2^L v_\Phi}{M_{E_L}}, \quad
c_{\mu_R}\define \frac{\la_V^E v_\phi }{M_{E_R}},\quad
s_{\mu_R} \define\frac{\la_2^E v_\Phi}{M_{E_R}},
\end{align}
and those for electron and tau,
\begin{align}
 \eps_{e_L} \define \frac{\la^L_1 v_\Phi}{\la_V^L v_\phi},\quad
 \eps_{e_R} \define \frac{\la^E_1 v_\Phi}{ \la_V^E v_\phi },\quad
 \eps_{\tau_L} \define \frac{\la^L_3 v_\Phi}{\la_V^L v_\phi},  \quad
 \eps_{\tau_R} \define \frac{\la^E_3 v_\Phi }{ \la_V^E v_\phi }.
\end{align}
We expect $\eps_{e}, \eps_\tau \ll 1$ in order to suppress the LFV processes.
With these parameterizations, the leading order unitary matrices are given by
\begin{align}
\label{eq-unit0}
 U^0_{e_L} =&\ 
\begin{pmatrix}
 1               &- \eps_{e_L} s_{\mu_L}&0                  &\eps_{e_L}c_{\mu_L}    & 0 \\
 0               & c_{\mu_L}                 &0                  & s_{\mu_L}                     & 0 \\
 0               & -\eps_{\tau_L}s_{\mu_L}&1                  &\eps_{\tau_L}c_{\mu_L} & 0 \\
-\eps_{e_L} & -s_{\mu_L}               &- \eps_{\tau_L}& c_{\mu_L}                    & 0 \\
 0 & 0 & 0& 0& 1
\end{pmatrix}, \\
 U^0_{e_R} =&\ 
\begin{pmatrix}
 1               &- \eps_{e_R} s_{\mu_R}&0                  &- \eps_{e_R}c_{\mu_R}    & 0 \\
 0               & c_{\mu_R}                 &0                     & s_{\mu_R}                     & 0 \\
 0               &- \eps_{\tau_R}s_{\mu_R}&1                  &- \eps_{\tau_R}c_{\mu_R} & 0 \\
 -\eps_{e_R} & -s_{\mu_R}               &-\eps_{\tau_R}& c_{\mu_R}                    & 0 \\
 0 & 0 & 0& 0& 1
\end{pmatrix}.
\end{align}
The diagonal structure of the upper-left $3\times 3$ block holds as far as
$\eps_{e_{L,R}}, \eps_{\tau_{L,R}} \ll 1 $.
The large mixing with the muon and VL leptons indicate that $\la_e \sim y_2^e \sim m_\mu/v_H$
to explain the muon mass without fine-tuning.
The Yukawa couplings in the off-diagonal block are given by
\begin{align}
 \ty^e_L =
\begin{pmatrix}
 c_{\mu_R} (-\la_e \eps_{e_L}+y^e_1 \eps_{e_R})  \\
c_{\mu_L} s_{\mu_R} y^e_2  -c_{\mu_R} s_{\mu_L} \la_e \\
c_{\mu_R} ( -\la_e \eps_{\tau_L}+y^e_3 \eps_{\tau_R})
\end{pmatrix},
\quad
 \ty^e_{R} =
\begin{pmatrix}
c_{\mu_L} (-\la_e \eps_{e_R}+y^e_1 \eps_{e_L})  \\
c_{\mu_R} s_{\mu_L }y^e_2 -\la_e c_{\mu_L} s_{\mu_R }                   \\
c_{\mu_L} ( -\la_e \eps_{\tau_R}+y^e_3 \eps_{\tau_L} )
\end{pmatrix}.
\end{align}
Their size is estimated as
\begin{align}
 \ty^e_{L_1, R_1}  \lesssim&\  1.0 \times 10^{-9} \times \left(\frac{\eps_e}{10^{-6}}\right)
                                                \left(\frac{\mathrm{max}\left(\la_e, y^e_1\right)}{  10^{-3}  } \right),  \\
 \ty^e_{L_2, R_2}  \lesssim&\  y^e_2 \sim  1.0 \times 10^{-3}, \\
 \ty^e_{L_3, R_3}  \lesssim&\  1.0 \times 10^{-4} \times \left(\frac{\eps_\tau}{10^{-2}}\right)
                                                \left(\frac{ y^e_3}{m_\tau / v_H} \right).
\end{align}
Hence, the perturbative corrections to the off-diagonal elements are at most,  
\begin{align}
\Delta m^e_{ij} \sim \frac{\ty^e_{L_i} \ty^e_{R_j} \la_f^\prime}{M_{E_L} M_{E_R}} v_H^3
\lesssim &\ \eps_\tau \frac{  m_\mu m_\tau}{M_{E_{L}}M_{E_R} } \la^\prime_e v_H   \\ \notag
\sim &\  1.3 \times 10^{-6} \ \mathrm{GeV}
\times      \left(\frac{\eps_{\tau}}{0.01}\right)
               \left(\frac{500\ \mathrm{GeV}}{ \sqrt{M_{E_{L}} M_{E_R}}  } \right)^2
                                            \left(\frac{\la^\prime_e}{1.0}\right).
\end{align}
Consequently, the basis defined in Eq.~\eqref{eq-appmass} is very close to the mass basis and
flavor violating couplings of the charged leptons to the SM bosons are strongly suppressed.

Using the above results, the $\Zp$ couplings to charged leptons are given by
\begin{align}
\label{eq-gZpe}
\!\!\!\!\!\! - \hg^{\Zp}_{e_L} \sim  \gp
\begin{pmatrix}
\eps_{e_L}^2  & s_{\mu_L} \eps_{e_L} & \eps_{e_L} \eps_{\tau_L}
 & - c_{\theta_L} c_{\mu_L} \eps_{e_L}  & s_{\theta_L} c_{\mu_L} \eps_{e_L} \\
s_{\mu_L} \eps_{e_L} & s_{\mu_L}^2 & s_{\mu_L} \eps_{\tau_L}
 & - c_{\theta_L} c_{\mu_L} s_{\mu_L} & s_{\theta_L}  c_{\mu_L} s_{\mu_L} \\
\eps_{\tau_L} \eps_{e_L}  & s_{\mu_L} \eps_{\tau_L} & \eps_{\tau_L}^2
 & - c_{\theta_L} c_{\mu_L} \eps_{\tau_L}  & s_{\theta_L} c_{\mu_L} \eps_{\tau_L} \\
-c_{\theta_L} c_{\mu_L} \eps_{e_L} & -c_{\theta_L} c_{\mu_L} s_{\mu_L}
& -c_{\theta_L} c_{\mu_L} \eps_{\tau_L}
& c_{\theta_L}^2 c_{\mu_L}^2 + s_{\theta_L}^2  & c_{\theta_L} s_{\theta_L}  s_{\mu_L}^2  \\
s_{\theta_L} c_{\mu_L} \eps_{e_L} & s_{\theta_L} c_{\mu_L} s_{\mu_L} & s_{\theta_L} c_{\mu_L} \eps_{\tau_L}
&  c_{\theta_L} s_{\theta_R} s_{\mu_L}^2 & c_{\theta_L}^2 + c_{\mu_L}^2 s_{\theta_L}^2
\end{pmatrix}, \\
 \!\!\!\!\!\! - \hg^{\Zp}_{e_R} \sim \gp
\begin{pmatrix}
\eps_{e_R}^2  & s_{\mu_R} \eps_{e_R} & \eps_{e_R} \eps_{\tau_R}
 & - s_{\theta_R} c_{\mu_R} \eps_{e_R}  & -c_{\theta_R} c_{\mu_R} \eps_{e_R} \\
s_{\mu_R} \eps_{e_R} & s_{\mu_R}^2 & s_{\mu_R} \eps_{\tau_R}
 & - s_{\theta_R} c_{\mu_R} s_{\mu_R} & - c_{\theta_R}  c_{\mu_R} s_{\mu_R} \\
\eps_{\tau_R} \eps_{e_R}  & s_{\mu_R} \eps_{\tau_R} & \eps_{\tau_R}^2
 & - s_{\theta_R} c_{\mu_R} \eps_{\tau_R}  & -c_{\theta_R} c_{\mu_R} \eps_{\tau_R} \\
-s_{\theta_R} c_{\mu_R} \eps_{e_R} & -s_{\theta_R} c_{\mu_R} s_{\mu_R}
& -s_{\theta_R} c_{\mu_R} \eps_{\tau_R}
&  c_{\theta_R}^2+s_{\theta_R}^2 c_{\mu_R}^2  & - c_{\theta_R} s_{\theta_R}  s_{\mu_R}^2  \\
-c_{\theta_R} c_{\mu_R} \eps_{e_R} & - c_{\theta_R} c_{\mu_R} s_{\mu_R}
&- c_{\theta_R} c_{\mu_R} \eps_{\tau_R}
&  - c_{\theta_R} s_{\theta_R} s_{\mu_R}^2 & s_{\theta_R}^2 + c_{\mu_R}^2 c_{\theta_R}^2
\end{pmatrix}.
\end{align}
The mixing effects from $U_{e_{L,R}}^2$ are neglected.
In the same approximation,
the off-diagonal Yukawa couplings of $\chi$ to VL and SM fermions are given by
\begin{align}
\label{eq-Ychie}
\left[\hY^\chi_e\right]_{aj} =&\ \frac{  c_{\mu_L} M_{E_L} }{v_\Phi}
\begin{pmatrix}
 c_{\theta_R}  \eps_{e_L} & c_{\theta_R} s_{\mu_L} & c_{\theta_R} \eps_{\tau_L} \\
- s_{\theta_R} \eps_{e_L} & -  s_{\theta_R} s_{\mu_L} & -s_{\theta_R}  \eps_{\tau_L} \\
\end{pmatrix},  \\
\left[\hY^\chi_e\right]_{ib} =&\  \frac{c_{\mu_R} M_{E_R}}{v_\Phi}
\begin{pmatrix}
  s_{\theta_L}  \eps_{e_R}     & c_{\theta_L} \eps_{e_R} \\
  s_{\theta_L}  s_{\mu_R}      & c_{\theta_L}  s_{\mu_R} \\
  s_{\theta_L}  \eps_{\tau_R} & c_{\theta_L}  \eps_{\tau_R}
\end{pmatrix}.
\end{align}
Unlike the $\Zp$ boson, $\chi$ does not couple to SM fermions unless $U^2_{e_{L,R}}$ is taken into account.
The Yukawa couplings to the SM leptons are estimated as
\begin{align}
 \left[\hY^\chi_e\right]_{ij} \lesssim &\
\frac{1}{{M}_{E_\mathrm{VL}}}
\begin{pmatrix}
 \eps_{e}^2  m_e & \eps_{e} m_\mu & \eps_e \eps_\tau m_\tau  \\
 \eps_{e} m_\mu  &   m_\mu             &  \eps_\tau m_\tau  \\
 \eps_{e} \eps_\tau m_\tau & \eps_\tau m_\tau & \eps_\tau^2 m_\tau
 \end{pmatrix}, 
\end{align}
where $\eps_e$, $\eps_\tau$ and ${M}_{E_\mathrm{VL}}$ are typical values of $\eps_{e_{L,R}}$, $\eps_{\tau_{L,R}}$ and VL lepton masses, respectively. 
Thus, the $\chi$ couplings to two SM fermions have an extra suppression factor $m_\ell/M_{E_{L,R}}$
compared with the $\Zp$ couplings, while those to one SM fermion and one VL fermion do not have this suppression.
The couplings of $\sigma$ are similar in structure than those of $\chi$.
However the mass of $\sigma$ is not bounded by $v_\Phi$ implying that 
its effects can be decoupled for large $v_\phi$.

Using above results, the leading contribution to $\Delta a_\mu$ from $\Zp$and $\chi$ boson can be 
estimated by Eq.~\eqref{eq:aMu_anal}. 
Here we want to give more details on the combination of loop functions $C_{LR}$ appearing there. 
The relevant combination of loop functions is defined as
\begin{align} 
\label{eq:CLRApp}
 C_{LR} \define&\   \frac{\tM_{E_L}}{\la^\prime_e v_H} \left( c_{\theta_L} s_{\theta_R} G_Z(x_L) 
                + s_{\theta_L}c_{\theta_R} \frac{M_{E_L}M_{E_R}}{2m_\chi^2} G_S(y_L)  \right) \\ \notag 
        &\        - \frac{\tM_{E_R}}{\la^\prime_e v_H} \left( s_{\theta_L} c_{\theta_R} G_Z(x_R) 
                + c_{\theta_L}s_{\theta_R} \frac{M_{E_L}M_{E_R}}{2m_\chi^2} G_S(y_R)  \right), 
\end{align}
where $x_{L,R}:=\tM_{E_{L,R}}^2/m_\Zp^2$,  $y_{L,R}:=\tM_{E_{L,R}}^2/m_\chi^2$. 
This is straightforwardly obtained from the sum of Eqs.~\eqref{eq:daMuZp} and \eqref{eq:daMuS}, while
using our approximations above.
For large enough mass splitting, $\la^\prime_e v_H \ll \abs{\Delta_E}$ we can use Eq.~\eqref{eq-thetaLR} to
simplify this expression to
\begin{align}
\label{eq-CLRdel}
 C_{LR}
=&\ 
\sqrt{x_Lx_R} \frac{G_Z(x_L)-G_Z(x_R)}{x_L-x_R} 
+ \frac{1}{2} \sqrt{y_L y_R} \frac{{y_L} G_{{S}}(y_L)-{y_R} G_{{S}}(y_R)}{y_L-y_R} + \order{\frac{\lambda^\prime_e v_H}{\Delta_E}}.
\end{align}
On the other hand, if the VL mass splitting is small,  $\abs{\Delta_E} \ll M_E$, 
we obtain the following formula by using Eq.~\eqref{eq-theta2}, 
\begin{align}
\label{eq-CLRDel}
 C_{LR} =&\ 
  x \frac{d\,G_Z(x)}{d x} + \frac{y}{2} \frac{d \left(y\,G_{S}(y)\right) }{d y}
   + \order{\frac{\mu_E}{M_E}},  
\end{align}
where $x \define M_E^2/m_\Zp^2$ and $y \define M_E^2/m_\chi^2 $ and $M_E$, $\Delta_E$ and $\mu_E$ are defined in Eq.~\eqref{eq-masscombi}. 
This expression is identical to the form which is obtained by taking a limit 
$M_{E_L} \to M_{E_R}$ or $\Delta_E \to 0$, in $C_{LR}$ of Eq.~\eqref{eq-CLRdel}.  
Hence, $C_{LR}$ in Eq.~\eqref{eq-CLRdel} 
is a good approximation even if $\lambda^\prime_e v_H \sim  \Delta_E$.
Formulae for $\mu\to e\gamma$ and $\tau\to\mu\gamma$, Eq.~\eqref{eq-muegam} and \eqref{eq-tamgam}, 
are obtained in an analogous way.

Relatively light VL leptons with large Higgs Yukawa couplings (necessary to explain $\Delta a_\mu$ by chiral enhancement) 
may contribute significantly to $h \to \gamma \gamma$.
The Higgs couplings to VL leptons are approximately given by
\begin{align}
 \left[\hY^h_e \right]_{ab} \sim \la^\prime_e
\begin{pmatrix}
 c_{\theta_R} s_{\theta_L} & c_{\theta_R} c_{\theta_L} \\
 -s_{\theta_R} s_{\theta_L} & -s_{\theta_R} c_{\theta_L}
\end{pmatrix},    
\end{align}
where $\la_e$ is neglected.
The amplitude of $h\to\gamma\gamma$ from the VL lepton loop is given by 
\begin{align}
 \sum_{X=L,R} \frac{y_{E_X} v_H}{\tilde{M}_{E_X}} A^H_{1/2}(\tau_{E_X})
\sim \frac{(\la^\prime_e v_H)^2}{\tM_{E_L}^2 - \tM_{E_R}^2 }
         \left(A^H_{1/2}(\tau_{E_L}) - A^H_{1/2}(\tau_{E_R})\right),
\end{align}
with $\tau_{E_{L,R}}:=m_H^2 / 4 \tilde{M}^2_{E_{L,R}}$. 
Again, the same result is obtained by using Eq.~\eqref{eq-thetaLR} or Eq.~\eqref{eq-theta2}, 
for small ($\abs{\lambda^\prime_e} v_H \ll \abs{\Delta_E}$) or large ($\abs{\lambda^\prime_e} v_H, \abs{\Delta_E} \ll M_E$) VL mass splitting, respectively.

\subsection{Quarks}
Before starting the analytical discussion of the quark couplings, let us obtain an estimate
for the necessary size of such couplings. Based on above results one finds that
\begin{align}
 \left[\hg^{\Zp}_{e_R}+\hg^{\Zp}_{e_L}\right]_{22} = \gp (s_{\mu_L}^2 + s_{\mu_R}^2  ) \sim \gp 
\end{align}
is required in order to explain $\Delta a_\mu$.
The $\Zp$ contribution to $C_9^\mu$ can then be estimated as
\begin{align}
 \abs{C_9^\mu} \sim 1.0 \times\left(\frac{0.3}{\gp}\right) \left(\frac{1\ \mathrm{TeV}}{v_\Phi} \right)^2
                                         \left( \frac{s_{\mu_L}^2 + s_{\mu_R}^2}{1.0}\right)
                                         \left(\frac{ \left[\hg^\Zp_{d_L}\right]_{23}}{0.001} \right).
\end{align}
Thus, the $\bsll$ anomalies can be explained even with the permille $\Zp$ couplings to quarks.

Let us start the discussion of quark mass diagonalization in a basis where the upper-left 
blocks of the quark mass matrices have already been diagonalized,
\begin{align}
\label{eq-matqd}
 \Mcal^{u} =
\begin{pmatrix}
  y_{i}^u  \delta_{ij} v_H  & 0_i &  \la^U_i v_\Phi \\
  0_j     & \la_u v_H & \la^U_V v_\phi \\
 \la^Q_j v_\Phi& \la^Q_V v_\phi & \la_u^\prime v_H \\
\end{pmatrix},
\quad
 \Mcal^{d} =
\begin{pmatrix}
  y_{i}^d \delta_{ij} v_H  & 0_i &  \la^D_i v_\Phi \\
  0_j     & \la_d v_H & \la^D_V v_\phi \\
 \la^Q_j v_\Phi& \la^Q_V v_\phi & \la_d^\prime v_H \\
\end{pmatrix}.
\end{align}
The quark couplings to the Higgs, $Z$, and $\Zp$ bosons is the same as in the gauge basis,
while the $W$ boson couplings \eqref{eq:Wcouplings} are changed to 
\begin{align}
\label{eq-Wgauge}
 \Lcal_{W} = \frac{g}{\sqrt{2}}W_\mu^+ \ol{u} \gamma^\mu
\left[
\begin{pmatrix}
 V_{ij} & 0_i & 0_i \\
 0_j  & 1 & 0 \\
 0_j  & 0 & 0
\end{pmatrix}
P_L
+
\begin{pmatrix}
  0_{ij} & 0_i & 0_i \\
 0_j  & 0 & 0 \\
 0_j  & 0 & 1
\end{pmatrix}
P_R
\right] d + \mathrm{h.c.}
 \end{align}
Here, $V_{ij} \sim V_\mathrm{CKM}$ is expected because of the small mixings between VL and SM quarks.
In general, the mass matrices can be diagonalized exactly in the same way as for 
the charged leptons in the previous section, but all mixing angles can be taken to be small.
We define the small mixing angles,
\begin{align}
 \eps_{Q_i} \define \frac{v_\Phi \la_i^Q}{\la_V^Q v_\phi},\quad
 \eps_{U_i} \define \frac{v_\Phi \la_i^U}{\la_V^U v_\phi},\quad
 \eps_{D_i} \define \frac{v_\Phi \la_i^D}{\la_V^D v_\phi}.
\end{align}
With this parametrization,
the unitary matrices are
\begin{align}
 U^0_{u_L} =   U^0_{d_L} =&\
\begin{pmatrix}
 1 & 0 & 0 & \eps_{Q_1} & 0 \\
 0 & 1-\eps_{Q_2}^2/2 & -\eps_{Q_2} \eps_{Q_3} & \eps_{Q_2} & 0 \\
 0 & 0 & 1-\eps_{Q_3}^2/2 & \eps_{Q_3} & 0 \\
 -\eps_{Q_1} &  -\eps_{Q_2} &  -\eps_{Q_3} & 1-(\eps_{Q_2}^2+\eps_{Q_3}^2)/2 & 0 \\
0&0&0&0&1
\end{pmatrix}, \\
 U^0_{u_R}  =&\
\begin{pmatrix}
 \delta_{ij} & \eps_{U_i} & 0_i \\
-\eps_{U_j}& 1 & 0 \\
0_j &  0 & 1
\end{pmatrix},\quad
 U^0_{d_R}  =
\begin{pmatrix}
 \delta_{ij} & \eps_{D_i} & 0_i \\
-\eps_{D_j}& 1 & 0 \\
0_j &  0 & 1
\end{pmatrix}.
\end{align}
Here, we keep terms of $\order{\eps_{Q_{2,3}}^2}$ and linear for the other angles.
The Yukawa couplings in the off-diagonal elements of \eqref{eq:m0} are given by
\begin{align}
  \ty^u_{R_i} \sim &\  \eps_{Q_i} y^u_i,\quad \ty^u_{L_i} \sim \eps_{U_i} y^u_i, 
  \quad 
  \ty^d_{R_i} \sim   \eps_{Q_i} y^d_i,\quad \ty^d_{L_i} \sim \eps_{D_i} y^d_i.
\end{align}
The perturbative correction to the SM up quark mass matrix is
\begin{align}
 \Delta m^u_{ij} \sim&\ - \eps_{U_i}\eps_{Q_j} y^u_i y^u_j
                                   \frac{\la^\prime_u v_H^2}{M_{U_L}M_{U_R}}  \\ \notag
                        \sim&\   1.0 \times 10^{-8}\ \mathrm{GeV} \times
                       \left(\frac{\sqrt{\eps_{U_i}\eps_{Q_j}}}{10^{-2}} \right)^2
                         \left(\frac{\la^\prime_u}{1.0}\right)
                         \left(\frac{\sqrt{m_{u_i}m_{u_j}} }{15\  \mathrm{GeV}} \right)^2
                           \left(\frac{1.5\ \mathrm{TeV}}{ \sqrt{M_{U_L}M_{U_R}} } \right)^2,
\end{align}
where the $M_{U_{L(R)}}$ is the left-handed (right-handed) VL quark mass $\sim \la_V^{Q(U)} v_\phi$.
Hence, the basis defined as in Eq.~\eqref{eq-appmass} is to a very good approximation the mass basis.
Perturbative corrections for the down quarks are even smaller due to their lighter masses.

The $Z^\prime$ boson couplings are estimated as
\begin{align}
 \left[\hg^\Zp_{u_L}\right]_{ij} = \left[\hg^\Zp_{d_L}\right]_{ij} \sim \gp \eps_{Q_i} \eps_{Q_j},
\quad
 \left[\hg^\Zp_{u_R}\right]_{ij}  \sim \gp \eps_{U_i} \eps_{U_j},
\quad
 \left[\hg^\Zp_{d_R}\right]_{ij} \sim \gp \eps_{D_i} \eps_{D_j}. 
\end{align}
Thus, the $\bsll$ anomaly requires $\eps_{Q_2} \eps_{Q_3} \sim 10^{-3}$.

We now estimate the couplings to $Z$ and Higgs boson. We focus on
the up quark sector, where mixing effects are larger due to the heavy top quark.
The Higgs Yukawa coupling are estimated as 
\begin{align}
\left[\mhc{U^u_L}Y^h_uU^u_R \right]_{ij} - y^u_{i}\delta_{ij}
\sim 1.3 \times 10^{-4} \times
\left(\frac{\sqrt{\eps_{Q_i}\eps_{Q_j}}}{0.1}\right)^2
\left(\frac{ \sqrt{m_{u_i}m_{u_j}}}{170\ \mathrm{GeV} }\right)^2
\left(\frac{1.5\ \mathrm{TeV} } {M_{U_\mathrm{VL}}  }\right)^2 
\end{align}
and the weak-isospin part of the $Z$ boson couplings are   
\begin{align}
\label{eq-ZLup}
\left[ \mhc{U^u_L} \Pfb U^u_L  \right]_{ij}- \delta_{ij}  \sim&\
1.3\times 10^{-6} \times
\left(\frac{\sqrt{\eps_{U_i}\eps_{U_j}}}{0.01}\right)^2
\left(\frac{ \sqrt{m_{u_i}m_{u_j}}}{170\ \mathrm{GeV} }\right)^2
\left(\frac{1.5\ \mathrm{TeV} } { M_{U_\mathrm{VL}}  }\right)^2,
\\
\left[ \mhc{U^u_R} P_5 U^u_R  \right]_{ij} \sim&\
1.3\times 10^{-4} \times
\left(\frac{\sqrt{\eps_{Q_i}\eps_{Q_j}}}{0.1}\right)^2
\left(\frac{ \sqrt{m_{u_i}m_{u_j}}}{170\ \mathrm{GeV} }\right)^2
\left(\frac{1.5\ \mathrm{TeV} } { M_{U_\mathrm{VL}}  }\right)^2, 
\label{eq-ZRup}
\end{align}
where $M_{U_\mathrm{VL}}$ is a typical VL up quark mass. 
Thus, the EW boson couplings to the SM up quarks do not significantly deviated from their SM values. 
Those for the down quarks are even more suppressed by the smaller down quark masses.

The $W$ boson couplings to the right-handed SM quarks are estimated as 
 \begin{align}
\left[ \mhc{U^u_R} P_5 U^d_R  \right]_{ij} \sim&\
1.3\times 10^{-4} \times
\left(\frac{\sqrt{\eps_{Q_i}\eps_{Q_j}}}{0.1}\right)^2
\left(\frac{ \sqrt{m_{u_i}m_{u_j}}}{170\ \mathrm{GeV} }\right)^2
\left(\frac{1.5\ \mathrm{TeV} } { M_{U_\mathrm{VL}} }\right)^2.  
\label{eq-WR}
\end{align}
An estimation of the non-unitarity of the extended CKM matrix has already been given in Eq.~\eqref{eq-exCKMunit}.  
In addition, the off-diagonal elements involving the VL quarks are 
\begin{align}
\label{eq-V5j}
 \left[\hat{V}_{\mathrm{CKM}} \right]_{4j} \sim&\  0,\quad 
\left[\hat{V}_{\mathrm{CKM}} \right]_{5j} \sim \sum_{k=1,2,3} \frac{\ty^{u*}_{L_k} v_H}{M_{U_R}} V_{kj}
                     \sim \eps_{U_3} \frac{ m_t}{M_{U_R}} V_{3j} , \\ 
\left[\hat{V}_{\mathrm{CKM}} \right]_{i4}  \sim&\ 0,\quad 
  \left[\hat{V}_{\mathrm{CKM}} \right]_{i5} \sim  \sum_{k=1,2,3}  \frac{\ty^{d}_{L_k} v_H}{M_{D_R}} V_{ik}    
                     \sim  \eps_{D_3} \frac{m_b}{M_{D_R}} V_{i3},  
\end{align}
where we have neglected terms in $U^1_{u_{L,R}}$, $U^1_{d_{L,R}}$ of $\order{s_{u_{L,R}}, s_{d_{L,R}}}$. 
These effects are much smaller for quarks as compared to charged leptons owing to the heavier VL quark masses.

\subsection{Neutrinos}
We consider the type-I seesaw mechanism with Majorana masses $M_\mathrm{Maj}\sim 10^{14}$ GeV.
In this model, 
there are three left-handed Majorana neutrinos, $m_{\nu_L} \sim v_H^2/M_\mathrm{Maj}$,
three right-handed Majorana neutrinos, $m_{\nu_R} \sim M_\mathrm{Maj}$
and two vector-like Dirac neutrinos, $m_{N} \sim v_\Phi$.
Here we want to show that the mixings among these three types of neutrinos 
are suppressed by the large Majorana mass terms. 

The $10\times 10$ mass matrix in Eq.~\eqref{eq-Mn10} is diagonalized as follows. 
At first, we rotate only the left-handed neutrinos as
\begin{align}
\begin{pmatrix}
\nu_{L_i}  \\ N_L^\prime \\ N_L 
\end{pmatrix}
=: V_L^n 
\begin{pmatrix}
\tilde{\nu}_{L_i}  \\ \tilde{N}_L^\prime \\ \tilde{N}_L 
\end{pmatrix},
\end{align}
with a unitary matrix $V^n_L$ which is given by 
\begin{align}
 V^n_L =  U^0_{n_L} U^1_{n_L},  
\quad 
 U^0_{n_L} =
\begin{pmatrix}
 \mathbf{1}_{ij} & 0_i &0_i  \\
 0_j  &  c_{N} & s_{N} e^{i \phi_N} \\
 0_j  & - s_{N} e^{-i \phi_N}  & c_{N}   \\
\end{pmatrix}, 
\quad
U^1_{n_L} =
\begin{pmatrix}
 \vz_{N_j} &  \vn_{N}  & \mathbf{0}  \\
 0_j    & 0       & 1
\end{pmatrix}.  
\end{align}
Here,  
\begin{align}
c_{N} \define  \frac{ \abs{\la^N_V} v_\phi}{\tM_N}, \quad
s_{N} \define   \frac{ \abs{\la_n} v_H }{\tM_N},\quad \text{with}\quad
\tM_N \define \sqrt{\abs{\la^{N}_V v_\phi }^2 + \abs{\la_n v_H} ^2},
\end{align}
and $\phi_N \define \mathrm{Arg} \left(\la_V^N\right) - \mathrm{Arg}\left(\la_n \right)$. 
The four-component vectors $\vz_{N_i}, \vn_N$ satisfy
\begin{align}
\label{eq:nN}
 \vz_{N_i}^\dag \vz_{N_j} = \delta_{ij},\quad \vz_{N_i}^\dag \vn_{N} = 0,\quad
\vn_N = \tM_L^{-1}
\begin{pmatrix}
\la^L_j v_\Phi, &  c_{N} \la_V^L v_\phi - s_{N} e^{-i\phi_N} \la'_n v_H
\end{pmatrix}^\mathrm{T},
\end{align}
where $\tM_L$ is defined as the normalization factor of $\vn_N^\dag \vn_N = 1$.
The Dirac matrix after this rotation is given by 
\begin{align}
\label{eq-MNt}
\tilde{\Mcal}^n \define  \Mcal^n V^n_{L} =:
\begin{pmatrix}
 \tilde{m}^n_{ij} &  \mu^n_{i} & \mu^\prime_{i}  \\
 0_j    & 0                                 & \tM_{N} \\
 0_j   &  \tM_{L} & \tilde{m}_{5}
\end{pmatrix}
=: 
\begin{pmatrix}
 \tilde{m}^n_{ij} &  \mu_{ib}   \\
 0_{a j}     & M^D_{ab}
\end{pmatrix},
\quad 
 M^D \define
\begin{pmatrix}
 0 & \tM_N \\
 \tM_L & \tilde{m}_5
\end{pmatrix}, 
\end{align}
where $a,b = 4,5$. 
Owing to the first rotation $U^0_{n_L}$, 
the first three columns in the 4th and 5th rows in $\tilde{\mathcal{M}}^n$ are all zero 
and there is no mixing between the singlet-like VL neutrinos $(\ol{N}_R,\ \ol{N}_R^\prime)$ 
and left-handed light neutrinos $\tilde{\nu}_L$.  

Back in the full $10\times10$ matrix (recall Eqs.~\eqref{eq-Mn10} and~\eqref{eq:Mneutino10})), 
we note that only the first three rows couple to $\nu_{R_i}$ 
and they will have $\order{M_\mathrm{Maj}}$ masses. 
Therefore, $\tilde{\nu}_{L_i}$ are approximately the light Majorana neutrinos, while $\tilde{N}^{\prime}_L, \tilde{N}_L$ together with $N^{\prime}_R, N_R$
form approximately Dirac neutrinos.    
After integrating out the $\order{M_\mathrm{Maj}}$ right-handed neutrinos,
the effective $7\times 7$ mass matrix is given by 
\begin{align}
\begin{pmatrix}
 \ol{N}_R & \ol{N}'_R & \ol{\tilde{\nu}}^c_L & \ol{\tilde{N}}^{'c}_L & \ol{\tilde{N}}^c_L
\end{pmatrix}
\begin{pmatrix}
 0_{2\times 2}& 0_{2\times3} & M^D \\
 0_{3\times 2}  &  m_\nu                  & \eps_{3\times2}  \\
 \left(M^D\right)^\mathrm{T}            &  \eps_{2\times 3} & \eps_{2\times2}  \\
\end{pmatrix}
\begin{pmatrix}
N^c_R \\  N^{'c}_R \\ \tilde{\nu}_L \\ \tilde{N}'_L \\ \tilde{N}_L
\end{pmatrix},
\end{align}
where $\eps_{2\times 3}$, $\eps_{2\times 2}$, $\eps_{3\times 2}$ are mass parameters 
suppressed by the Majorana mass term such as $(\mu^{n})^T M_\mathrm{Maj}^{-1} \mu^n$.  
These terms are at most of $\mathcal{O}(v_\Phi^2/ M_\text{Maj})$ and, therefore, negligible
compared to $M^D$. 
The $3\times 3$ effective Majorana mass matrix $m_\nu$ for the active neutrinos $\tilde{\nu}_L$ is given by
\begin{align}
m_\nu = - \left(\tilde{m}^n\right)^\mathrm{T} M_\mathrm{Maj}^{-1} \tilde{m}^n. 
\end{align}
Altogether, the mass terms for the neutrinos are decomposed as
\begin{align}
 \lag_\text{neutrino} =&\ \lag_{R}+\lag_{\nu} +\lag_{D} \notag \\
 =&\ \frac{1}{2} \ol{\nu}_{R} M_\mathrm{Maj} \nu^c_{R} + \frac{1}{2}  \ol{\tilde{\nu}}^c_{L} m_{\nu }  \tilde{\nu}_{L}
+\left[
\begin{pmatrix}
 \ol{N}_R & \ol{N}'_R
\end{pmatrix}
M^D
\begin{pmatrix}
 \tilde{N}'_L \\ \tilde{N}_L
\end{pmatrix}
+ \mathrm{h.c.} \right],
\end{align}
where the family indices are omitted.
The mixing terms among the three types of neutrinos, 
namely $\tilde{\nu}_L, \nu_R$ and the Dirac neutrinos, are all suppressed by $M_\text{Maj}$ and not stated.

The remaining mass matrices then are diagonalized as 
\begin{align}
\left( u^N_R\right)^{\dag} M_D u^N_L = \text{diag} \left( M_{N_1}, M_{N_2}\right), 
\end{align}
and 
\begin{align}
 \left(u^n_R\right)^\mathrm{T} M_\mathrm{Maj} u^n_R = \text{diag} {(M_1, M_2, M_3)},\quad
\left(u^n_L\right)^\mathrm{T} m_\nu u^n_L = \text{diag}{(m_{\nu_1}, m_{\nu_2}, m_{\nu_3})}.
\end{align}
The mass basis for the active neutrinos $\hat{\nu}_L$,
heavy right-handed neutrinos $\hat{\nu}_R$
and Dirac neutrinos $\hat{N}^a \define \left(\hat{N}^a_L, \hat{N}^a_R \right)$ 
are given by
\begin{align}
&\left[ \hat{\nu}_L \right]_i = \left[ \left(u^n_L\right)^\dag \right]_{ij}
                                             \left[ \left(V^n_L\right)^\dag \right]_{jA} \left[ \nu_L \right]_A, \quad
& \left[ \hat{N}_L\right]_a   =& \left[\left(u^N_{L}\right)^\dagger \right]_{ab}
                                                \left[ \left(V^n_L\right)^\dag \right]_{bA}
                                          \left[\nu_L\right]_{A}, \\
& \left[ \hat{\nu}_R \right]_i = \left[ \left(u^n_R\right)^\dag \right]_{ij} \left[ \nu_R \right]_j,
& \left[ \hat{N}_R\right]_a   =& \left[ \left(u^N_{R} \right)^\dagger \right]_{ab}  \left[\nu_R \right]_{b},
\end{align}
where $i,j = 1,2,3$\,, $a,b = 4,5$\,, and $A=1,2,3,4,5$.
Altogether, the relation between the mass and gauge basis is 
\begin{align}
n_L = U^n_L \hat{n}_L \define 
V^n_L
\begin{pmatrix}
 u^n_L             & 0_{3\times 2} \\
 0_{2\times 3} & u^N_{L}
\end{pmatrix}
\begin{pmatrix}
 \hat{\nu}_L \\ \hat{N}_L 
\end{pmatrix}
,\;\;
n_R = U^n_R \hat{n}_R \define 
\begin{pmatrix}
 u^n_R             & 0_{3\times 2} \\
 0_{2\times 3} & u^N_{R}
\end{pmatrix}
\begin{pmatrix}
\hat{\nu}_R \\ \hat{N}_R 
\end{pmatrix}.
 \end{align}
The mixing between the left- and right-handed neutrinos 
is suppressed by the heavy Majorana masses, and therefore negligible.

The $W$ boson coupling to the SM leptons is given by 
\begin{align}
\Lcal_{W} = \frac{g}{\sqrt{2}} W_\mu^- \ol{\hat{e}}_i \gamma^\mu P_L
                              \left[ U_{e_L}^\dagger \Pfb V^n_L  \right]_{ik} \left[u^n_L\right]_{kj} 
                              \left[\hat{\nu}_L\right] _{j}+\mathrm{h.c.}, 
\end{align}
where $k=1,2,3$. Therefore, we define flavor neutrino states $\nu_f$ and the $3\times3$ PMNS matrix as 
\begin{align}
\label{eq-nuf}
 \nu_f \define  V_\mathrm{PMNS} \ \hat{\nu}_L,\quad 
 \left[V_\mathrm{PMNS}\right]_{ij} 
\define \sum_{k=1,2,3}   \left[ U_{e_L}^\dagger \Pfb V^n_L  \right]_{ik} \left[u^n_L\right]_{kj}.  
\end{align}  
The PMNS matrix is not unitary here. 
The non-unitarity comes from 
a misalignment between $\vn_{N}$ and $\vn_{L}^e $ (defined in Eqs.~\eqref{eq:nN} and \eqref{eq-fvcond}, respectively), 
and the small mixing $U_{e_L}^2$ defined in Eq.~\eqref{eq-U2fL}.  
The mixing with the $\order{\mathrm{TeV}}$ Dirac neutrino 
can be substantial if $\la_n \sim \order{1}$. 
This would be an interesting possibility but is beyond the scope of the present paper.  
We restrict ourselves to the case where $\la_n$ is as small as $\la_e$ 
and the mixing between active and new Dirac neutrinos is negligible. 
In this case, the $\Zp$ boson couplings to the flavor neutrinos defined in Eq.~\eqref{eq-nuf} is given by 
\begin{align}
 \Lcal_{\Zp\nu_f} =&\
                 \gp Z^\prime_\rho \ol{\hat{\nu}}_{i} \left[ U_{n_L}^\dag Q^\prime_{n_L} U_{n_L}  \right]_{ij} 
                                                  \gamma^\rho  P_L  \hat{\nu}_{j} 
               =  \gp Z^\prime_\rho \ol{\nu}_{f_i} \left[ U_{e_L}^\dag Q^\prime_{n_L} U_{e_L}  \right]_{ij} 
                                                  \gamma^\rho  P_L  \nu_{f_j}.
\end{align}
Hence, the coupling of $\Zp$ to muon neutrinos can be estimated as
\begin{equation}
 \Lcal_{\Zp\nu_\mu}\sim\ - s_{\mu_L}^2 \gp Z^\prime_\rho \ol{\nu}_{\mu}  \gamma^\rho P_L  \nu_{\mu} + \order{\eps_{e_L}, \eps_{\tau_L}},
\end{equation}   
where we have omitted couplings involving the Dirac neutrinos. 
The electron and tau neutrinos have tiny couplings to $\Zp$ due to the tiny mixing angles $\eps_{e_L}$, $\eps_{\tau_L}$.

We can find parameters consistent with the experimental results on 
neutrino mass differences and PMNS mixing~\cite{Tanabashi:2018oca}. 
This can be done by fitting the Yukawa couplings involving the three generations of active neutrinos and 
the Majorana masses. The explicit values of the neutrino Yukawa couplings at 
the best fit points are shown in the subsequent section of the appendix. 
For simplicity, we always assume $\la^N_i = 0_i$, $M_\mathrm{Maj}^{ij} = M_0 \delta_{ij}$ 
and take the lightest neutrino to be massless. However, nothing in our analysis 
really depends on these assumptions. All of our fit points realize the 
experimental values of the mass differences, 
\begin{align}
 \Delta m_{21}^2 = 7.37 \times 10^{-5}\,\mathrm{eV}^{2},\quad 
 \Delta m_{31}^2 = 2.56 \times 10^{-3}\,\mathrm{eV}^{2},
\end{align}
mixing angles 
\begin{align}
\sin^2\theta_{12}=0.297,\quad 
\sin^2 \theta_{23} = 0.425,\quad 
\sin^2\theta_{13} = 0.0215,  
\end{align}
and the rephasing invariant 
\begin{align}
 J_\mathrm{CP} \define \mathrm{Im} 
\left( V_\mathrm{PMNS}^{23} V_\mathrm{PMNS}^{13*} V_\mathrm{PMNS}^{12} V_\mathrm{PMNS}^{22*} \right)
= -0.03,
\end{align}
within numerical errors. 
The Majorana mass is set to $M_0 = 10^{14}$ GeV.

\clearpage 
\section{Input Parameters at Best Fit Points}
\label{sec-inputs}

\subsection{Best Fit Point A}

Values of the inputs parameters for the boson sector are given by 
\begin{align}
m_{\Zp}=277.608,\quad v_\phi=4079.3,\quad \gp=0.250042,\quad 
\lambda_\chi=0.689454 
,\quad 
\lambda_\sigma=0.210518 
\;.
\notag 
\end{align}
Values for fermion mass matrices are 
\small 
\begin{align}
& M_e = \notag \\ \notag &
\begin{pmatrix}
0.000486575 & 0.000000322078 & -0.0000009971 & 0 & 0.000201232  \\
0.0000000453521 & 0.159775 & 0.00162206 & 0 & -153.074  \\
-0.0000614248 & -0.00512644 & -1.74616 & 0 & -0.0409467  \\
0 & 0 & 0 & 0.0000361209 & 448.074  \\
-0.00000237863 & -312.626 & 0.0547758 & 289.432 & -174.104  \\
\end{pmatrix}, 
\end{align}
\begin{align}
& M_n = \notag \\ \notag &
\begin{pmatrix}
0. & 0. & 0. & 0 & 0  \\
-15.7947 & 28.3788\cdot e^{-0.0735218i} & 15.4093\cdot e^{0.107535i} & 0 & 0  \\
10.4292\cdot e^{1.19397i} & 67.3777\cdot e^{0.0000000228655i} & -53.4556 & 0 & 0  \\
0 & 0 & 0 & 1.54426 & -454.964  \\
-0.00000237863 & -312.626 & 0.0547758 & 289.357 & -21.7762  \\
\end{pmatrix},
\end{align}
\begin{align}
& M_u = \notag \\ \notag &
\begin{pmatrix}
0.000893504 & 0.00562655 & 0.688382\cdot e^{1.52313i} & 0 & -0.0228009  \\
-0.000172924 & 0.631189 & -0.119538 & 0 & -40.8575  \\
0.535431\cdot e^{1.72288i} & 4.4472 & -171.657 & 0 & -8.7671  \\
0 & 0 & 0 & -0.000301965 & -3596.52  \\
0.0051151 & 214.302 & -96.8583 & 3445.76 & 5.50646  \\
\end{pmatrix},
\end{align}
\begin{align}
& M_d = \notag \\ \notag &
\begin{pmatrix}
0.0121338 & 0.0527148 & 0.0199472\cdot e^{-3.08081i} & 0 & -0.0254167  \\
-0.00528222 & -0.0409769 & -2.58755 & 0 & -41.2307  \\
0.00189847\cdot e^{-1.62491i} & -0.0198056 & -1.21195 & 0 & 6.1041  \\
0 & 0 & 0 & -0.122713 & -1571.54  \\
0.0051151 & 214.302 & -96.8583 & 3445.76 & -1.99223  \\
\end{pmatrix}.
\end{align}
\normalsize 

\subsection{Best Fit point B}

Values of the inputs parameters for the boson sector are given by 
\begin{align}\notag
m_{\Zp}=535.334,\quad v_\phi=3121.37,\quad \gp=0.340407,\quad 
\lambda_\chi=0.0062973 
,\quad \lambda_\sigma= 0.00109477
\;.
\end{align}
The fermion mass matrices are 
\small 
\begin{align}
& M_e = \notag \\ \notag &
\begin{pmatrix}
0.000486573 & 0.00000134295 & -0.0000688018 & 0 & 0.0000061799  \\
-0.000000884741 & 0.174184 & 0.00486889 & 0 & 684.361  \\
-0.00000034416 & 0.00029932 & 1.74617 & 0 & -0.0829304  \\
0 & 0 & 0 & 0.0000177937 & 1115.66  \\
0.000428239 & 926.678 & 0.0199536 & 887.77 & -174.061  \\
\end{pmatrix}, 
\end{align} 
\begin{align}
& M_n = \notag \\ \notag &
\begin{pmatrix}
0. & 0. & 0. & 0 & 0  \\
-15.7951 & 27.8222\cdot e^{3.06781i} & 15.4457\cdot e^{-3.03449i} & 0 & 0  \\
10.43\cdot e^{1.19378i} & -66.2876 & 53.3662\cdot e^{0.0000000376777i} & 0 & 0  \\
0 & 0 & 0 & -1.21561 & -1069.88  \\
0.000428239 & 926.678 & 0.0199536 & 887.769 & 0.540635  \\
\end{pmatrix},
\end{align}
\begin{align}
& M_u = \notag \\ \notag &
\begin{pmatrix}
-0.00312202 & -0.0252024 & 0.0072623\cdot e^{-1.88131i} & 0 & -0.649726  \\
0.17765 & -5.43688 & 167.589 & 0 & -1.65829  \\
0.0045274\cdot e^{-1.07337i} & 2.00559 & -41.9368 & 0 & -8.95522  \\
0 & 0 & 0 & -0.00177389 & -1852.98  \\
0.0376789 & 93.9529 & -311.597 & 2728.52 & -0.157394  \\
\end{pmatrix},
\end{align}
\begin{align}
& M_d = \notag \\ \notag &
\begin{pmatrix}
-0.00346891 & -0.000644624 & 2.63889\cdot e^{-1.76496i} & 0 & 0.00777153  \\
0.00266115 & 0.00621731 & -0.0159 & 0 & 34.7376  \\
0.0116163\cdot e^{1.39463i} & 0.0575321 & 1.13518 & 0 & -5.3196  \\
0 & 0 & 0 & -0.00422283 & 2130.63  \\
0.0376789 & 93.9529 & -311.597 & 2728.52 & 0.0672149  \\
\end{pmatrix}.
\end{align}
\normalsize 

\subsection{Best Fit point C}
Values of the inputs parameters for the boson sector are given by 
\begin{align}\notag
 m_{\Zp}=486.709,\quad v_\phi=4980.22,\quad \gp=0.322661,\quad 
\lambda_\chi=1., 
\quad 
\lambda_\sigma=0.674008 
\;.
\end{align}
Values for fermion mass matrices are 
\small 
\begin{align}
& M_e = \notag \\ \notag &
\begin{pmatrix}
-0.000486574 & -0.000000218258 & 0.0000503735 & 0 & 0.00061118  \\
0.00000012205 & 0.367187 & -0.000101055 & 0 & -305.194  \\
-0.00000174729 & -0.000547026 & 1.74617 & 0 & 0.0455051  \\
0 & 0 & 0 & 0.00551033 & 406.321  \\
-0.000421884 & -500.988 & 0.00636737 & -193.583 & -174.104  \\
\end{pmatrix},
\end{align}
\begin{align}
& M_n = \notag \\ \notag &
\begin{pmatrix}
0. & 0. & 0. & 0 & 0  \\
15.7952\cdot e^{-0.0000000265697i} & 53.3929\cdot e^{3.06779i} & 15.4488\cdot e^{0.107064i} & 0 & 0  \\
10.4299\cdot e^{-1.94779i} & -127.249& -53.3586 & 0 & 0  \\
0 & 0 & 0 & 0.362325 & -4980.22  \\
-0.000421884 & -500.988 & 0.00636737 & -193.583 & -0.0644098  \\
\end{pmatrix},
\end{align}
\begin{align}
& M_u = \notag \\ \notag &
\begin{pmatrix}
0.00527096 & 0.251612 & 0.000517685\cdot e^{-1.90963i} & 0 & -0.00571705  \\
0.0154127 & 0.567102 & -0.200693 & 0 & -90.1228  \\
0.151639\cdot e^{1.15375i} & -7.16893 & -172.466 & 0 & -0.116493  \\
0 & 0 & 0 & -0.000772971 & -4980.22  \\
0.019648 & 100.573 & 338.918 & 3708.84 & -0.271052  \\
\end{pmatrix},
\end{align}
\begin{align}
& M_d = \notag \\ \notag &
\begin{pmatrix}
0.00985885 & -0.0404963 & 0.0721585\cdot e^{1.99027i} & 0 & -1.03475  \\
0.00481504 & -0.0343246 & 0.0324285 & 0 & -0.48716  \\
0.00766034\cdot e^{1.63225i} & 0.000571412 & 2.86587 & 0 & -13.2473  \\
0 & 0 & 0 & -0.000462012 & 4803.69  \\
0.019648 & 100.573 & 338.918 & 3708.84 & 0.873924  \\
\end{pmatrix}.
\end{align}
\normalsize 

\subsection{Best Fit point D}
Values of the inputs parameters for the boson sector are given by 
\begin{align}\notag
 m_{\Zp}=758.743,\quad v_\phi=4820.94,\quad \gp=0.348599,\quad 
\lambda_\chi= 0.00892191
,\quad \lambda_\sigma= 0.999999  
\;.
\end{align}
Values for fermion mass matrices are 
\small 
\begin{align}
& M_e = \notag \\ \notag &
\begin{pmatrix}
0.000486575 & -0.00000000602545 & 0.00000019239 & 0 & -0.0000102812  \\
0.00000112006 & 0.146358 & -0.0000795331 & 0 & -407.638  \\
-0.00000448652 & -0.0000111978 & -1.74616 & 0 & 0.0247399  \\
0 & 0 & 0 & -0.0735254 & 635.847  \\
-0.00010118 & -488.24 & 0.0995171 & -337.63 & -173.973  \\
\end{pmatrix},
\end{align}
\begin{align}
& M_n = \notag \\ \notag &
\begin{pmatrix}
0. & 0. & 0. & 0 & 0  \\
-15.7954 & 33.8318\cdot e^{3.06777i} & 15.455\cdot e^{0.106991i} & 0 & 0  \\
10.4297\cdot e^{1.19386i} & -80.6485 & -53.3437 & 0 & 0  \\
0 & 0 & 0 & -0.358095 & -4820.94  \\
-0.00010118 & -488.24 & 0.0995171 & -337.63 & -0.575107  \\
\end{pmatrix},
\end{align}
\begin{align}
& M_u = \notag \\ \notag &
\begin{pmatrix}
-0.00226803 & -0.00526038 & 0.00673866\cdot e^{2.97063i} & 0 & -0.00737306  \\
0.119732 & 0.618476 & -0.00585875 & 0 & -315.604  \\
0.00911865\cdot e^{0.802338i} & 0.0270238 & -172.537 & 0 & 1.13048  \\
0 & 0 & 0 & 0.295538 & 4498.79  \\
-0.15859 & 223.445 & 430.866 & 4039.17 & 110.941  \\
\end{pmatrix},
\end{align}
\begin{align}
& M_d = \notag \\ \notag &
\begin{pmatrix}
-0.000626484 & -0.0522184 & 0.124081\cdot e^{0.546067i} & 0 & 0.00126174  \\
-0.00330599 & 0.0273122 & 0.00102239 & 0 & -0.107688  \\
0.0214224\cdot e^{2.66418i} & -0.115298 & 2.86442 & 0 & 2.97535  \\
0 & 0 & 0 & 0.577941 & -2442.58  \\
-0.15859 & 223.445 & 430.866 & 4039.17 & -1.82777  \\
\end{pmatrix}.
\end{align}
\normalsize 
\clearpage

\section{Full List of Observables at Best Fit Points}
\label{sec-fullobs}

The CKM matrices at the best fit points are as follows:

\scriptsize
\begin{align} 
& \hat{V}_\mathrm{CKM}^A= \\ \notag 
&\begin{pmatrix} 
0.974468&0.224499&0.003595\cdot e^{-1.227622i}&0.000000&0.000000 \\  
0.224355\cdot e^{-3.140947i}&0.973615&0.041688&0.000003\cdot e^{1.366139i}&0.000000 \\  
0.008820\cdot e^{-0.382954i}&0.040903\cdot e^{-3.122994i}&0.999124&0.000042\cdot e^{-1.775654i}&0.000074\cdot e^{1.418692i} \\  
0.000001&0.000003\cdot e^{-1.560707i}&0.000076\cdot e^{1.562550i}&0.000707\cdot e^{2.928496i}&0.999823\cdot e^{-0.160351i} \\  
0.000001\cdot e^{-0.344253i}&0.000007\cdot e^{3.077755i}&0.000116\cdot e^{-0.076994i}&0.000013\cdot e^{-1.852631i}&0.018809\cdot e^{1.341708i} \\  
\end{pmatrix} 
\end{align} 
\begin{align} 
& \hat{V}_\mathrm{CKM}^B= \\ \notag 
&\begin{pmatrix} 
0.974466&0.224508&0.003584\cdot e^{-1.226648i}&0.000000&0.000000 \\  
0.224366\cdot e^{-3.140954i}&0.973626&0.041369&0.000000&0.000000 \\  
0.008750\cdot e^{-0.384851i}&0.040591\cdot e^{-3.122912i}&0.999137&0.000002\cdot e^{2.471251i}&0.000452\cdot e^{3.135974i} \\  
0.000001&0.000003\cdot e^{0.131789i}&0.000028\cdot e^{-3.020122i}&0.000000&0.000107\cdot e^{0.115817i} \\  
0.000004\cdot e^{-0.379008i}&0.000018\cdot e^{-3.117107i}&0.000451\cdot e^{0.005807i}&0.000059\cdot e^{-0.661943i}&1.000000\cdot e^{0.000188i} \\  
\end{pmatrix} 
\end{align} 
\begin{align} 
& \hat{V}_\mathrm{CKM}^C= \\ \notag 
&\begin{pmatrix} 
0.974464&0.224515&0.003607\cdot e^{-1.228436i}&0.000000&0.000000 \\  
0.224372\cdot e^{-3.140948i}&0.973621&0.041465&0.000000&0.000000 \\  
0.008778\cdot e^{-0.386411i}&0.040686\cdot e^{-3.122824i}&0.999133&0.000197\cdot e^{-1.153121i}&0.000002\cdot e^{-2.377782i} \\  
0.000002\cdot e^{-2.344296i}&0.000008\cdot e^{1.202460i}&0.000197\cdot e^{-1.957901i}&1.000000\cdot e^{0.030571i}&0.000353\cdot e^{-1.192357i} \\  
0.000000&0.000002\cdot e^{1.151528i}&0.000002\cdot e^{1.142344i}&0.000092\cdot e^{-0.011322i}&0.000000 \\  
\end{pmatrix} 
\end{align} 
\begin{align} 
& \hat{V}_\mathrm{CKM}^D= \\ \notag 
&\begin{pmatrix} 
0.974469&0.224495&0.003583\cdot e^{-1.226283i}&0.000002\cdot e^{2.166735i}&0.000000 \\  
0.224353\cdot e^{-3.140955i}&0.973629&0.041373&0.000013\cdot e^{2.159161i}&0.000000 \\  
0.008749\cdot e^{-0.384634i}&0.040596\cdot e^{-3.122928i}&0.999137&0.000026\cdot e^{-0.956519i}&0.000190\cdot e^{2.280798i} \\  
0.000002\cdot e^{-2.598386i}&0.000010\cdot e^{0.880650i}&0.000191\cdot e^{-2.281000i}&0.000829\cdot e^{3.045676i}&0.993172\cdot e^{3.141399i} \\  
0.000002\cdot e^{-2.249810i}&0.000013\cdot e^{0.886543i}&0.000002\cdot e^{0.865773i}&0.000097\cdot e^{-0.095778i}&0.116659\cdot e^{-0.000055i} \\  
\end{pmatrix} 
\end{align}

\begin{table}[th] 
\centering 
\footnotesize 
\caption{Observables for charged leptons at the benchmark points.}
\begin{tabular}{c|cccc|cc}\hline 
Name &  Point (A) & Point (B) & Point (C) & Point (D) &Data & Unc. \\  \hline\hline 
$m_e(m_Z)$ [GeV] $\times 10^{4}$&4.8658&4.8658&4.8658&4.8658&4.8658&0.00049  \\ 
$m_\mu(m_Z)$ [GeV]&0.102719&0.102719&0.102719&0.102719&0.102719&0.000010  \\ 
$m_\tau(m_Z)$ [GeV]&1.7462&1.7462&1.7462&1.7462&1.7462&0.00017  \\ 
\hline 
$\text{Br}\left(\mu\to e\nu\overline{\nu}\right)$&0.999965&0.999970&0.999971&0.999971&0.999971&0.000100  \\ 
$\text{Br}\left(\mu  \to e  \gamma \right)$ $\times 10^{13}$&0.147&1.597&6.1$\times10^{-2}$&0.822&0&2.6  \\ 
$\text{Br}\left(\mu^- \to e^-e^+e^-           \right)$ $\times 10^{13}$&0.000&0.000&0.000&0.000&0&6.1  \\ 
\hline 
$\text{Br}\left(\tau\to e \nu\overline{\nu}\right)$&0.178510&0.178510&0.178510&0.178510&0.178510&0.000018  \\ 
$\text{Br}\left(\tau\to\mu\nu\overline{\nu}\right)$&0.173611&0.173612&0.173612&0.173612&0.173612&0.000017  \\ 
$\text{Br}\left(\tau\to e  \gamma \right)$ $\times 10^{8}$&0.000&0.000&0.000&0.000&0&2.0  \\ 
$\text{Br}\left(\tau\to \mu\gamma \right)$ $\times 10^{8}$&3.3$\times10^{-4}$&3.6$\times10^{-4}$&3.3$\times10^{-6}$&8.5$\times10^{-7}$&0&2.7  \\ 
$\text{Br}\left(\tau^-\to e^-e^+  e^-        \right)$ $\times 10^{8}$&0.000&0.000&0.000&0.000&0&1.6  \\ 
$\text{Br}\left(\tau^-\to   e^-\mu^+\mu^-    \right)$ $\times 10^{8}$&0.000&0.000&0.000&0.000&0&1.6  \\ 
$\text{Br}\left(\tau^-\to \mu^-e^+\mu^-      \right)$ $\times 10^{8}$&0.000&0.000&0.000&0.000&0&1.0  \\ 
$\text{Br}\left(\tau^-\to \mu^-  e^+  e^-    \right)$ $\times 10^{8}$&0.000&0.000&0.000&0.000&0&1.1  \\ 
$\text{Br}\left(\tau^-\to \mu^-e^+\mu^-      \right)$ $\times 10^{8}$&0.000&0.000&0.000&0.000&0&1.0  \\ 
$\text{Br}\left(\tau^-\to \mu^-\mu^+\mu^-    \right)$ $\times 10^{8}$&7.0$\times10^{-3}$&4.8$\times10^{-4}$&6.6$\times10^{-5}$&4.4$\times10^{-6}$&0&1.3  \\ 
\hline 
$\Delta a_e$ $\times 10^{13}$&0.000&-8.8$\times10^{-9}$&0.000&0.000&-8.700&3.6  \\ 
$\Delta a_\mu$ $\times 10^{9}$&2.62&2.52&2.52&2.45&2.68&0.76  \\ 
\hline
\end{tabular}  
\end{table}    

\begin{table}[th] 
\centering 
\footnotesize 
\caption{Observables for SM bosons at the benchmark points.}
\begin{tabular}{c|cccc|cc}\hline 
Name &  Point (A) & Point (B) & Point (C) & Point (D) &Data & Unc. \\  \hline\hline 
$\text{Br}\left(W^+  \to     e^+ \nu \right)$&0.10862&0.10862&0.10862&0.10862&0.10862&0.00011  \\ 
$\text{Br}\left(W^+  \to   \mu^+ \nu \right)$&0.10862&0.10862&0.10862&0.10862&0.10862&0.00011  \\ 
$\text{Br}\left(W^+  \to \tau^+ \nu \right)$&0.10855&0.10855&0.10855&0.10855&0.10855&0.00011  \\ 
$\text{Br}\left(W   \to \text{had}   \right)$&0.652&0.652&0.652&0.652&0.666&0.025  \\ 
$\text{Br}\left(W^+ \to c\overline{s} \right)$&0.309&0.309&0.309&0.309&0.324&0.032  \\ 
\hline 
$\text{Br}\left(Z    \to    e^+    e^- \right)$ $\times 10^{2}$&3.333&3.333&3.333&3.333&3.333&0.0062  \\ 
$\text{Br}\left(Z    \to  \mu^+  \mu^- \right)$ $\times 10^{2}$&3.333&3.333&3.333&3.333&3.333&0.0062  \\ 
$\text{Br}\left(Z    \to\tau^+\tau^- \right)$ $\times 10^{2}$&3.326&3.326&3.326&3.326&3.326&0.0062  \\ 
$\text{Br}\left(Z   \to \text{had}   \right)$&0.676&0.676&0.676&0.676&0.677&0.025  \\ 
$\text{Br}\left(Z   \to u\overline{u}+c\overline{c} \right)/2$&0.1157&0.1157&0.1157&0.1157&0.1157&0.0043  \\ 
$\text{Br}\left(Z  \to d\overline{d}+s\overline{s}+b\overline{b}\right)/3$&0.1483&0.1483&0.1483&0.1483&0.1483&0.0056  \\ 
$\text{Br}\left(Z   \to c\overline{c} \right)$&0.1157&0.1157&0.1157&0.1157&0.1157&0.0043  \\ 
$\text{Br}\left(Z   \to b\overline{b} \right)$&0.1479&0.1479&0.1479&0.1479&0.1479&0.0056  \\ 
\hline 
$\text{Br}\left(Z    \to    e  \mu \right)$ $\times 10^{7}$&0.000&0.000&0.000&0.000&0&3.8  \\ 
$\text{Br}\left(Z    \to    e \tau \right)$ $\times 10^{6}$&0.000&0.000&0.000&0.000&0&5.0  \\ 
$\text{Br}\left(Z    \to  \mu \tau \right)$ $\times 10^{6}$&0.000&0.000&0.000&0.000&0&6.1  \\ 
\hline 
$A_e$&0.1469&0.1469&0.1469&0.1469&0.1469&0.0015  \\ 
$A_\mu$&0.147&0.147&0.147&0.147&0.147&0.015  \\ 
$A_\tau$&0.1469&0.1469&0.1469&0.1469&0.1469&0.0015  \\ 
$A_s$&0.941&0.941&0.941&0.941&0.941&0.094  \\ 
$A_c$&0.6949&0.6949&0.6949&0.6949&0.6949&0.0069  \\ 
$A_b$&0.9406&0.9406&0.9406&0.9406&0.9406&0.0094  \\ 
\hline 
$\mu_{\mu\mu}$&0.977&0.977&0.976&0.977&0&1.3  \\ 
$\mu_{\tau\tau}$&0.981&0.981&0.981&0.981&1.12&0.23  \\ 
$\mu_{bb}$&0.843&0.842&0.843&0.842&0.950&0.22  \\ 
$\mu_{\gamma\gamma}$&1.00&1.00&1.00&1.00&1.16&0.18  \\ 
$\text{Br}\left(h    \to  e^+ e^-      \right)$ $\times 10^{4}$&4.8$\times10^{-5}$&4.8$\times10^{-5}$&4.8$\times10^{-5}$&4.8$\times10^{-5}$&0&9.7  \\ 
$\text{Br}\left(h \to e    \mu \right)$ $\times 10^{4}$&0.000&0.000&0.000&0.000&0&1.8  \\ 
$\text{Br}\left(h \to e  \tau \right)$ $\times 10^{3}$&0.000&0.000&0.000&0.000&0&3.1  \\ 
$\text{Br}\left(h \to \mu \tau \right)$ $\times 10^{3}$&0.000&0.000&0.000&0.000&0&1.3  \\ 
\hline
\end{tabular}  
\end{table}    
 
\begin{table}[th] 
\centering 
\footnotesize 
\caption{SM quark masses and CKM matrix at the benchmark points.} 
\begin{tabular}{c|cccc|cc}\hline 
Name &  Point (A) & Point (B) & Point (C) & Point (D) &Data & Unc. \\  \hline\hline 
$m_u(m_Z)$ [GeV] $\times 10^{3}$&1.27&1.29&1.32&1.22&1.29&0.39  \\ 
$m_c(m_Z)$ [GeV]&0.627&0.627&0.627&0.626&0.627&0.019  \\ 
$m_t(m_Z)$ [GeV]&171.64&171.71&171.88&171.57&171.68&1.5  \\ 
$m_d(m_Z)$ [GeV] $\times 10^{3}$&2.76&2.76&2.76&2.72&2.75&0.29  \\ 
$m_s(m_Z)$ [GeV] $\times 10^{3}$&54.30&54.47&54.14&55.11&54.32&2.9  \\ 
$m_b(m_Z)$ [GeV]&2.86&2.85&2.86&2.85&2.85&0.026  \\ 
\hline 
$\left|V_{ud}\right|$&0.97447&0.97447&0.97446&0.97447&0.97420&0.00021  \\ 
$\left|V_{us}\right|$&0.22450&0.22451&0.22451&0.22450&0.22430&0.00050  \\ 
$\left|V_{ub}\right|$ $\times 10^{3}$&3.60&3.58&3.61&3.58&3.94&0.36  \\ 
$\left|V_{cd}\right|$&0.2244&0.2244&0.2244&0.2244&0.2180&0.0040  \\ 
$\left|V_{cs}\right|$&0.974&0.974&0.974&0.974&0.997&0.017  \\ 
$\left|V_{cb}\right|$ $\times 10^{2}$&4.17&4.14&4.15&4.14&4.22&0.080  \\ 
$\left|V_{td}\right|$ $\times 10^{3}$&8.82&8.75&8.78&8.75&8.10&0.50  \\ 
$\left|V_{ts}\right|$ $\times 10^{2}$&4.09&4.06&4.07&4.06&3.94&0.23  \\ 
$\left|V_{tb}\right|$&0.999&0.999&0.999&0.999&1.02&0.025  \\ 
\hline 
$\alpha$&1.53&1.53&1.53&1.53&1.47&0.097  \\ 
$\sin{2\beta}$&0.694&0.697&0.699&0.697&0.691&0.017  \\ 
$\gamma$&1.23&1.23&1.23&1.23&1.28&0.081  \\ 
\hline
\end{tabular}  
\end{table}    

\begin{table}[th] 
\centering 
\footnotesize 
\caption{Observables for quarks at the benchmark points.} 
\begin{tabular}{c|cccc|cc}\hline 
Name &  Point (A) & Point (B) & Point (C) & Point (D) &Data & Unc. \\  \hline\hline 
$\Delta M_K$ [ps$^{-1}$] $\times 10^{3}$&6.886&5.012&4.633&4.622&5.293&2.2  \\ 
$\epsilon_K$ $\times 10^{3}$&2.23&2.23&2.17&2.22&2.23&0.21  \\ 
\hline 
$\Delta M_{B_d}$ [ps$^{-1}$]&0.561&0.610&0.598&0.590&0.506&0.081  \\ 
$S_{\psi K_s}$&0.697&0.696&0.692&0.695&0.695&0.019  \\ 
\hline 
$\Delta M_{B_s}$ [ps$^{-1}$]&19.61&19.75&19.44&19.95&17.76&2.5  \\ 
$S_{\psi \phi}$ $\times 10^{2}$&3.659&3.742&3.730&3.791&2.100&3.1  \\ 
\hline 
$\left|x^D_{12}\right|$ $\times 10^{3}$&1.7$\times10^{-3}$&5.4$\times10^{-3}$&0.195&0.285&0&5.0  \\ 
\hline 
$R_K^{\nu\nu}$&1.046&1.053&1.060&1.057&1.000&2.6  \\ 
$R_{K^*}^{\nu\nu}$&1.046&1.053&1.060&1.057&1.000&2.7  \\ 
$R_{B_d\to\mu\mu}$&0.985&0.888&0.867&0.971&1.509&1.4  \\ 
$R_{B_s\to\mu\mu}$&0.841&0.890&0.850&0.861&0.750&0.16  \\ 
$\Gamma_t$&1.49&1.49&1.49&1.49&1.41&0.17  \\ 
$\text{Br}\left(t   \to Z q\right)$ $\times 10^{4}$&0.000&0.000&0.000&0.000&0&2.6  \\ 
$\text{Br}\left(t   \to Z u\right)$ $\times 10^{4}$&0.000&0.000&0.000&0.000&0&9.7  \\ 
$\text{Br}\left(t   \to Z c\right)$ $\times 10^{4}$&0.000&0.000&0.000&0.000&0&8.2  \\ 
\hline
\end{tabular}  
\end{table}    
\begin{table}[th] 
\centering 
\footnotesize 
\caption{Wilson coefficients relevant to $b\to s\ell\ell$ processes and $\br{B}{K\tau^+\tau^-}$ at the benchmark points.} 
\begin{tabular}{c|cccc|cc}\hline 
Name &  Point (A) & Point (B) & Point (C) & Point (D) &Data & Unc. \\  \hline\hline 
$\text{Re}C^e_9$&0.000&0.000&0.000&0.000&0&0.10  \\ 
$\text{Im}C^e_9$&0.000&0.000&0.000&0.000&0&0.10  \\ 
$\text{Re}C^e_{10}$&0.000&0.000&0.000&0.000&0&0.10  \\ 
$\text{Im}C^e_{10}$&0.000&0.000&0.000&0.000&0&0.10  \\ 
$\text{Re}C^{'e}_9$&0.000&0.000&0.000&0.000&0&0.10  \\ 
$\text{Im}C^{'e}_9$&0.000&0.000&0.000&0.000&0&0.10  \\ 
$\text{Re}C^{'e}_{10}$&0.000&0.000&0.000&0.000&0&0.10  \\ 
$\text{Im}C^{'e}_{10}$&0.000&0.000&0.000&0.000&0&0.10  \\ 
$\text{Re}C^\mu_9$&-0.548&-0.806&-0.838&-0.808&-0.700&0.30  \\ 
$\text{Im}C^\mu_9$&-1.0$\times10^{-2}$&6.2$\times10^{-4}$&-6.8$\times10^{-3}$&5.4$\times10^{-3}$&0&0.10  \\ 
$\text{Re}C^\mu_{10}$&0.370&0.252&0.347&0.322&0.400&0.20  \\ 
$\text{Im}C^\mu_{10}$&6.9$\times10^{-3}$&-1.9$\times10^{-4}$&2.8$\times10^{-3}$&-2.1$\times10^{-3}$&0&0.10  \\ 
$\text{Re}C^{'\mu}_9$&1.0$\times10^{-3}$&8.9$\times10^{-5}$&2.2$\times10^{-4}$&-2.6$\times10^{-6}$&0&0.10  \\ 
$\text{Im}C^{'\mu}_9$&8.5$\times10^{-4}$&-4.1$\times10^{-5}$&-4.4$\times10^{-5}$&-4.1$\times10^{-6}$&0&0.10  \\ 
$\text{Re}C^{'\mu}_{10}$&-7.1$\times10^{-4}$&-2.8$\times10^{-5}$&-9.2$\times10^{-5}$&1.1$\times10^{-6}$&0&0.10  \\ 
$\text{Im}C^{'\mu}_{10}$&-5.8$\times10^{-4}$&1.3$\times10^{-5}$&1.8$\times10^{-5}$&1.6$\times10^{-6}$&0&0.10  \\ 
$\text{Br}\left(B \to K\tau^+\tau^-\right)$ $\times 10^{3}$&1.2$\times10^{-4}$&1.2$\times10^{-4}$&1.2$\times10^{-4}$&1.2$\times10^{-4}$&0&1.4  \\ 
\hline
\end{tabular}  
\end{table}    
\clearpage

\begin{table}[hp]
\centering
\caption{
\label{tab-massdecayA}
Masses, total widths, and branching fractions (BR) at point (A).
Decay~1(2) denote the (next to) dominant decay mode.}
\begin{tabular}[t]{c|cc|cccc} \hline
              & Mass [GeV] & Width [GeV]       & Decay~1 & BR          & Decay~2   & BR \\ \hline\hline
 $Z'$ & 277.6 & 0.1361 & $\mu\mu$ & 0.5091 & $\nu$ $\nu$ & 0.4907 \\
 $\chi$ & 651.9 & 0.669538 & $E_1\mu$ & 0.4391 & $N_1\nu$ & 0.4227 \\
 $\sigma$ & 1871.7 & 0.9049 & $N_2 N_2$ & 0.2988 & $E_2 E_2$ & 0.1473 \\
 $E_1$ & 367.9 & 0.0354639 & $Z'\mu$ & 1. & $h\mu$ & 0. \\
 $N_1$ & 422.2 & 0.0817534 & $Z'\nu$ & 0.9995 & $W\mu$ & 0.0003 \\
 $N_2$ & 459. & 0.113389 & $W E_1$ & 0.8792 & $Z' \nu$ & 0.1179 \\
 $E_2$ & 548.3 & 4.07452 & $WN_1$ & 0.4799 & $Z E_1$ & 0.4415 \\
 $D_1$ & 1572.1 & 0.0371 & $Z'b$ & 0.4117 & $\chi b$ & 0.2831 \\
 $U_1$ & 3453.7 & 3.0221 & $Z'c$ & 0.4117 & $\chi c$ & 0.3829 \\
 $D_2$ & 3453.8 & 3.0228 & $Z's$ & 0.4063 & $\chi s$ & 0.3779 \\
 $U_2$ & 3596.8 & 0.1085 & $Z'c$ & 0.4504 & $\chi c$ & 0.4213 \\ \hline 
 \end{tabular}
\vspace{3.0cm}
\caption{
\label{tab-massdecayB}
Masses, total widths, and branching fractions (BR) at point (B).
Decay~1(2) denote the (next to) dominant decay mode.}
\begin{tabular}[t]{c|cc|cccc} \hline
              & Mass [GeV] & Width [GeV]  & Decay~1 & BR           & Decay~2 & BR \\ \hline\hline
 $Z'$ & 535.3 & 0.5097 & $\mu\mu $ & 0.5595 & $\nu\nu$ & 0.4388 \\
 $\chi$ & 88.2 & 1.6$\times10^{-8}$ & $\mu\mu$ & 0.6045 & $bb$ & 0.36 \\
 $\sigma$ & 103.3 & 2.3$\times10^{-9}$ & $\mu\mu $ & 0.6041 & $bb$ & 0.3604 \\
 $N_1$ & 1069.9 & 0.000572547 & $W\mu$ & 0.4724 & $Z\nu$ & 0.2362 \\
 $E_1$ & 1211.1 & 3.27601 & $\chi\mu$ & 0.5235 & $Z'\mu$ & 0.4765 \\
 $N_2$ & 1283.3 & 4.05237 & $\chi \nu$ & 0.5185 & $Z'\nu$ & 0.4814 \\
 $E_2$ & 1386.9 & 8.53551 & $Z E_1$ & 0.2959 & $\chi\mu$ & 0.2563 \\
 $U_1$ & 1853. & 0.0013 & $\chi c$ & 0.4647 & $Z'c$ & 0.4576 \\
 $D_1$ & 2130.9 & 0.0223 & $\chi d$ & 0.4696 & $Z'd$ & 0.4659 \\
 $D_2$ & 2747.9 & 2.8004 & $\chi b$ & 0.3747 & $Z'b$ & 0.3739 \\
 $U_2$ & 2747.9 & 2.8081 & $\chi t$ & 0.3827 & $Z't$ & 0.3818 \\
\hline
 \end{tabular}
\end{table}
\begin{table}[hp]
\centering
\caption{
\label{tab-massdecayC}
Masses, total widths, and branching fractions (BR) at best fit point C.
Decay~1(2) denote the (next to) dominant decay mode.}
\begin{tabular}[t]{c|cc|cccc}\hline
                & Mass [GeV] & Width [GeV]    & Decay~1   & BR        & Decay~2   & BR \\ \hline\hline
 $\Zp$ & 486.7 & 1.1178 & $\mu\mu $ & 0.5335 & \text{$\nu\nu$} & 0.4552 \\
 $\chi$ & 1066.6 & 1.587 & $E_1\mu $ & 0.3284 & \text{$E_2\mu$} & 0.2668 \\
 $\sigma$ & 4088.7 & 0.8017 & \text{$E_2\mu $} & 0.2006 & \text{$E_1 E_1$} & 0.1768 \\
 $E_1$ & 441.7 & 6.613$\times10^{-6}$ & \text{$h\mu $} & 0.6816 & \text{$Z\mu $} & 0.2172 \\
 $N_1$ & 537.1 & 0.851545 & \text{$WE_1$} & 0.9924 & \text{$\Zp\nu$} & 0.0076 \\
 $E_2$ & 618. & 2.17089 & \text{$ZE_1$} & 0.8738 & \text{$WN_1$} & 0.0772 \\
 $N_2$ & 4980.2 & 0.000430776 & \text{$W\mu $} & 0.4334 & \text{$Z\nu$} & 0.2167 \\
 $D_1$ & 3725.6 & 4.0242 & \text{$\Zp b$} & 0.46 & \text{$\chi b$} & 0.3881 \\
 $U_1$ & 3725.7 & 4.0254 & \text{$\Zp t$} & 0.4705 & \text{$\chi t$} & 0.3967 \\
 $D_2$ & 4803.7 & 0.0086 & \text{$\Zp b$} & 0.4274 & \text{$\chi b$} & 0.3864 \\
 $U_2$ & 4981. & 0.3388 & \text{$\Zp c$} & 0.4381 & \text{$\chi c$} & 0.3989 \\
\hline
 \end{tabular}
\vspace{3.0cm}
\caption{
\label{tab-massdecayD}
Masses, total widths, and branching fractions (BR) at best fit point D.
Decay~1(2) denote the (next to) dominant decay mode.} 
\begin{tabular}[t]{c|cc|cccc}\hline
                 & Mass [GeV] & Width [GeV]    & Decay~1 & BR            & Decay~2    & BR \\ \hline\hline
 $\Zp$ & 758.7 & 1.481 & \text{$\mu\mu $} & 0.448 & \text{$\nu\nu$} & 0.3779 \\
 $\chi$ & 145.4 & 6.0$\times 10^{-9}$ & \text{$b b$} & 0.5551 & \text{$\mu\mu $} & 0.2299 \\
 $\sigma$ & 4820.9 & 2.8529 & \text{$E_2 E_2$} & 0.2771 & \text{$E_2\mu$} & 0.1646 \\
 $E_1$ & 561.3 & 0.0765212 & \text{$\chi\mu $} & 1. & \text{$h\mu $} & 0. \\
 $N_1$ & 593.6 & 0.0849371 & \text{$\chi\nu$} & 1. & \text{$W\mu $} & 0. \\
 $E_2$ & 798.8 & 8.70794 & \text{$WN_1$} & 0.526 & \text{$ZE_1$} & 0.3138 \\
 $N_2$ & 4820.9 & 0.00162908 & \text{$WE_1$} & 0.3378 & \text{$ZN_1$} & 0.2083 \\
 $D_1$ & 2442.6 & 0.0001 & \text{$\chi b$} & 0.46 & \text{$\Zp b$} & 0.4511 \\
 $U_1$ & 4061.7 & 4.3534 & \text{$\chi t$} & 0.3504 & \text{$\Zp t$} & 0.3501 \\
 $D_2$ & 4068.2 & 4.4075 & \text{$\chi b$} & 0.3372 & \text{$\Zp b$} & 0.3368 \\
 $U_2$ & 4517.1 & 14.0401 & \text{$WD_2$} & 0.4354 & \text{$ZU_1$} & 0.2226 \\
\hline
 \end{tabular}
\end{table}

\clearpage
{\small
\bibliographystyle{JHEP}
\bibliography{reference_vectorlike}
}
\end{document}